\documentclass{bmvc2k}

\title{Unsupervised Image Denoising with Frequency Domain Knowledge}

\addauthor{Nahyun Kim*}{nhkim21@kaist.ac.kr}{0}
\addauthor{Donggon Jang*}{jdg900@kaist.ac.kr}{0}
\addauthor{Sunhyeok Lee}{sunhyeok.lee@kaist.ac.kr}{0}
\addauthor{Bomi Kim}{by5747@kaist.ac.kr}{0}
\addauthor{Dae-Shik Kim}{daeshik@kaist.ac.kr}{0}

\addinstitution{
Korea Advanced Institute of Science \\
and Technology (KAIST),\\
Daejeon, Korea
}

\runninghead{N. KIM, D. JANG ET AL}{Image Denoising with Frequency Domain Knowledge}

\def\etal{\emph{et al}\bmvaOneDot}

\begin{document}

\maketitle

\begin{abstract}
Supervised learning-based methods yield robust denoising results, yet they are inherently limited by the need for large-scale clean/noisy paired datasets. The use of unsupervised denoisers, on the other hand, necessitates a more detailed understanding of the underlying image statistics. In particular, it is well known that apparent differences between clean and noisy images are most prominent on high-frequency bands, justifying the use of low-pass filters as part of conventional image preprocessing steps. However, most learning-based denoising methods utilize only one-sided information from the spatial domain without considering frequency domain information. To address this limitation, in this study we propose a frequency-sensitive unsupervised denoising method. To this end,  a generative adversarial network (GAN) is used as a base structure. Subsequently, we include spectral discriminator and frequency reconstruction loss to transfer frequency knowledge into the generator. Results using natural and synthetic datasets indicate that our unsupervised learning method augmented with frequency information achieves state-of-the-art denoising performance, suggesting that frequency domain information could be a viable factor in improving the overall performance of unsupervised learning-based methods.
\end{abstract}


\section{Introduction}
\label{intro}
Based on clean and noisy image pairs, supervised learning-based image denoisers have shown impressive performance compared to prior-based approaches. A large number of high-quality image pairs play an important role in the performance of supervised learning-based methods. However, constructing large-scale paired datasets may be unavailable or expensive in real-world situations. For this reason, image denoising methods that do not require clean and noisy image pairs have recently drawn attention. 

\begin{figure*}[t]
    \centering
    \subfigure[]{\includegraphics[width=.49\textwidth]{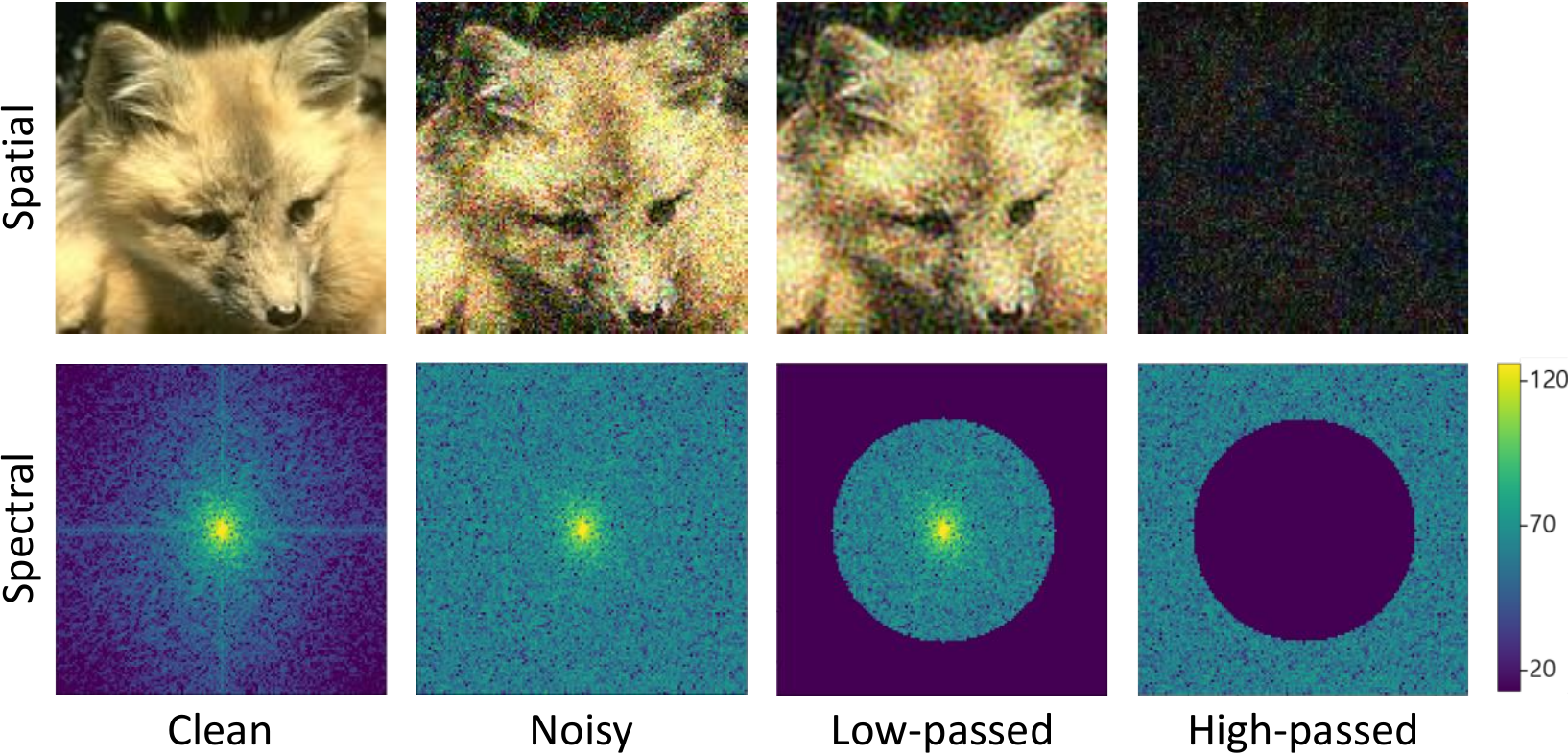}}
    \subfigure[]{\includegraphics[width=.4\textwidth]{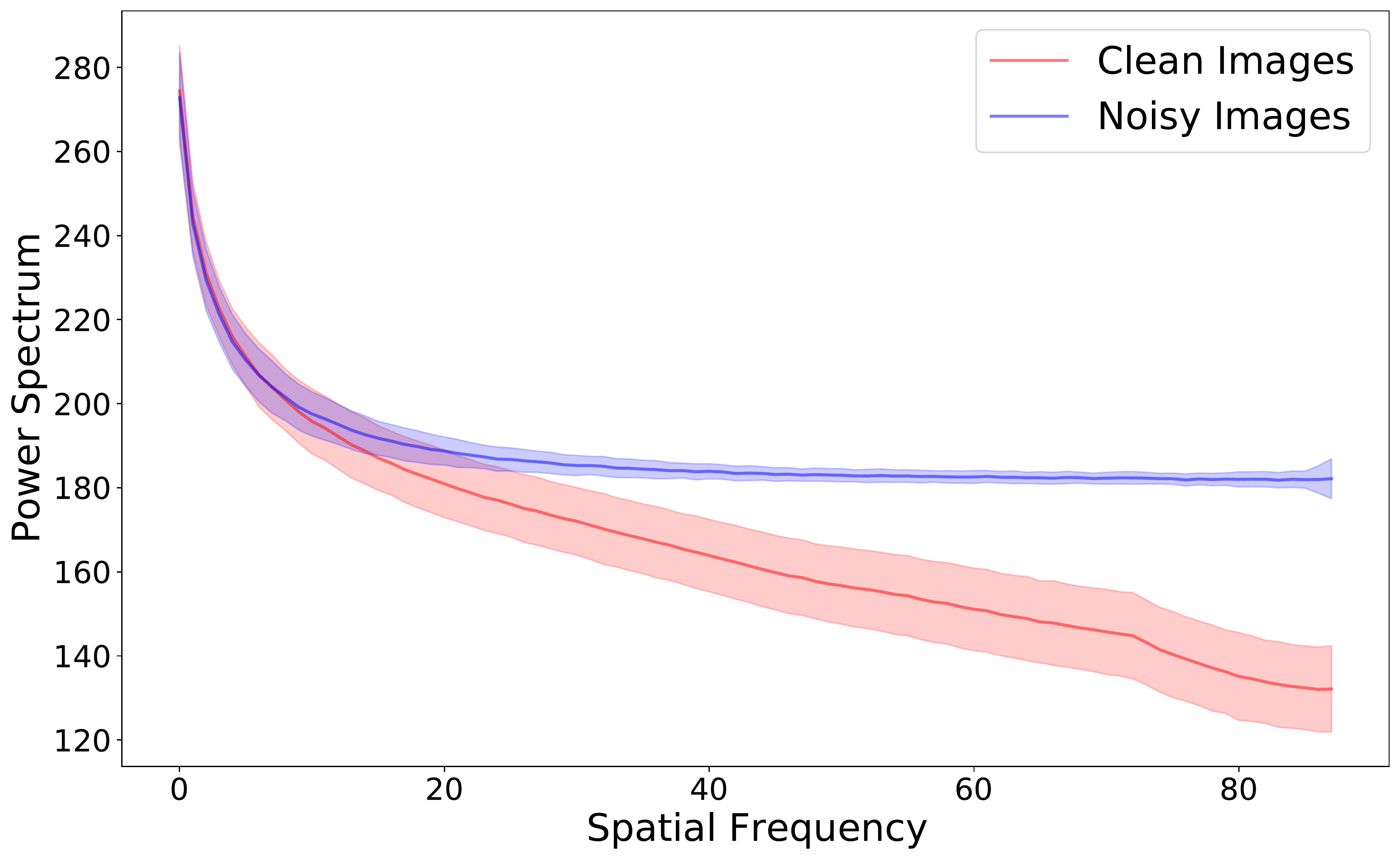}}
    \caption{The spectrum analysis in the frequency domain. (a) Visualization of images in the spatial domain and corresponding spectrum maps in the frequency domain. (b) The statistics (mean and variance) after azimuthal integral over the power spectrum on clean and noisy images of \textit{CBSD68}. We use AWGN with a noise level $\sigma=50$ to yield noisy images.}
    \label{fig:filtering}
\end{figure*}

A noisy image $x$ is usually modeled as the sum of clean background $y$ and noise $n$: $x = y+n$. Subsequently, noise corrupts the benign pixels, which makes it hard to distinguish the pixels of noise and content in the spatial domain. However, in the frequency domain, noise and content can be easily identified. As shown in Figure \ref{fig:filtering} (a), we observe that the noise lies in the high-frequency bands and semantic information lies in the low-frequency bands. Furthermore, in Figure \ref{fig:filtering} (b), we note that apparent differences between clean and noisy images are most prominent on high-frequency bands. It may indicate that the frequency domain provides useful evidence for noise removal. However, the recent learning-based denoisers overlook the frequency domain information and use only one-sided information from the spatial domain.

Motivated by these observations, we propose the unsupervised denoising method that reflects frequency domain information. Specifically, with a generative adversarial network as a base structure, we introduce the spectral discriminator and frequency reconstruction loss to transfer frequency knowledge to the generator. The spectral discriminator distinguishes the differences between denoised and clean images on high-frequency bands. By propagating this knowledge to the generator for noise removal, the generator considers the frequency domain and thus produces visually more plausible denoised images to fool the spectral discriminator. The frequency reconstruction loss, combined with the cycle consistency loss, improves the image quality and preserves the content of images while narrowing the gap between clean and denoised images in the frequency domain.

The main contributions of our method are summarized as follows: 1) We propose the GAN-based unsupervised image denoising method that preserves semantic information and produces a high-quality noise-free image. 2) To the best of our knowledge, it is the first approach to explore the potential of the frequency domain with Fourier transform in the field of noise removal tasks. The proposed spectral discriminator and frequency reconstruction loss make the generator concentrate on the noise and produce satisfying results. Denoised images recovered by our method are close to clean reference images in both spatial and frequency domain. 3) The proposed method outperforms existing unsupervised image denoisers by a considerable margin. Moreover, our performance is even comparable with supervised learning-based approaches trained with paired datasets.

\section{Related Work}
\subsection{Image Denoising}
Non-learning based image denoisers \cite{bm3d,nlm,wnnm1,wnnm2,osher2005iterative,xu2007iterative,aharon2006k,mairal2009non,zoran2011learning,xu2018trilateral} have tried to reconstruct clean images using pre-defined priors which model the distribution of noise. Specifically, a widely used prior in image denoising is non-local self-similarity prior \cite{bm3d,nlm,wnnm1,wnnm2}. Assuming that similar patches exist in a single image, the methods based on non-local self-similarity \cite{bm3d,nlm} remove the noise using these patches.

Recently, with the advent of deep neural networks, supervised learning-based image denoisers \cite{dncnn,zhang2018ffdnet,red30} show promising performance on a set of clean and noisy image pairs. However, it is challenging to construct clean and noisy image pairs in a real-world scenario. To address the above issues, denoisers that do not rely on clean and noisy image pairs have been proposed \cite{lehtinen2018noise2noise,krull2019noise2void,gcbd,lir,ulyanov2018deep}. N2N \cite{lehtinen2018noise2noise} learns reconstruction using only noisy image pairs without ground-truth clean images. N2V \cite{krull2019noise2void} estimates a corrupted pixel from its neighboring pixels based on a blind-spot mechanism. GCBD \cite{gcbd} generates the noisy images while modeling the real-world noise distribution through the GAN \cite{goodfellow2014generative} and trains the denoiser with pseudo clean and noisy image pairs. LIR \cite{lir} trains an image denoiser by disentangling invariant representations from noisy images with an unpaired dataset.

\subsection{Frequency Domain in CNNs}
In traditional image processing, analyzing images in the frequency domain is known to be effective by transforming the image from the spatial domain to the frequency domain. Inspired by this idea, several works attempt to utilize the information from the frequency domain in deep neural networks. Xu \etal \cite{lfd} accelerate the training of neural networks utilizing the discrete cosine transform. Dzanic \etal \cite{fsd} observe that discrepancy exists between the images generated by the GAN \cite{goodfellow2014generative} and the real images through the analysis of high-frequency Fourier modes. In addition, attempts to utilize the frequency domain information in the various fields, including image forensics \cite{durall2020watch,frank2020leveraging,zhang2019detecting}, image generation \cite{chen2020ssd,cai2020frequency,jiang2020focal}, and domain adaptation \cite{yang2020phase,yang2020fda} are gradually increasing. However, image denoising methods combining the frequency domain analysis with DNN remain much less explored.

\begin{figure}[t!]
    \centering
    \includegraphics[width=0.9\textwidth]{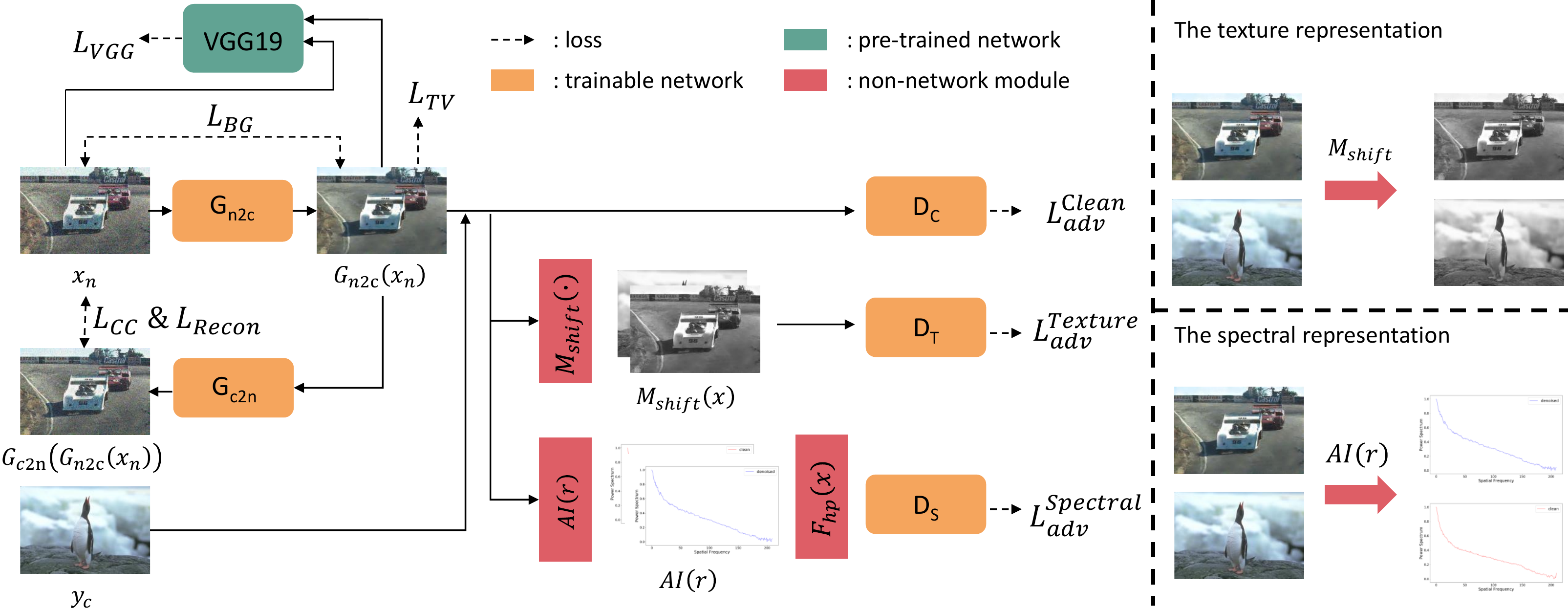}
    
    \caption{An overview of the proposed framework. Given an unpaired clean $y_{c}$ and noisy image $x_{n}$, the generator $G_{n2c}$ for image denoising takes the noisy image $x_{n}$ as an input and learns the mapping for noise removal. Additional network $G_{c2n}$ is used to impose the cycle consistency. Three discriminators $D_{C}$, $D_{T}$, and $D_{S}$ try to distinguish the denoised image $G_{n2c}(x_{n})$ from real clean image $x_{c}$ in terms of both spatial domain and frequency domain. The whole framework is end-to-end trainable.}
    \label{fig:network}
\end{figure}

\section{Method}
In this section, we first introduce the spectral discriminator and frequency reconstruction loss that use information from the frequency domain. Then, we present an unsupervised framework for image denoising, integrating the proposed discriminator and loss with the GAN. The proposed framework is illustrated in Figure \ref{fig:network}.

\subsection{Frequency Domain Constraints}\label{section3.1}
\paragraph{Spectral Discriminator}
The simple way for the generator to consider the frequency domain is that the discriminator transfers the frequency domain knowledge to the generator. To this end, we propose the spectral discriminator similar to that introduced by \cite{chen2020ssd} to measure spectral realness. We compute the discrete Fourier transform on 2D image data $f(w,h)$ in size $W \times H$ to feed frequency representations to the discriminator.
\begin{equation}
    F(k,l) = \sum^{W-1}_{w=0}\sum^{H-1}_{h=0}f(w,h)e^{-2{\pi}i\frac{kw}{W}}e^{-2{\pi}i\frac{lh}{H}}
\end{equation}
for spectral coordinates $k=0,...,W-1$ and $l=0,...,H-1$. 

Recent studies \cite{chen2020ssd,durall2020watch} show that the 1D representation of the Fourier power spectrum is sufficient to highlight spectral differences. Following their works, we transform the result of Fourier transform to polar coordinate and compute azimuthal integration over $\theta$.
\begin{equation}
    F(r, \theta) = F(k,l) : r = \sqrt{k^2 + l^2}, \quad \theta = \arctan{\frac{l}{k}}, \quad AI(r) = \frac{1}{2\pi}\int^{2\pi}_{0}{|F(r, \theta)|d\theta}
\end{equation}
where $AI(r)$ means the average intensity of the image signal about radial distance $r$.

We propose the spectral discriminator that allows the generator to focus on noise using high-frequency spectral information. To learn the differences on high-frequency bands, we pass the 1D spectral vector into the high-pass filter $F_{hp}$ and input it to the spectral discriminator.
\begin{equation}\label{eq3}
    v_{I} = F_{hp}(AI(r)), \quad F_{hp}(x) = \begin{cases} x, \quad r>r_{\tau}, \\ 0, \quad otherwise \end{cases}
\end{equation}
where $r_{\tau}$ is a threshold radius for high-pass filtering and $v_{I}$ is a high-pass filtered 1D spectral vector of an input $I$.

Generally, the most distinct characteristics between clean and noisy images exist on high-frequency bands. Thus, if there is some remained noise on denoised images, the spectral discriminator easily distinguishes the difference between the clean and denoised images on high-frequency bands. By transferring this knowledge to the generator, the generator for noise removal learns to yield visually more plausible images to fool the spectral discriminator.
\paragraph{Frequency Reconstruction Loss}
Cai \etal \cite{cai2020frequency} demonstrate the existence of a gap between the real and generated image in the frequency domain, which leads to artifacts in the spatial domain. Motivated by this observation, we propose to use frequency reconstruction loss with cycle consistency loss to ameliorate the quality of denoised images while reducing the gap. We aim that the frequency reconstruction loss which is complementary to cycle consistency loss enables the generator to consider the frequency domain. Furthermore, we expect that it can serve as an assistant in generating high-quality denoised images. To compute the frequency reconstruction loss, we map an input $x_{n}$ and reconstructed image $G_{c2n}(G_{n2c}(x_{n}))$ to the frequency domain using Fourier transform. Then, we calculate the frequency reconstruction loss by measuring the difference between the two results of the Fourier transform and taking a logarithm to normalize it. Finally, we minimize the following objective:
\begin{equation}\label{eq4}
    L_{Freq} = log(1 + \frac{1}{WH}\sum^{W-1}_{k=0}\sum^{H-1}_{l=0}|F_{x_{n}}(k,l) - F_{G_{c2n}(G_{n2c}(x_{n}))}(k,l)|)
\end{equation}

\subsection{Unsupervised Framework for Image Denoising}\label{section3.2}
Our goal is to learn a mapping from a noise domain $X_{N}$ to a clean domain $Y_{C}$ given unpaired training images $x_{n} \in X_{N}$ and $y_{c} \in Y_{C}$. To learn this mapping, we use the CycleGAN-like framework consisting of two generators, $G_{n2c}$ and $G_{c2n}$, and three discriminators, $D_{C}$, $D_{T}$, and $D_{S}$. Given a noisy image $x_{n}$, the generator $G_{n2c}$ learns to generate a denoised image $G_{n2c}(x_{n})$. While distinguishing the denoised image $G_{n2c}(x_{n})$ from the real clean image $y_{c}$, the discriminator $D_{C}$ makes the generator produce the denoised images closer to the real clean domain $Y_{C}$. To stablize training, we use the Least Squares GAN (LSGAN) loss \cite{lsgan} for adversarial loss. The LSGAN loss for $G_{n2c}$ and $D_{C}$ is:
\begin{equation}\label{eq5}
    L^{Clean}_{adv} = E_{y_{c} \sim P_{c}}[(D_{C}(y_{c}))^2] + E_{x_{n} \sim P_{n}}[(1 - D_{C}(G_{n2c}(x_{n})))^2]
\end{equation}
where $P_{n}$ and $P_{c}$ are the data distributions of the domain $X_{N}$ and domain $Y_{C}$, respectively. 

As introduced in \cite{wang2020learning}, we adopt the texture discriminator $D_T$ in order to guide the generator to produce clean contour and preserve texture while removing the noise. Following the scheme of \cite{wang2020learning}, a random color shift algorithm $M_{shift}$ is applied to the denoised image $G_{n2c}(x_{n})$. The texture loss for $G_{n2c}$ and $D_{T}$ is:
\begin{equation}
    L^{Texture}_{adv} = E_{y_{c} \sim P_{c}}[(D_{T}(M_{shift}(y_{c})))^2] + E_{x_{n} \sim P_{n}}[(1 - D_{T}(M_{shift}(G_{n2c}(x_{n}))))^2]
\end{equation}
As discussed in Section \ref{section3.1}, we use the spectral discriminator $D_{S}$ to guide the generator to generate more realistic images by reducing the gap between the clean and denoised image in the frequency domain. The spectral loss for $G_{n2c}$ and $D_{S}$ is:
\begin{equation}
    L^{Spectral}_{adv} = E_{y_{c} \sim P_{c}}[(D_{S}(v_{y_{c}}))^2] + E_{x_{n} \sim P_{n}}[(1 - D_{S}(v_{G_{n2c}(x_{n})}))^2]
\end{equation}
where $v$ denotes the high-pass filtered 1D spectral vector in Eq. \ref{eq3}.

CycleGAN \cite{zhu2017cyclegan} imposes the two-sided cycle consistency constraint to learn the one-to-one mappings between two domains. On the other hand, we use only one-sided cycle consistency to maintain the content between noisy and denoised images. By incorporating a network $G_{c2n}$, we let $G_{c2n}(G_{n2c}(x_n))$ be identical to the noisy image $x_{n}$. The cycle consistency loss is expressed as:
\begin{equation}
    L_{CC} = || x_{n} - G_{c2n}(G_{n2c}(x_{n})) ||_{1}
\end{equation}
where $||\cdot||_{1}$ is the L1 norm.

Furthermore, we add the reconstruction loss between the $G_{c2n}(G_{n2c}(x_n))$ and $x_n$ to stabilize the training. We employ the negative SSIM loss \cite{ssim} and combine it with the frequency reconstruction loss $L_{Freq}$ in Eq. \ref{eq4}. The reconstruction loss is expressed as:
\begin{equation}
    L_{Recon} = L_{Freq}(x_{n}, G_{c2n}(G_{n2c}(x_{n}))) + L_{SSIM}(x_{n}, G_{c2n}(G_{n2c}(x_{n})))
\end{equation}
where $L_{SSIM}(a, b)$ denotes the negative SSIM loss, $-SSIM(a,b)$.

To impose the local smoothness and mitigate the artifacts in the restored image, we adopt the total variation loss \cite{totalvariation}. The total variation loss is expressed as:
\begin{equation}
    L_{TV} = \sum_{w,h}(||{\bigtriangledown_{w}G_{n2c}(x_{n})}||_{2}+ {||\bigtriangledown_{h}G_{n2c}(x_{n})}||_{2})
\end{equation}
where $||\cdot||_{2}$ denotes the L2 norm, $\bigtriangledown_{w}$ and $\bigtriangledown_{h}$ are the operations to compute the gradients in terms of horizontal and vertical directions, respectively.

Inspired by \cite{lir, wang2020learning}, we use the perceptual loss \cite{johnson2016perceptual} to ensure that extracted features from the noisy and denoised image are semantically invariant. This allows the image to keep its semantics even after the noise has been removed. The perceptual loss is expressed as: 
\begin{equation}
    L_{VGG} = ||\phi_{l}(x_{n}) - \phi_{l}(G_{n2c}(x_{n}))||_{2}
\end{equation}
where $\phi_{l}(\cdot)$ denotes the pre-trained VGG-19 \cite{vgg} on ImageNet \cite{deng2009imagenet}, $l$ denotes $l\textrm{th}$ layer from VGG-19, and we use the \textit{Conv5-4} layer of VGG-19 model in our experiments.

Moreover, we employ the background loss to preserve background consistency between the noisy and denoised image. The background loss constrains the L1 norm between blurred results of the noisy and denoised image. As a blur operator, we adopt a guided filter \cite{guided} that smooths the image while preserving the sharpness such as edges and details. The background loss is expressed as:
\begin{equation}
    L_{BG} = ||GF(x_{n}) - GF(G_{n2c}(x_{n}))||_{1}
\end{equation}
where $GF(\cdot)$ denotes the guided filter.

Our full objective for the two generators and the three discriminators is expressed as:
\begin{equation}
\begin{split}
    \min_{G_{n2c}, G_{c2n}}\max_{D_{C}, D_{T}, D_{S}} L^{Clean}_{adv} + L^{Texture}_{adv} + L^{Spectral}_{adv} + L_{CC} + \\ \lambda_{VGG}L_{VGG} + \lambda_{BG}L_{BG} + \lambda_{TV}L_{TV} + \lambda_{Recon}L_{Recon}
\end{split}
\end{equation}
We empirically define the weights in the full objective as: $\lambda_{VGG}=2$, $\lambda_{BG}=2$ , $\lambda_{TV}=0.2$, and $\lambda_{Recon}=0.2$. 

\section{Experiment}
In this section, we provide the implementation details of the proposed method. Then, we present extensive experiments on synthetic and real-world noisy images. Lastly, we conduct an ablation study to show the effectiveness of the proposed method. For synthetic noise, we use Additive White Gaussian Noise (AWGN) to synthesize the noisy images. We adopt the CBSD68 \cite{martin2001database} for evaluation. For real noise, we use the Low-Dose Computed Tomography dataset \cite{moen2021low} and real photographs SIDD \cite{abdelhamed2018sidd} to demonstrate the generalization capacity of the proposed method. We employ PSNR and SSIM \cite{ssim} to evaluate the results. 

\begin{figure*}[t]
    \centering
    \subfigure[Input]{\includegraphics[width=.16\textwidth]{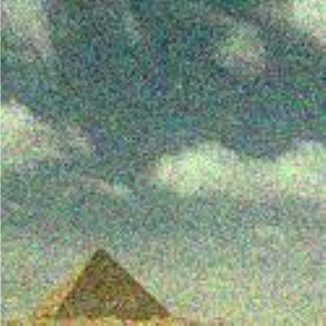}}
    \subfigure[LPF]{\includegraphics[width=.16\textwidth]{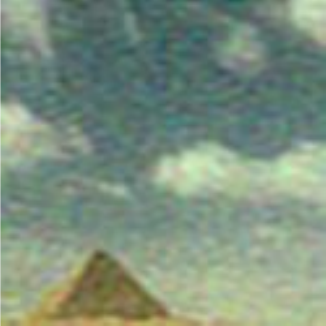}}
    \subfigure[CBM3D]{\includegraphics[width=.16\textwidth]{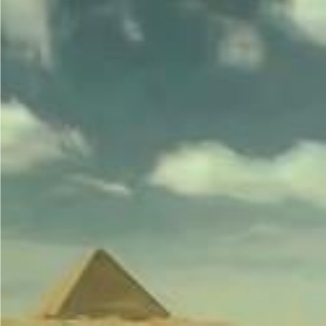}}
    \subfigure[DnCNN]{\includegraphics[width=.16\textwidth]{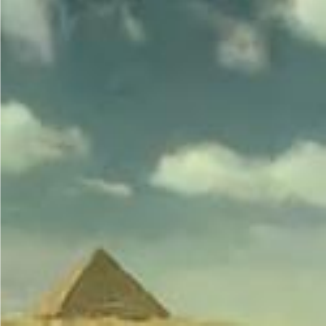}}
    \subfigure[FFDNet]{\includegraphics[width=.16\textwidth]{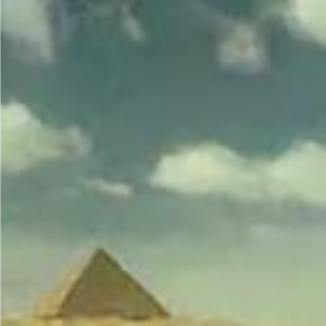}}
    \subfigure[RedNet-30]{\includegraphics[width=.16\textwidth]{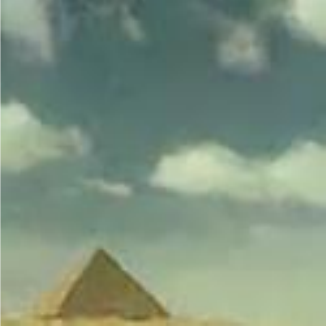}}
    \subfigure[N2N]{\includegraphics[width=.16\textwidth]{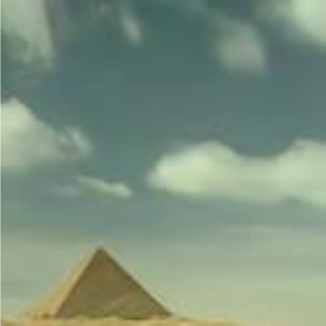}}
    \subfigure[DIP]{\includegraphics[width=.16\textwidth]{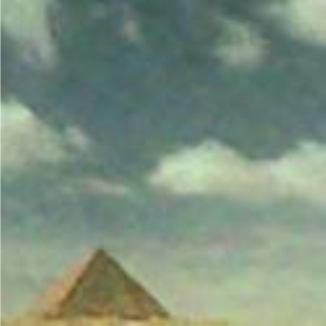}}
    \subfigure[N2V]{\includegraphics[width=.16\textwidth]{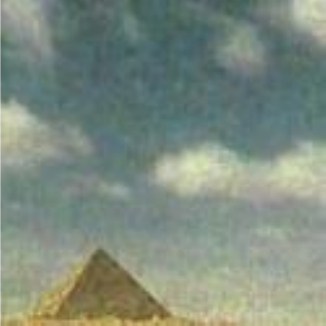}}
    \subfigure[LIR]{\includegraphics[width=.16\textwidth]{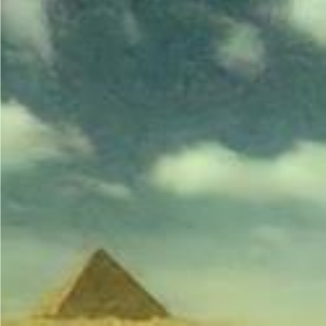}}
    \subfigure[Ours]{\includegraphics[width=.16\textwidth]{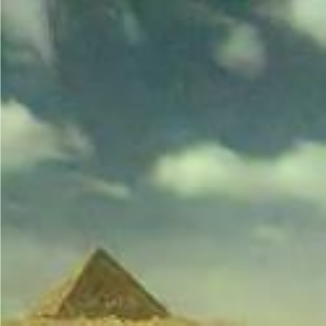}}
    \subfigure[GT]{\includegraphics[width=.16\textwidth]{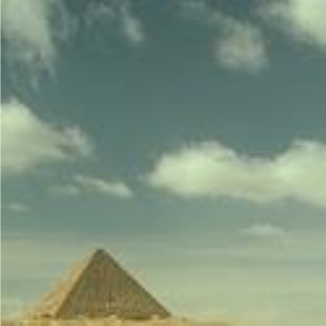}}

    \caption{Qualitative results of our method and other baselines on \textit{CBSD68} corrupted by AWGN with a noise level $\sigma=25$.}
    \label{fig:CBSD68}
\end{figure*}

\subsection{Implementation Details}
We implement our method with Pytorch \cite{paszke2017automatic}. The generator and discriminator architectures are detailed in the supplementary material. We train our method up to 100 epochs on Nvidia TITAN RTX GPU and RTX A6000 in experiments. We adopt ADAM \cite{kingma2014adam} for optimization. The initial learning rate is set to 0.0001, and we keep the same learning rate for the first 70 epochs and linearly decay the rate to zero over the last 30 epochs. We set the batch size to 16 in all experiments. We randomly crop $128 \times 128$ patches for synthetic noise removal and use input patches of size $256 \times 256$ for real-world noise removal. We randomly flip the images horizontally for data augmentation. For high-pass filter on spectral discriminator, $r_{\tau}$ is set to $\lfloor H/2\sqrt{2} \rfloor$ where $H$ is the height of an image and $\lfloor{\:} \rfloor$ is a floor operator. Loss weights are described in Section \ref{section3.2}. Our model is evaluated with three random seeds, and we report its average values for rigorous evaluation.

\begin{table}[t]
\begin{center}
\resizebox{\textwidth}{!}
{%
\begin{tabular}{|c|cc|cccc|cccc|}
\hline
   & \multicolumn{2}{c|}{Traditional} & \multicolumn{4}{c|}{Paired setting} &\multicolumn{4}{c|}{Unpaired setting} \\ \hline
\hline
Methods & LPF & CBM3D \cite{bm3d} & DnCNN \cite{dncnn} & FFDNet \cite{zhang2018ffdnet} & RedNet-30 \cite{red30} & N2N \cite{lehtinen2018noise2noise} & DIP \cite{ulyanov2018deep} & N2V \cite{krull2019noise2void} & LIR \cite{lir} & Ours \\ \hline \hline
Noise level   & \multicolumn{10}{c|}{PSNR (dB)} \\ \hline
$\sigma=15$ &25.93 &33.55 &33.72 &29.68 &33.60 &33.92 &28.51 &28.66 &30.44 &\textbf{32.21} \\
$\sigma=25$ &24.61 &30.91 &30.85 &28.71 &30.68 &31.31 &27.26 &27.20 &29.08 &\textbf{29.37}  \\
$\sigma=50$ &21.49 &27.47 &27.19 &26.79 &26.42 &28.10 &23.66 &24.52 &25.69 &\textbf{26.03}  \\ \hline
Noise level  & \multicolumn{10}{c|}{SSIM} \\ \hline
$\sigma=15$ &0.7079 &0.9619 &0.9254 &0.8616 &0.9620 &0.9301 &0.8851 &0.9024 &0.9414 &\textbf{0.9502} \\
$\sigma=25$ &0.6102 &0.9331 &0.8724 &0.8254 &0.9308 &0.8857 &0.8613 &0.8684 &0.9126 &\textbf{0.9124} \\
$\sigma=50$ &0.4266 &0.8722 &0.7490 &0.7463 &0.8502 &0.7973 &0.7510 &0.7927 &0.8435 &\textbf{0.8375}  \\ \hline
\end{tabular}}
\end{center}
\caption{The average PSNR and SSIM results of our method and other baselines on \textit{CBSD68} corrupted by AWGN with noise levels $\sigma=\{15, 25, 50\}$. Our results are marked in \textbf{bold}.}
\label{table1}
\end{table}

\subsection{Synthetic Noise Removal}
We train the model with DIV2K \cite{martin2001database} that contains 800 images with 2K resolution. For the unpaired training, we randomly divide the dataset into two parts without intersection. To construct a noise set, we add the AWGN with noise levels $\sigma = \{15, 25, 50\}$ to images in one part using the other part as a clean set. For a fair comparison, we use only the noise set and their corresponding ground-truth when training other supervised learning-based methods. We select unsupervised methods, i.e. DIP \cite{ulyanov2018deep}, N2N \cite{lehtinen2018noise2noise}, N2V \cite{krull2019noise2void}, and LIR \cite{lir}, and supervised methods, i.e. DnCNN \cite{dncnn}, FFDNet \cite{zhang2018ffdnet}, and RedNet-30 \cite{red30}, to compare the performance. Traditional Low-Pass Filtering (LPF) and BM3D \cite{bm3d} are also evaluated. As shown in Figure \ref{fig:CBSD68}, the unsupervised methods tend to shift the color and leave apparent visual artifacts in the sky. Especially, LIR removes the noise but fails to preserve the texture. With frequency domain information, our method successfully eliminates noise and preserves the texture. The classical LPF using Fourier transform alleviates the noise, but our framework that reflects not only the frequency domain knowledge but also spatial domain knowledge shows superior results. As shown in Table \ref{table1}, our model outperforms other unsupervised methods, i.e. DIP, N2V, and LIR, by at least +0.29 dB in PSNR. Although our model is trained on unpaired images, it achieves superior performance in the SSIM than DnCNN and FFDNet trained on paired datasets. We conjecture that the reason for better noise removal is the use of the extra domain information that other previous methods do not consider.

\begin{figure*}[t]
    \centering
    \subfigure[LDCT]{\includegraphics[width=.135\textwidth]{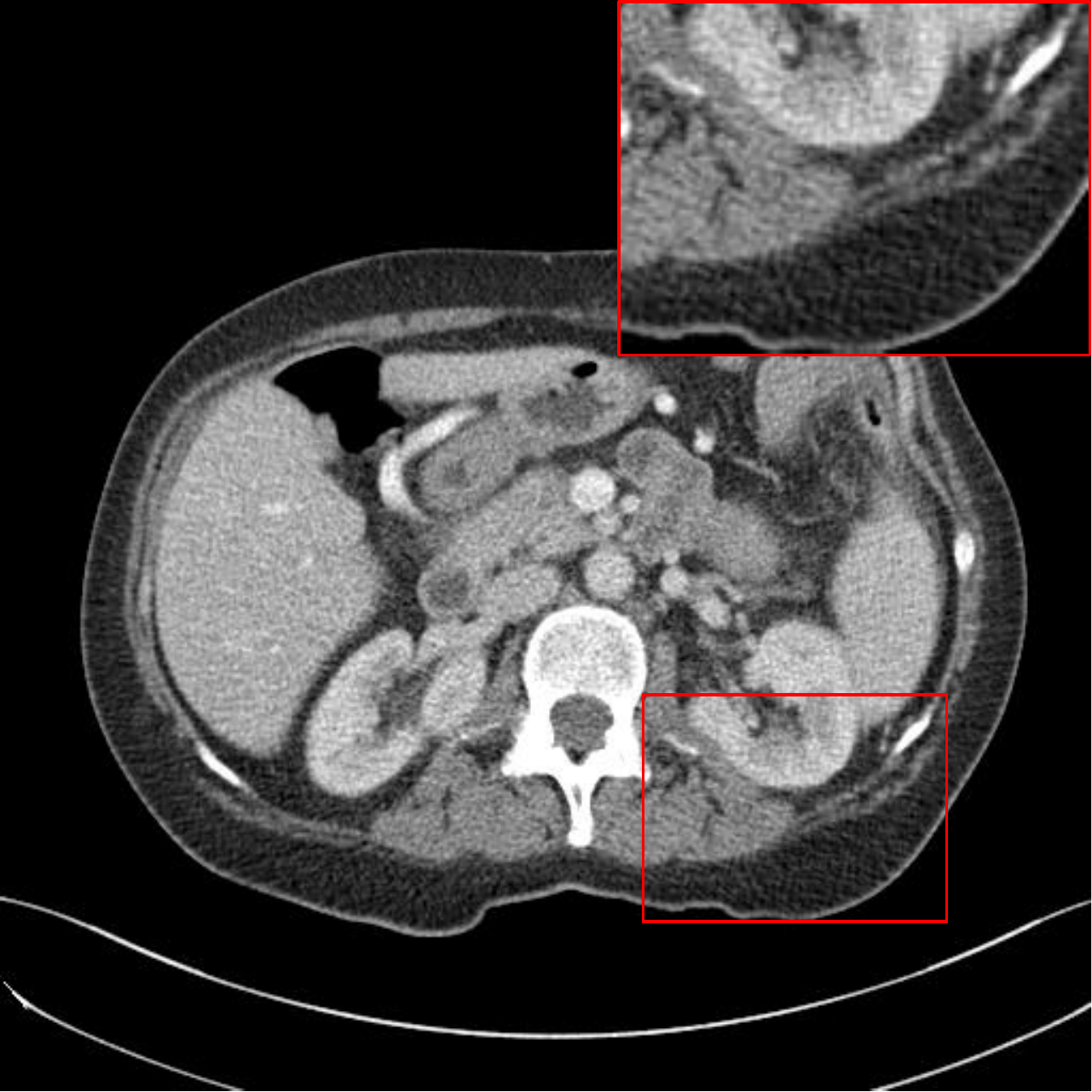}}
    \subfigure[BM3D]{\includegraphics[width=.135\textwidth]{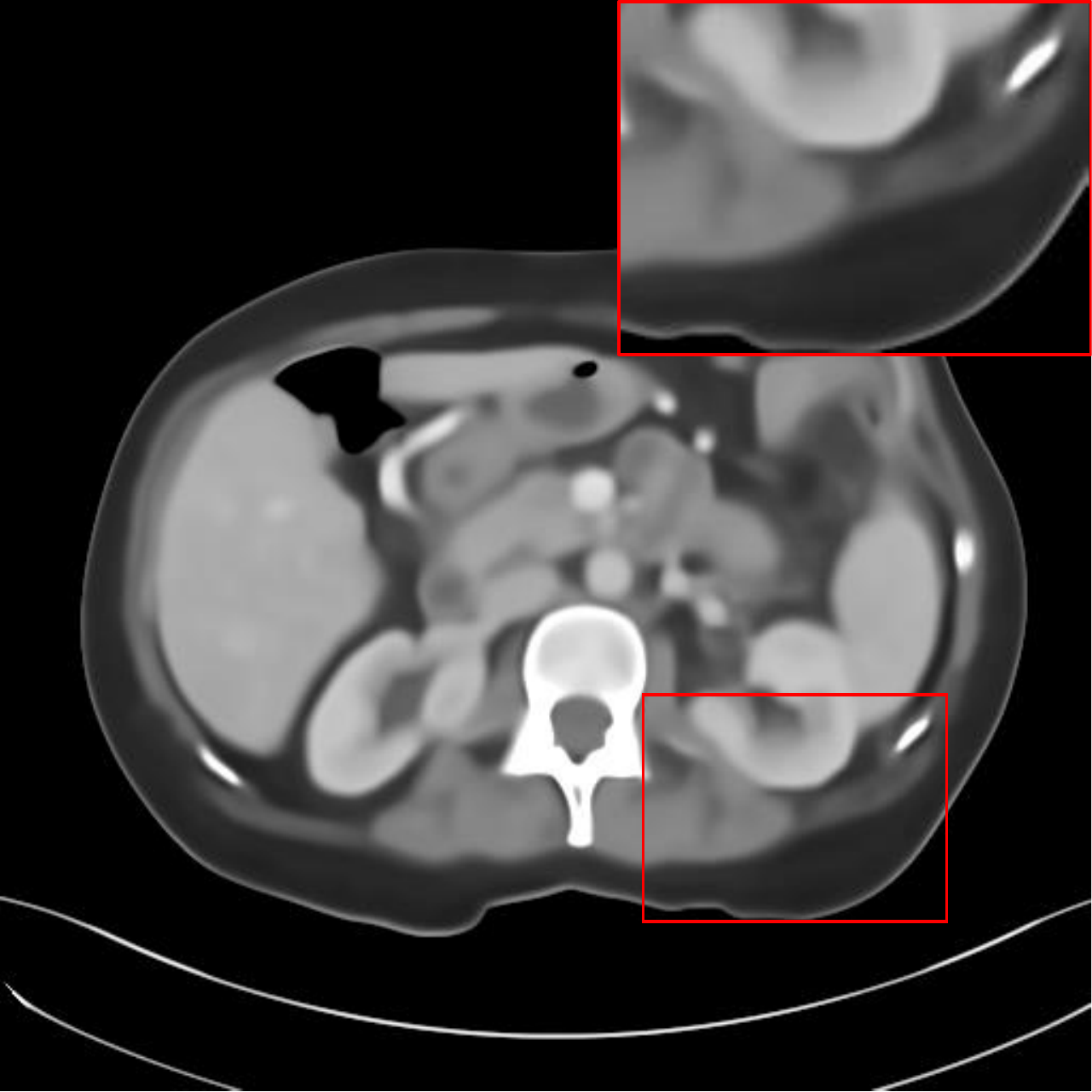}}
    \subfigure[RED-CNN]{\includegraphics[width=.135\textwidth]{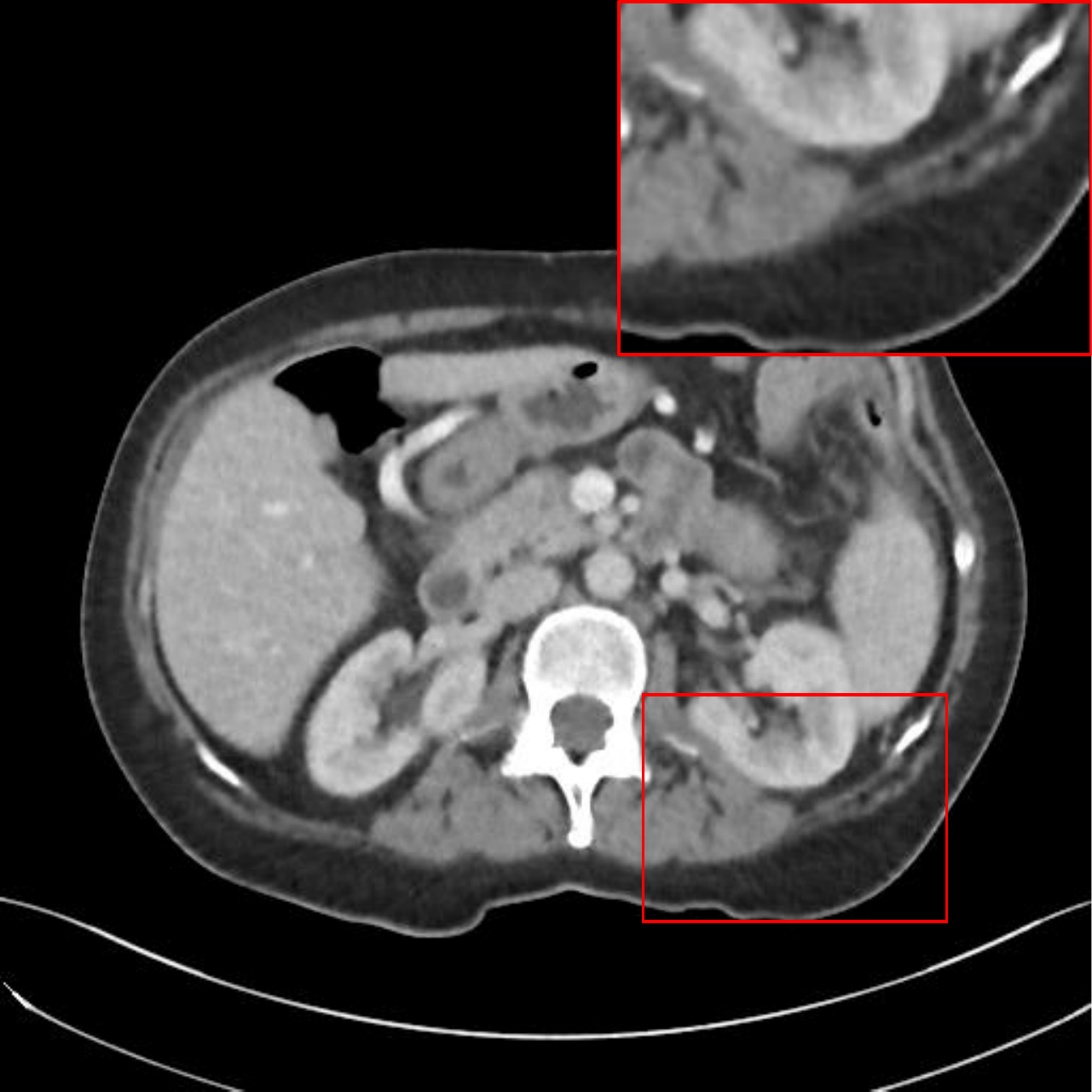}}
    \subfigure[DIP]{\includegraphics[width=.135\textwidth]{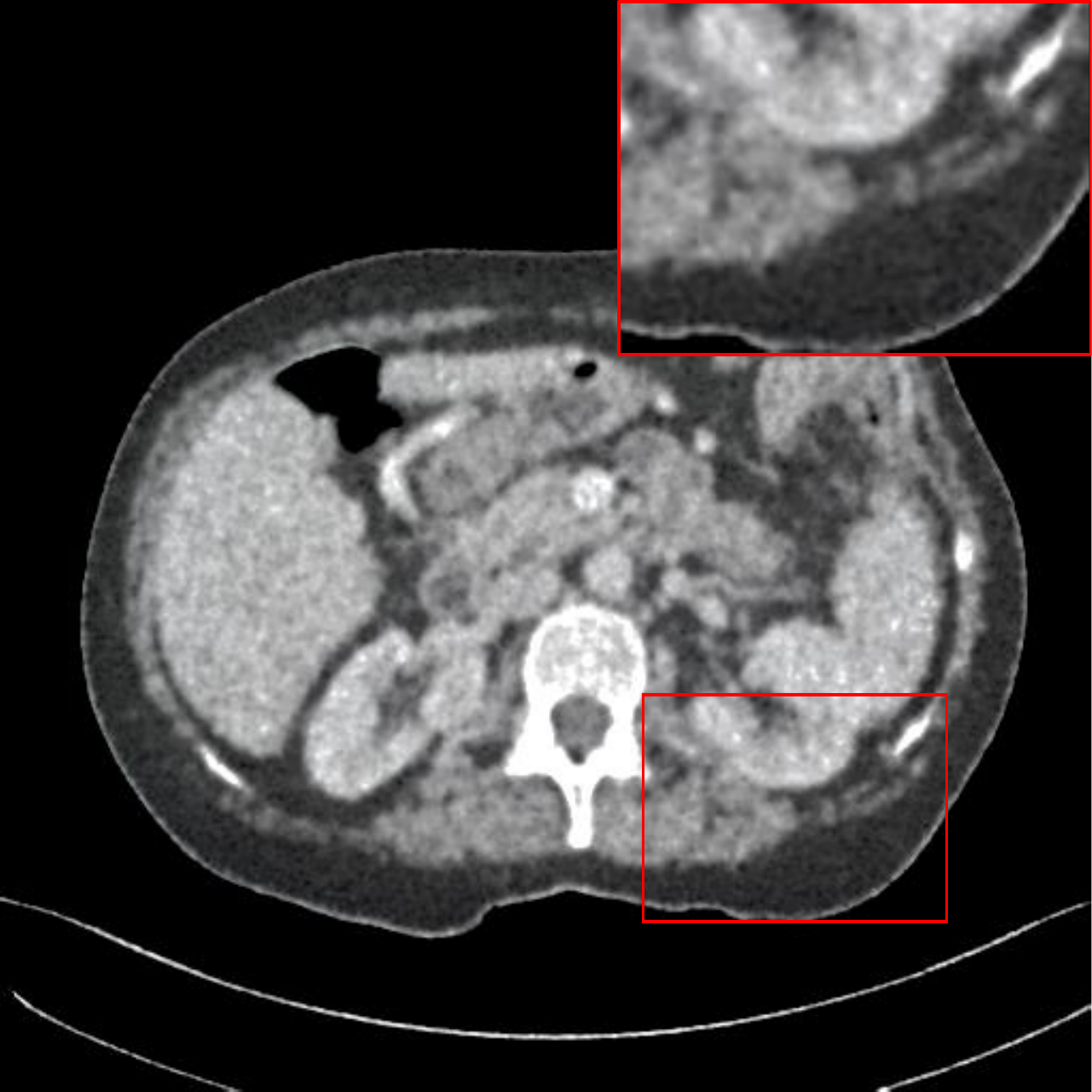}}
    \subfigure[LIR]{\includegraphics[width=.135\textwidth]{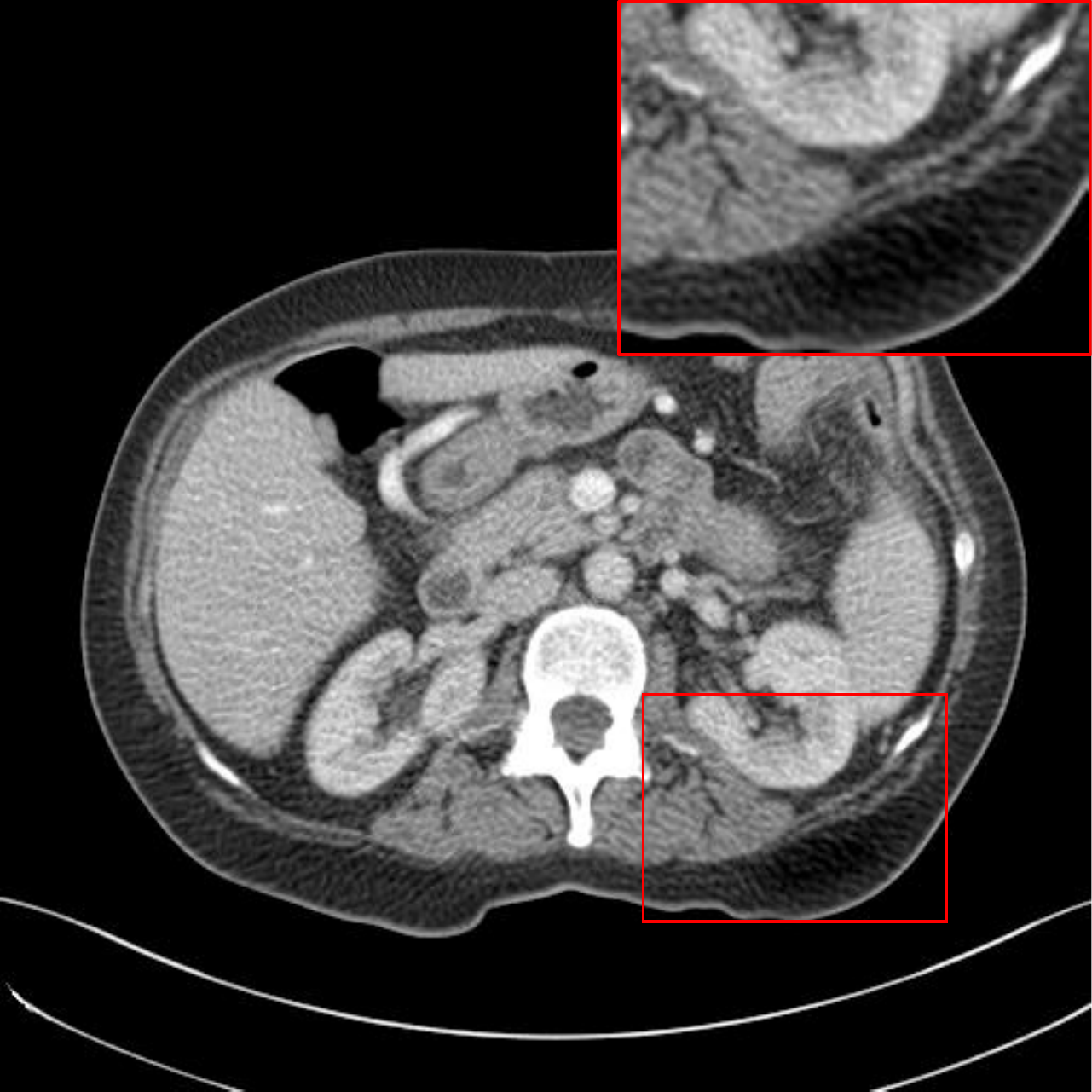}}
    \subfigure[Ours]{\includegraphics[width=.135\textwidth]{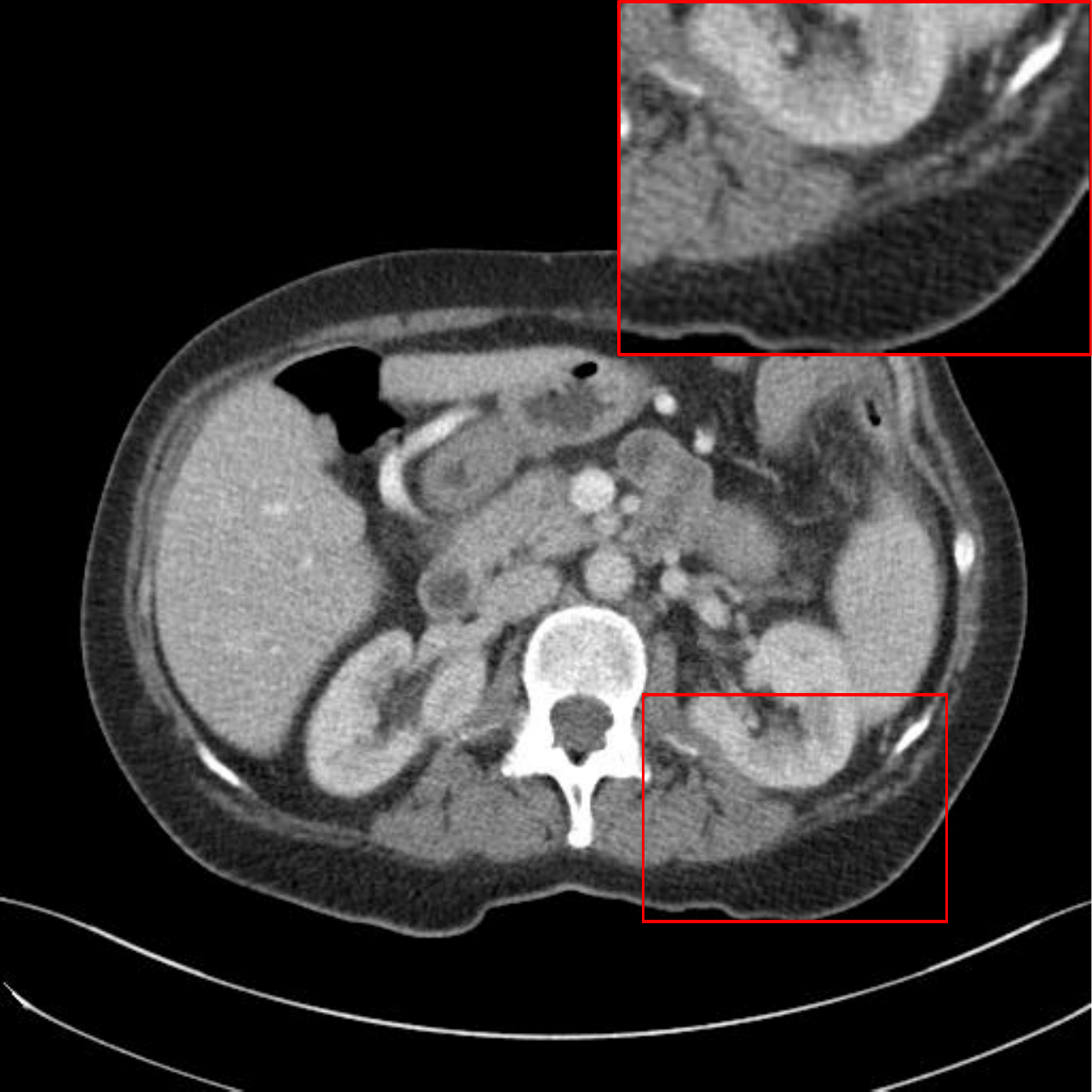}}
    \subfigure[NDCT]{\includegraphics[width=.135\textwidth]{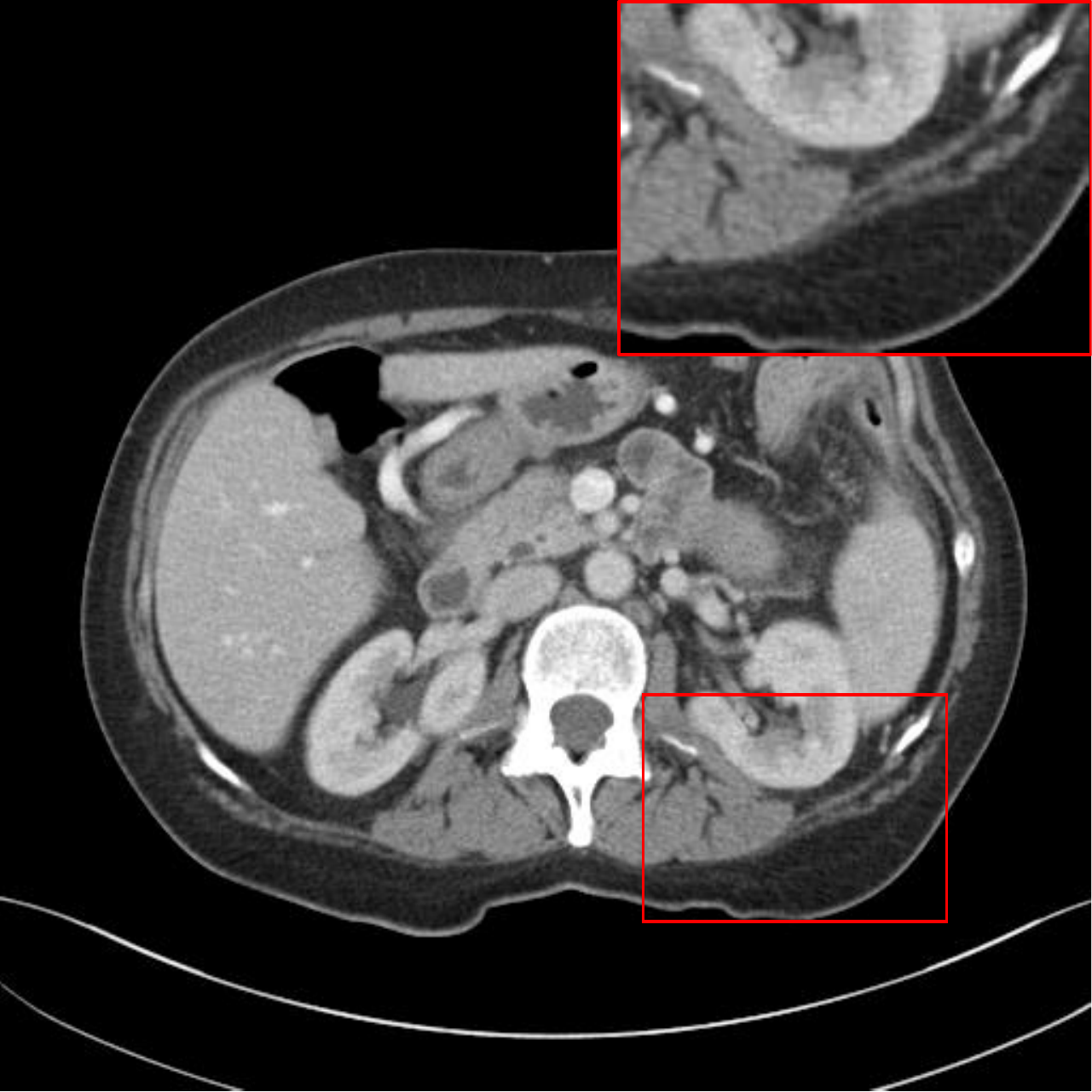}}
    \caption{Qualitative results of our method and other baselines on \textit{Mayo Clinic Low Dose CT dataset}. (a) Real low-dose. (b)-(f) Results of each methods. (g) Real normal-dose. As shown in the highlighted red box, the reconstructed image by our method has few noise and artifacts. The display window is $[160, 240]$ HU.}
    \label{fig:CT}
\end{figure*}

\begin{table}[t]
\begin{center}
\resizebox{0.65\textwidth}{!}
{%
\begin{tabular}{|c|c|c|ccc|}
\hline
   & \multicolumn{1}{c|}{Traditional} & \multicolumn{1}{c|}{Paired setting} &\multicolumn{3}{c|}{Unpaired setting} \\ \hline
\hline
Methods  & BM3D \cite{bm3d} & RED-CNN \cite{redcnnCT}& DIP \cite{ulyanov2018deep} & LIR \cite{lir} & Ours \\ \hline \hline
PSNR (dB)&29.16 &29.39   &26.97  &27.26 &\textbf{30.11}    \\ \hline
SSIM     &0.8514 &0.9078 &0.8267 &0.8452 &\textbf{0.8728}    \\ \hline
\end{tabular}}
\end{center}
\caption{The average PSNR and SSIM results of different methods on \textit{Mayo Clinic Low Dose CT dataset}. Our results are marked in \textbf{bold}.}
\label{table2}
\end{table}

\subsection{Real-World Noise Removal}
In this section, we evaluate the generalization ability of the proposed method on real-world noise, i.e. Low-Dose Computed Tomography (CT) and real photographs. For the comparison of the Low-Dose CT, we adopt BM3D \cite{bm3d}, DIP \cite{ulyanov2018deep}, RED-CNN \cite{redcnnCT}, and LIR \cite{lir} as baselines. For the comparison of the real photographs, BM3D \cite{bm3d}, DIP \cite{ulyanov2018deep}, RedNet-30 \cite{red30}, and LIR \cite{lir} are selected as baselines.
\paragraph{Denoising on Low-Dose CT}
Since Computed Tomography (CT) helps to diagnose abnormalities of organs, CT is widely used in medical analysis. Reducing the radiation dose in order to decrease health risks causes noise and artifacts in the reconstructed images. Like the real-world noise, the noise distributions of the reconstructed image are difficult to model analytically. Therefore, we adopt a CT dataset authorized by Mayo Clinic \cite{moen2021low} to evaluate the generalization ability of our method on real-world noise. Mayo Clinic dataset consists of paired normal-dose and lose-dose CT images for each patient. The Normal-Dose CT (NDCT) and the Low-Dose CT (LDCT) images correspond to clean and noisy images, respectively. For the training, we obtain 2,850 images in $512 \times 512$ resolution from 20 different patients. We construct 1,422 LDCT images from randomly selected 10 patients as a noise set and 1,428 NDCT images from the remaining patients as a clean set for unpaired training. For the test, we obtain 865 images from 5 different patients. As shown in Table \ref{table2}, our method achieves the best and the second-best performance in PSNR and SSIM, respectively. Note that our model trained on the unpaired dataset outperforms the RED-CNN trained on the paired dataset in PSNR. It indicates that our method can be more practical in medical analysis where obtaining paired datasets is challenging. We also compare the qualitative results with other baselines. As shown in Figure \ref{fig:CT}, other methods tend to generate artifacts or lose details. On the other hand, our method shows a reasonable balance between noise removal and image quality. More qualitative results are provided in the supplementary material.

\begin{figure*}[t]
    \centering
    \subfigure[Input]{\includegraphics[width=.135\textwidth]{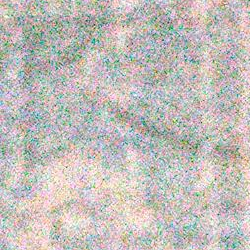}}
    \subfigure[CBM3D]{\includegraphics[width=.135\textwidth]{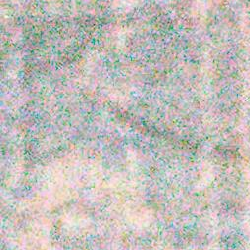}}
    \subfigure[RedNet-30]{\includegraphics[width=.135\textwidth]{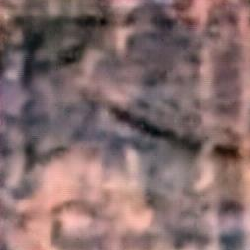}}
    \subfigure[DIP]{\includegraphics[width=.135\textwidth]{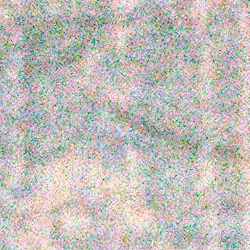}}
    \subfigure[LIR]{\includegraphics[width=.135\textwidth]{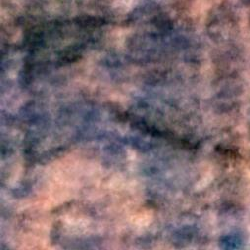}}
    \subfigure[Ours]{\includegraphics[width=.135\textwidth]{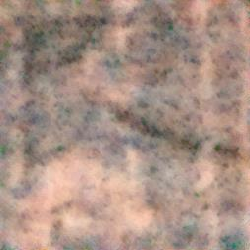}}
    \subfigure[GT]{\includegraphics[width=.135\textwidth]{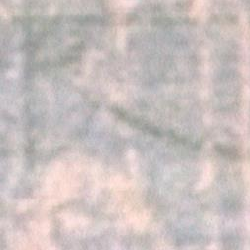}}
    \caption{Qualitative results of our method and other baselines on \textit{SIDD}.}
    \label{fig:sidd}
\end{figure*}

\paragraph{Denoising on Real Photographs}
To demonstrate the effectiveness of our method on real noisy photographs, we evaluate our method on SIDD \cite{abdelhamed2018sidd} which is obtained from smartphone cameras. Because the images of the SIDD comprise various noise levels and brightness, this dataset is the best appropriate to validate the generalization capacity of the denoisers. The SIDD includes 320 pairs of noisy images and corresponding clean images with 4K or 5K resolutions for the training. For the unpaired training, we divide the dataset into 160 clean and 160 noisy images without intersection. The other training settings are the same as implementation details. For evaluation, we use 1280 cropped patches of size $256 \times 256$ in the SIDD validation set.
As show in Figure \ref{fig:sidd}, other baselines tend to leave the noise or fail to preserve the color of images. In contrast, our method removes the intense noise while keeping the color compared to other baselines. We also report the quantitative results in Table \ref{table:sidd}. More qualitative results are provided in the supplementary material.

\begin{table}[t]
\begin{center}
\resizebox{0.65\textwidth}{!}
{%
\begin{tabular}{|c|c|c|ccc|}
\hline
& \multicolumn{1}{c|}{Traditional} & \multicolumn{1}{c|}{Paired setting} &\multicolumn{3}{c|}{Unpaired setting} \\ \hline
\hline
Methods     & CBM3D \cite{bm3d} & RedNet-30 \cite{red30} & DIP \cite{ulyanov2018deep} & LIR \cite{lir} & Ours        \\ \hline
PSNR (dB)   & 28.32 & 38.02          & 24.68     & 33.79       & \textbf{34.30}       \\ \hline
SSIM        & 0.6784 & 0.9619         & 0.5901    & 0.9466      & \textbf{0.9334}      \\ \hline
\end{tabular}}
\end{center}
\caption{The average PSNR and SSIM results of different methods on \textit{SIDD}. Our results are marked in \textbf{bold}.}
\label{table:sidd}
\end{table}

\subsection{Ablation Study}
We conduct an ablation study to demonstrate the validity of our key components: the texture discriminator $D_{T}$, the spectral discriminator $D_{S}$, and the frequency reconstruction loss $L_{Freq}$. 
We employ an additional evaluation metric LFD \cite{jiang2020focal} to measure the difference between denoised images and reference images in the frequency domain. The small LFD value indicates that the denoised images are close to the reference images. First, to verify the effectiveness of the $D_{S}$, we only add the $D_{S}$ to the base structure. As shown in Table \ref{table3}, when the $D_{S}$ is integrated, both PSNR and SSIM increase by 0.2 dB and 0.0014, respectively. It demonstrates that the spectral discriminator leads the generator to remove high-frequency related noise effectively by transferring the difference between noisy and clean images on the high-frequency bands. Also, we see that the spectral discriminator makes the denoised images close to clean domain images in the frequency domain, resulting in the decrease of LFD. Next, to verify the effectiveness of the $D_{T}$, we integrate it with the $D_{S}$. Distinguishing the texture representations helps restore clean contours and fine details related to image quality, which improves the SSIM metric. A curious phenomenon is that the texture discriminator increases the LFD. We conjecture that the introduction of $D_{T}$ causes a bias to the spatial domain in maintaining the balance between the spatial and frequency domains, thus increasing the distance in the frequency domain. Adding the $L_{Freq}$ shows results validating our hypothesis that narrowing the gap in the frequency domain is crucial to generate the high-quality denoised image. In addition, through the decrease of LFD, the frequency reconstruction loss may help to maintain the balance between the spatial and frequency domain.

\begin{table}[t]
\begin{center}
\resizebox{0.45\textwidth}{!}
{%
\begin{tabular}{|ccc|ccc|}
\hline
$D_{S}$ & $D_{T}$ & $L_{Freq}$ & PSNR (dB) & SSIM & LFD \\ \hline \hline
\xmark &\xmark &\xmark &25.59 &0.8290 &6.5955\\ 
\cmark &\xmark &\xmark &25.79 &0.8304 &\textbf{6.5649}\\
\cmark &\cmark &\xmark &25.82 &0.8334 &6.5874 \\
\cmark &\cmark &\cmark &\textbf{26.03} &\textbf{0.8375} &6.5795 \\\hline
\end{tabular}}
\end{center}
\caption{Ablation study. Quantitative results of our method with and without the texture discriminator $D_{T}$, spectral discriminator $D_{S}$, and frequency reconstruction loss $L_{Freq}$ on CBSD68 corrupted by AWGN with a noise level $\sigma=50$. we report the PSNR, SSIM (higher is better) and LFD (lower is better). The best results are marked in \textbf{bold}.}
\label{table3}
\end{table}

\section{Conclusion}
In this paper, we propose an unsupervised learning-based image denoiser that enables the image denoising without clean and noisy image pairs. To the best of our knowledge, it is the first approach that aims to recover a noise-free image from a corrupted image using frequency domain information. To this end, we introduce the spectral discriminator and frequency reconstruction loss that can propagate the frequency knowledge to the generator. By reflecting the information from the frequency domain, our method successfully focuses on high-frequency components to remove noise. Experiments on synthetic and real noise removal show that our method outperforms other unsupervised learning-based denoisers and generates more visually pleasing images with fewer artifacts. We believe that considering the frequency domain can be advantageous in other low-level vision tasks as well.

\section{Acknowledgements}
This work was supported by the Engineering Research Center of Excellence (ERC) Program supported by National Research Foundation (NRF), Korean Ministry of Science \& ICT (MSIT) (Grant No. NRF-2017R1A5A1014708).
\bibliography{arxiv}

\newpage

\appendix

\section{Supplementary Material}
\label{overview}
In this supplementary material, we describe the architecture details and show the additional experiments as follows:
\begin{itemize}
    \item In Section \ref{architecture}, we describe the architectures of two generators, i.e. $G_{n2c}$ and $G_{c2n}$, and three discriminators, i.e. $D_{C}$, $D_{T}$, and $D_{S}$, in our framework.
    \item In Section \ref{qual}, we show the additional results on CBSD68 \cite{martin2001database} corrupted by AWGN with a noise level $\sigma=25$.
    \item In Section \ref{real}, we show the additional qualitative results on real-world noise, i.e. Low-Dose CT authorized by Mayo Clinic \cite{moen2021low} and SIDD \cite{abdelhamed2018sidd}.
    \item In Section \ref{ablation}, we show the results of an additional ablation study to demonstrate the validity of the perceptual loss $L_{VGG}$, the cycle consistency loss $L_{CC}$, and the reconstruction loss $L_{Recon}$.
    \item In Section \ref{several}, we show the results on several noise types, such as structured noise and Poisson noise, to evaluate the generalization ability of our method.

\end{itemize}

\section{The Details of Architectures}
\label{architecture}
\begin{figure}[htbp]
    \centering
    \includegraphics[width=\textwidth]{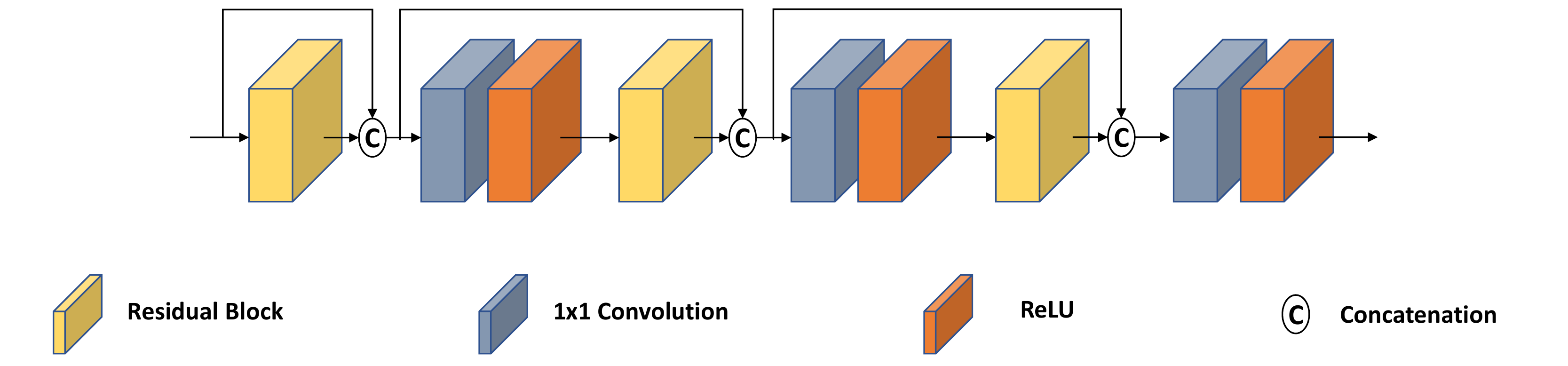}
    
    \caption{The architecture of Cascading Block used as the basic component in the $G_{n2c}$. We use the Residual Block proposed by \cite{he2016resnet} and the ReLU.}
    \label{fig:cascading block}
\end{figure}

\begin{figure}[htbp]
    \centering
    \includegraphics[width=\textwidth]{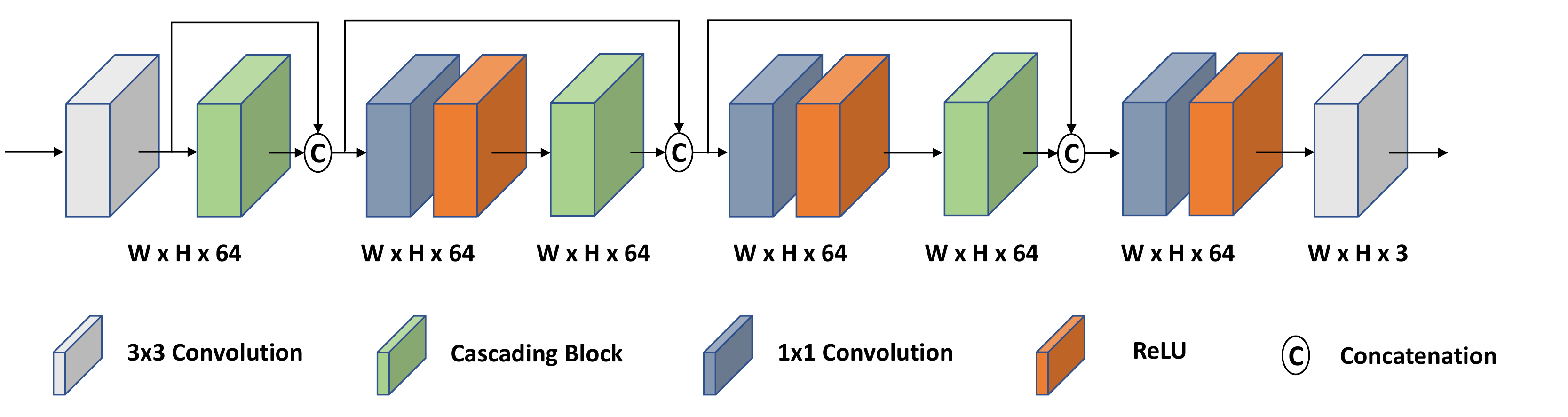}
    
    \caption{The architecture of generator $G_{n2c}$ for noise removal. We use the convolution with kernel size=3, stride=1, and padding=1.}
    \label{fig:n2c_generator}
\end{figure}

\paragraph{Generator $G_{n2c}$}
For the noise removal generator $G_{n2c}$, we adopt the network introduced by \cite{ahn2018carn}. The main idea of this architecture is multiple cascading connections at global and local levels which help to propagate low-level information to later layers and remove noise. The details of $G_{n2c}$ are illustrated in Figure \ref{fig:cascading block} and \ref{fig:n2c_generator}.

\begin{figure}[htbp]
    \centering
    \includegraphics[width=\textwidth]{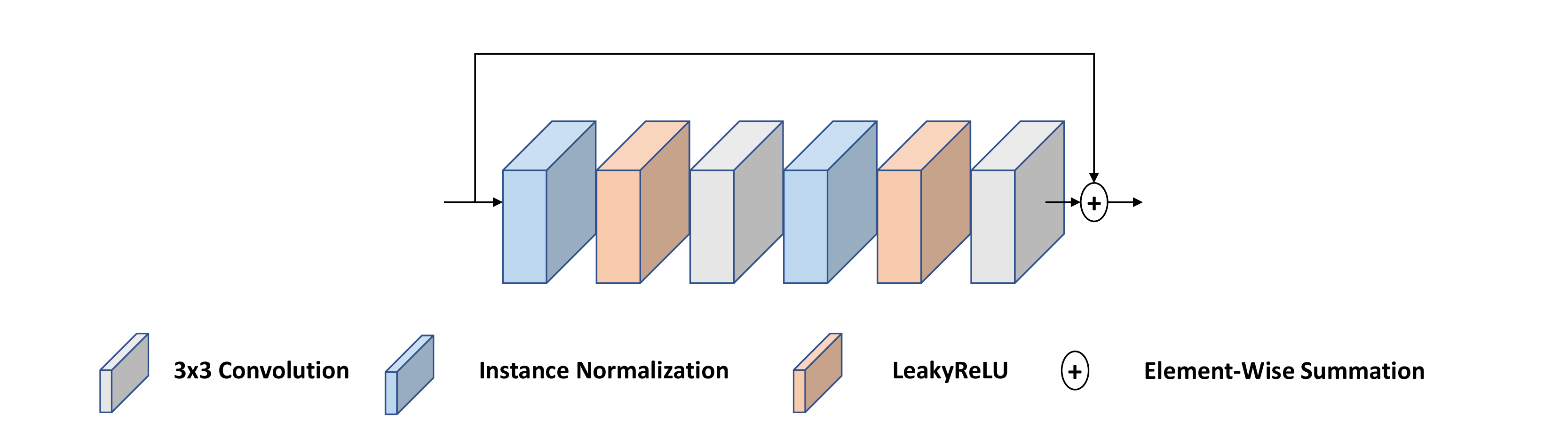}
    
    \caption{The architecture of Instance Residual Block used as the basic component in the $G_{c2n}$. We use the convolution with kernel size=3, stride=1, and padding=1 and the LeakyReLU with a slope of $0.2$.}
    \label{fig:inresblock}
\end{figure}

\begin{figure}[ht]
    \centering
    \includegraphics[width=\textwidth]{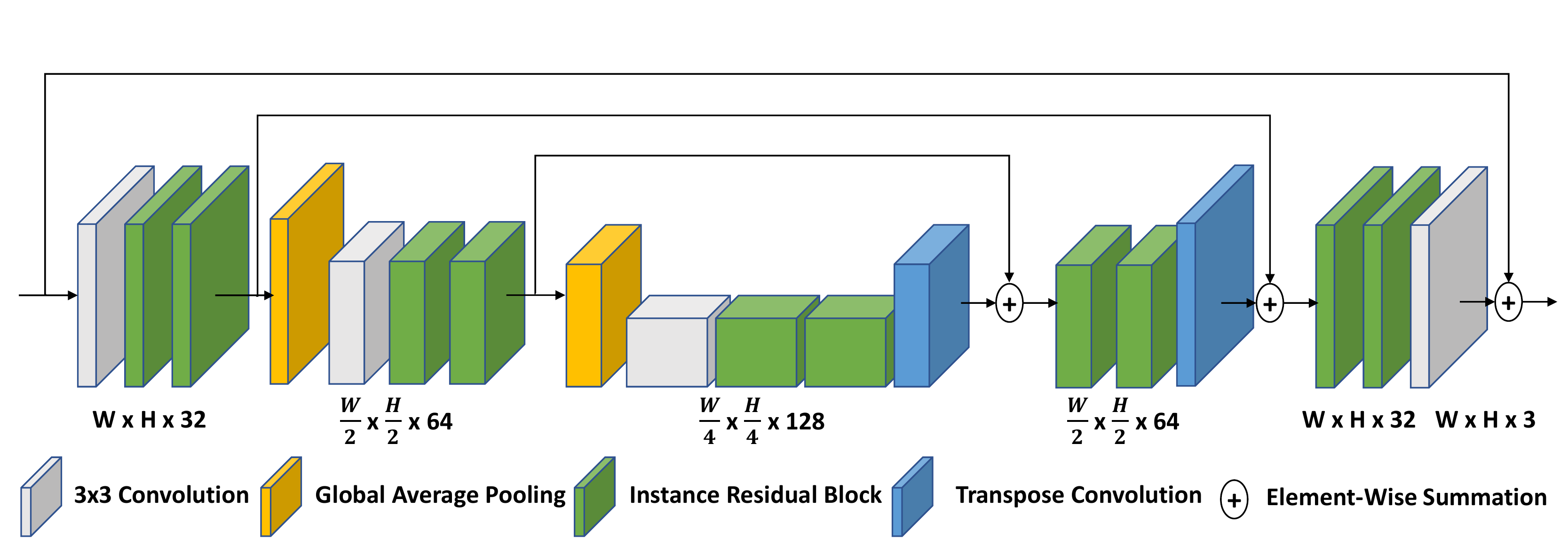}
    
    \caption{The architecture of generator $G_{c2n}$. We use the convolution with kernel size=3, stride=1, and padding=1 and transposed convolution with kernel size=3, stride=2, padding=1, and output padding=1.}
    \label{fig:c2n_generator}
\end{figure}

\paragraph{Generator $G_{c2n}$}
For the generator $G_{c2n}$, we adopt the U-Net based network that is similar to the architecture introduced by \cite{kim2020transfer}. The role of this network is to translate images from the noise domain to the clean domain. The details of $G_{c2n}$ are illustrated in Figure \ref{fig:inresblock} and \ref{fig:c2n_generator}.

\begin{figure}[ht]
    \centering
    \includegraphics[width=\textwidth]{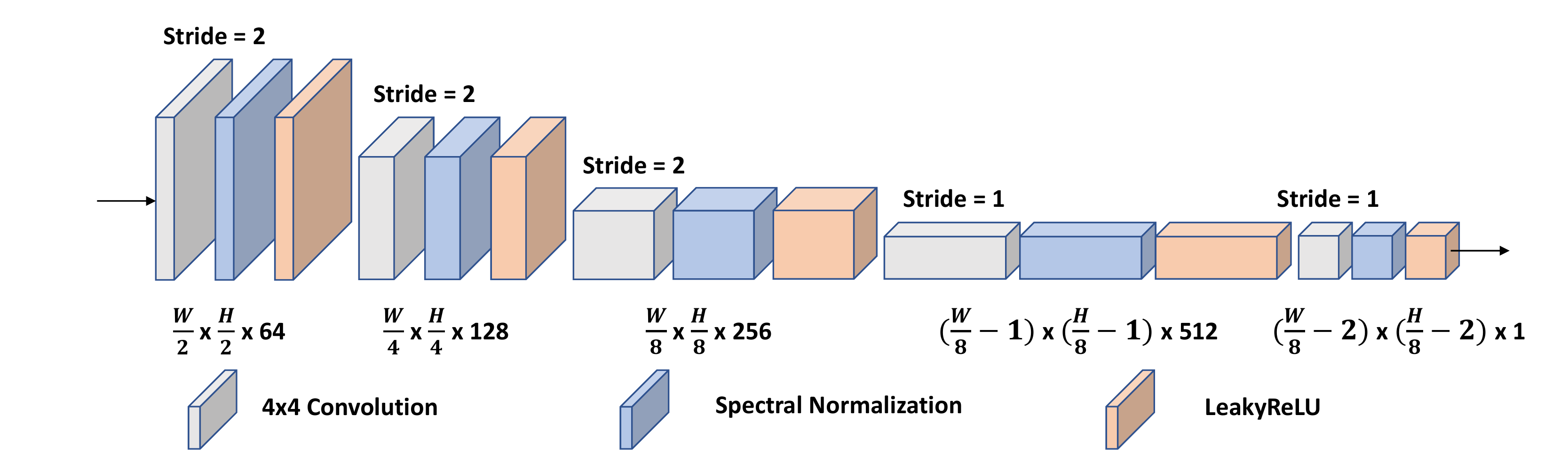}
    
    \caption{The architecture of discriminators $D_{C}$ and $D_{T}$. We use the convolution with kernel size=4 and padding=1. Followed by the convolution, we use the spectral normalization \cite{miyato2018spectral} and the LeakyReLU with a slope of $0.2$.}
    \label{fig:discriminator}
\end{figure}

\paragraph{Discriminators $D_{C}$ and $D_{T}$}
For the discriminators $D_{C}$ and $D_{T}$, we employ the $70\times70$ PatchGAN discriminator \cite{isola2017image} which classifies whether $70\times70$ image patches are real or fake. The details of $D_{C}$ and $D_{T}$ are illustrated in Figure \ref{fig:discriminator}. 

\paragraph{Discriminator $D_{S}$}
For the spectral discriminator $D_{S}$, we employ the single linear unit as the spectral discriminator. The $D_{S}$ takes a high-pass filtered 1D spectral vector and aims to classify whether the spectral vector is real or fake.

\section{Additional Results on AWGN}
\label{qual}

\begin{figure*}[hp]
    \centering
    \subfigure[Input (21.21/0.55)]{\includegraphics[width=.3\textwidth]{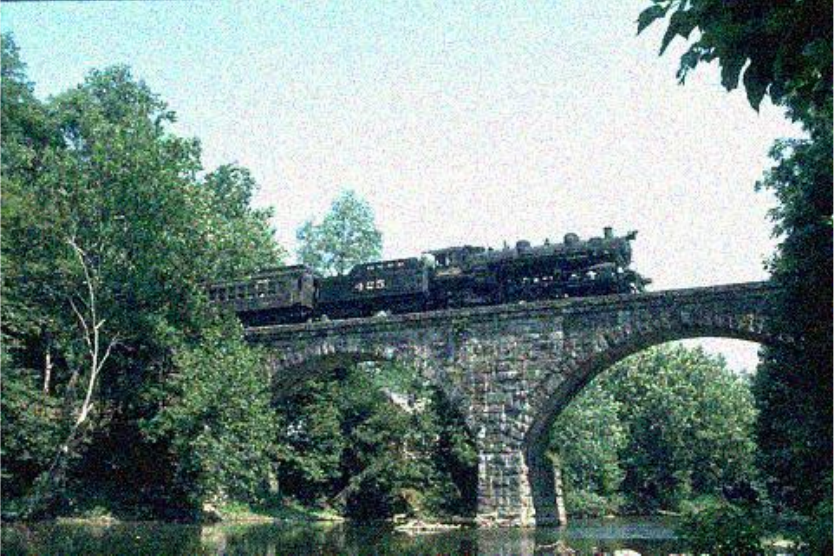}}
    \subfigure[LPF (21.43/0.60)]{\includegraphics[width=.3\textwidth]{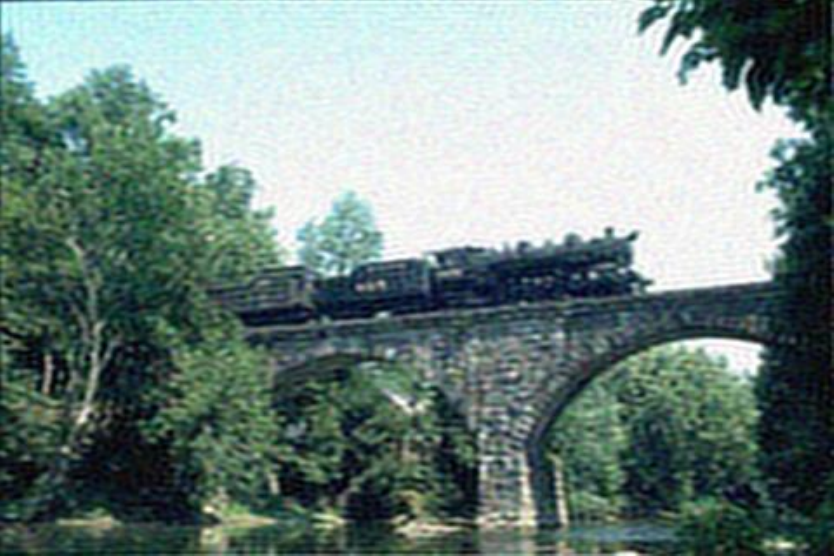}}
    \subfigure[CBM3D (28.55/0.91)]{\includegraphics[width=.3\textwidth]{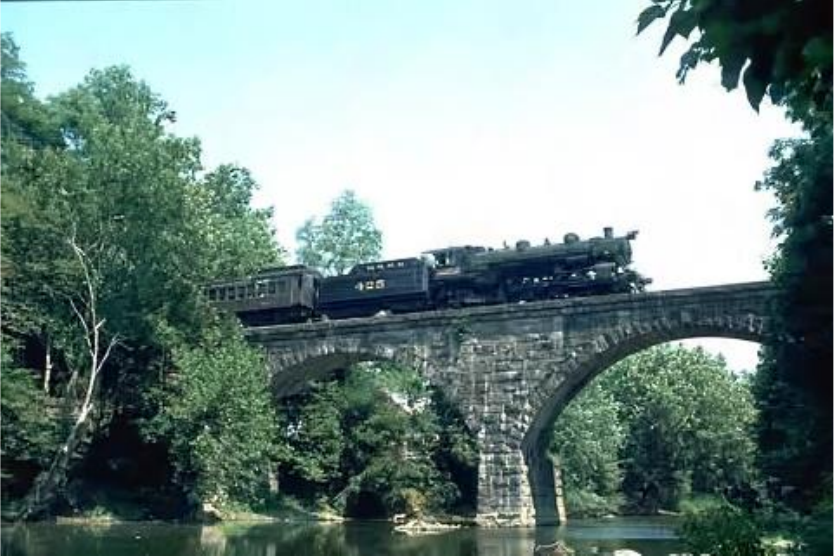}}
    \subfigure[DnCNN (28.64/0.92)]{\includegraphics[width=.3\textwidth]{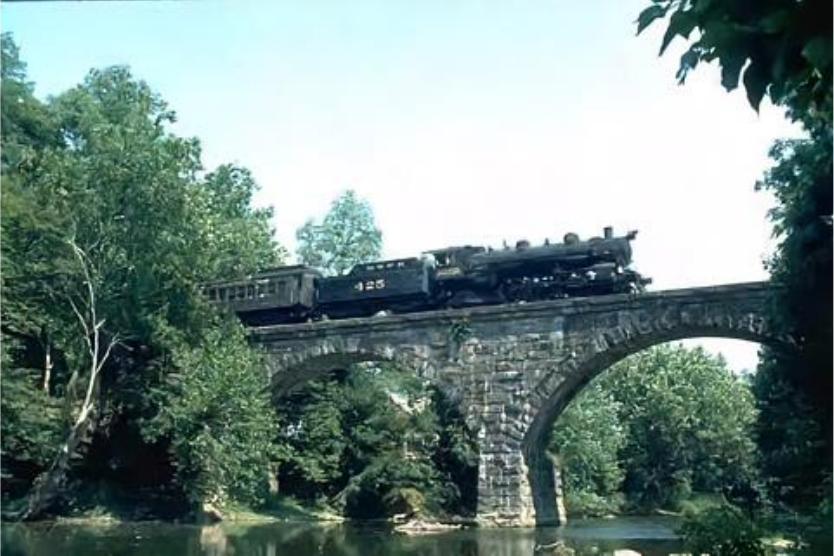}}
    \subfigure[FFDNet (24.56/0.83)]{\includegraphics[width=.3\textwidth]{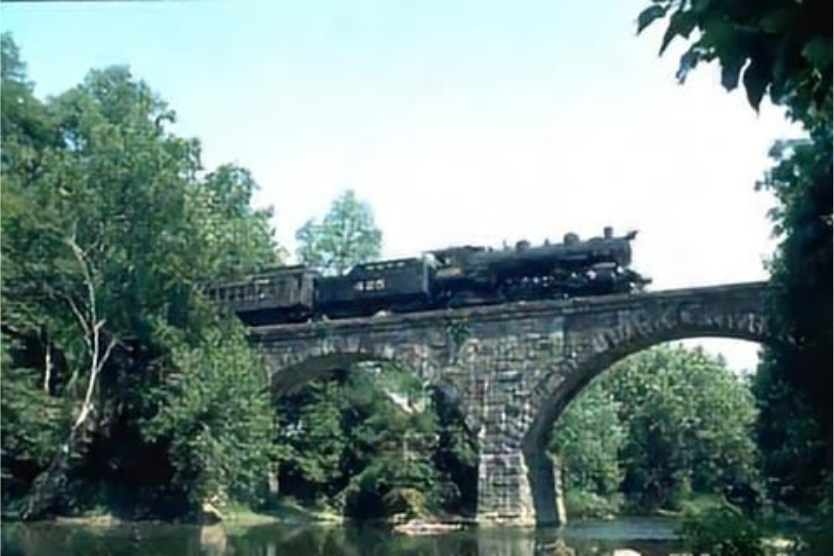}}
    \subfigure[RedNet-30 (27.46/0.91)]{\includegraphics[width=.3\textwidth]{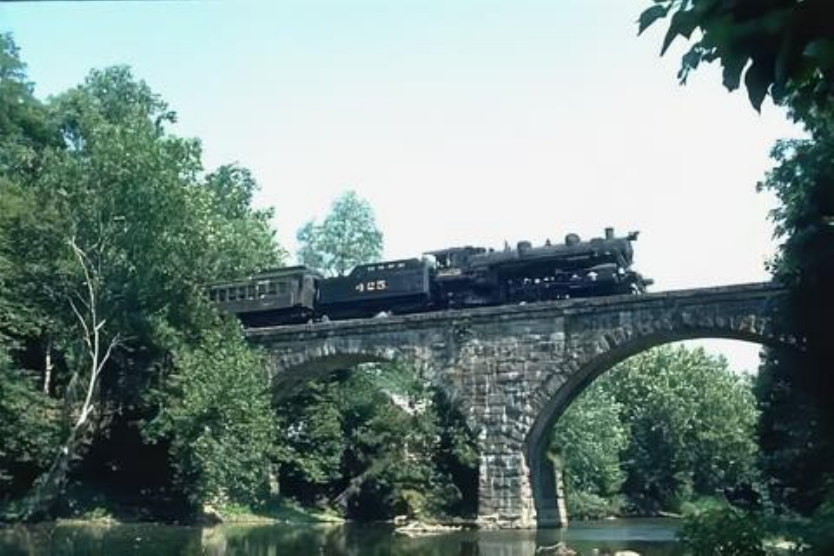}}
    \subfigure[N2N (28.92/0.92)]{\includegraphics[width=.3\textwidth]{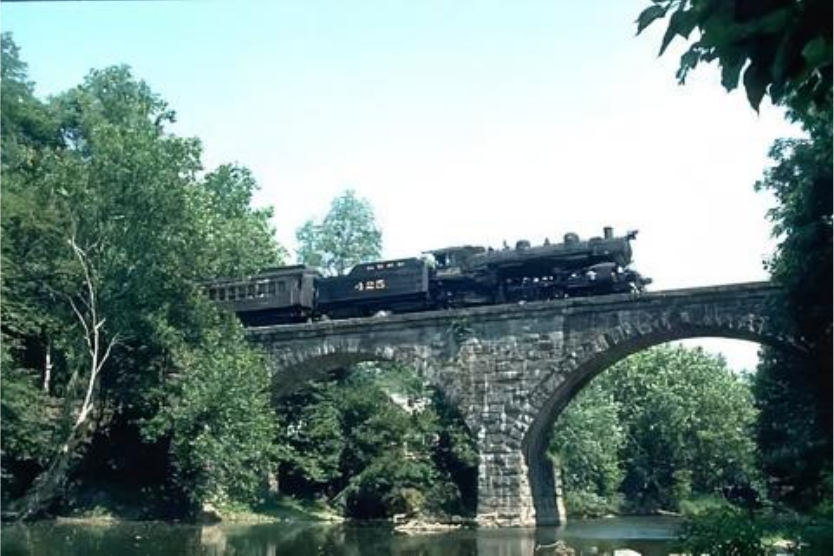}}
    \subfigure[DIP (22.90/0.74)]{\includegraphics[width=.3\textwidth]{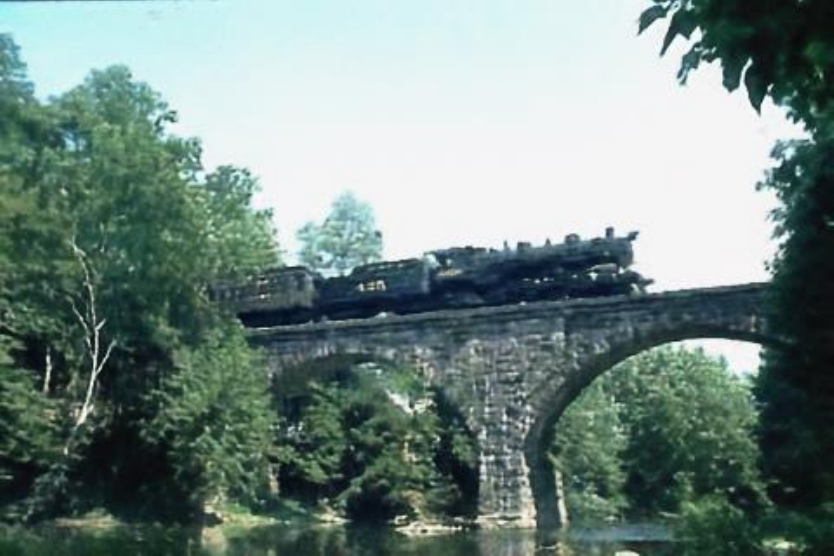}}
    \subfigure[N2V (24.05/0.81)]{\includegraphics[width=.3\textwidth]{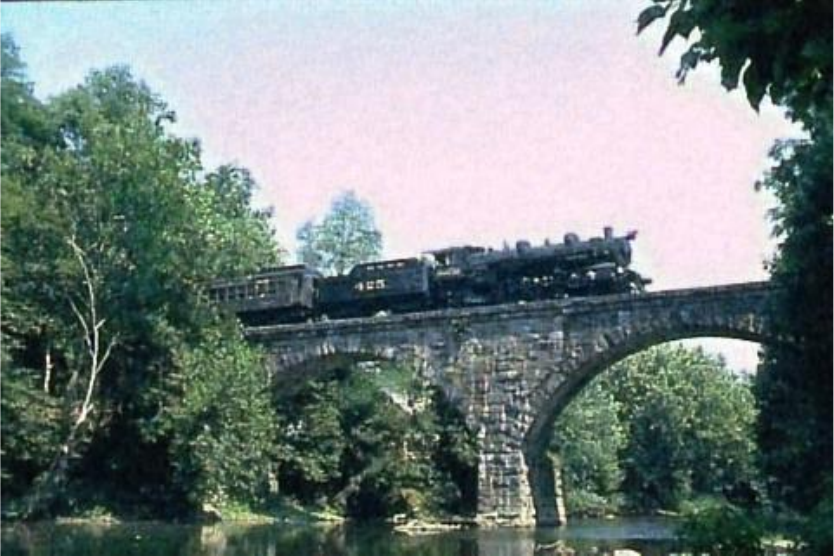}}
    \subfigure[LIR (19.48/0.82)]{\includegraphics[width=.3\textwidth]{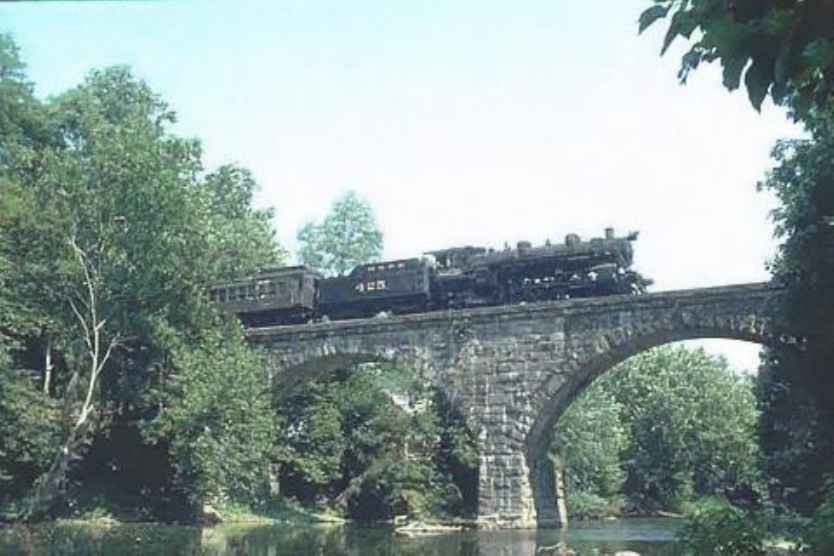}}
    \subfigure[Ours (27.49/0.90)]{\includegraphics[width=.3\textwidth]{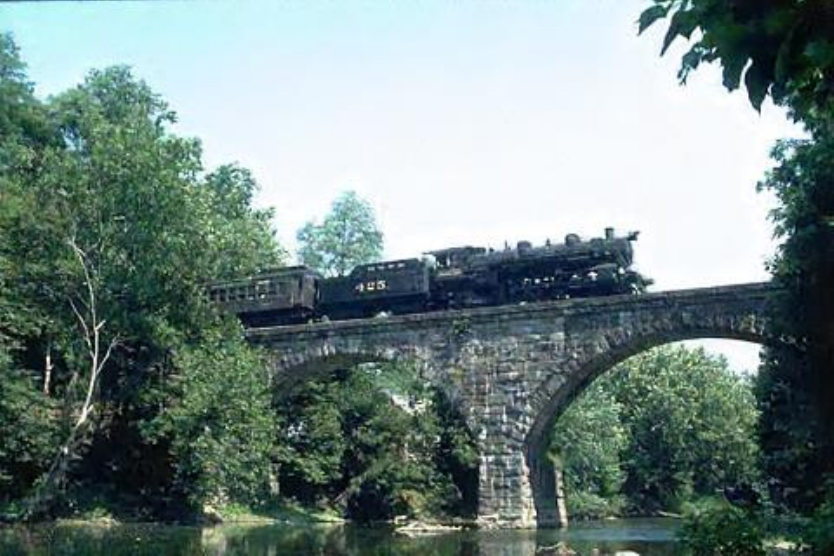}}
    \subfigure[GT (PSNR/SSIM)]{\includegraphics[width=.3\textwidth]{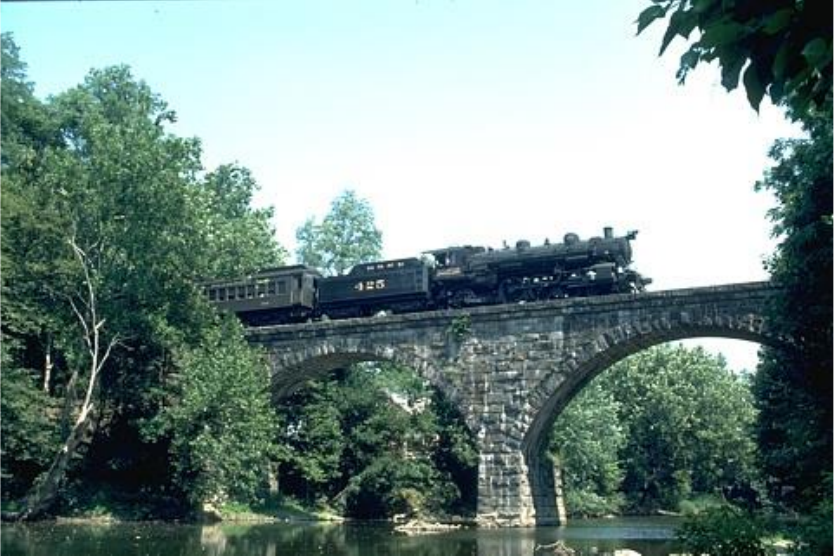}}

    \caption{Qualitative results of our method and other baselines on \textit{CBSD68} corrupted by AWGN with a noise level $\sigma=25$.}
    \label{fig:CBSD68_qual1}
\end{figure*}

\begin{figure*}[hp]
    \centering
    \subfigure[Input (20.27/0.44)]{\includegraphics[width=.3\textwidth]{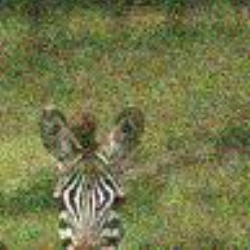}}
    \subfigure[LPF (21.29/0.58)]{\includegraphics[width=.3\textwidth]{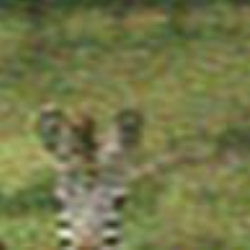}}
    \subfigure[CBM3D (29.99/0.85)]{\includegraphics[width=.3\textwidth]{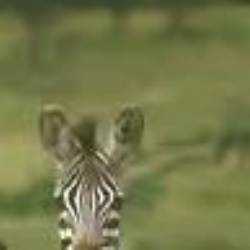}}
    \subfigure[DnCNN (30.32/0.87)]{\includegraphics[width=.3\textwidth]{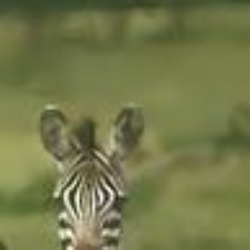}}
    \subfigure[FFDNet (26.29/0.81)]{\includegraphics[width=.3\textwidth]{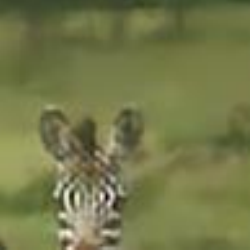}}
    \subfigure[RedNet-30 (30.50/0.88)]{\includegraphics[width=.3\textwidth]{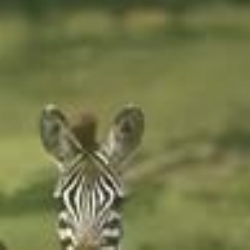}}
    \subfigure[N2N (30.71/0.88)]{\includegraphics[width=.3\textwidth]{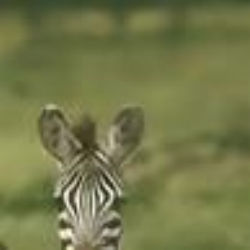}}
    \subfigure[DIP (27.40/0.77)]{\includegraphics[width=.3\textwidth]{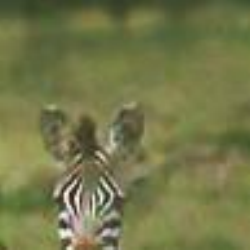}}
    \subfigure[N2V (26.73/0.75)]{\includegraphics[width=.3\textwidth]{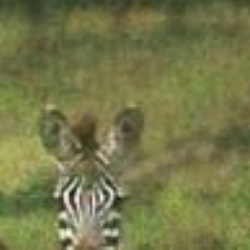}}
    \subfigure[LIR (26.04/0.83)]{\includegraphics[width=.3\textwidth]{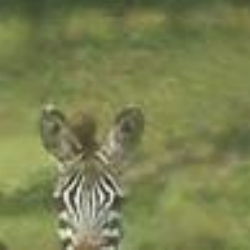}}
    \subfigure[Ours (28.35/0.83)]{\includegraphics[width=.3\textwidth]{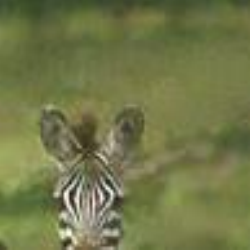}}
    \subfigure[GT (PSNR/SSIM)]{\includegraphics[width=.3\textwidth]{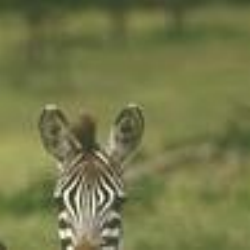}}

    \caption{Qualitative results of our method and other baselines on \textit{CBSD68} corrupted by AWGN with a noise level $\sigma=25$.}
    \label{fig:CBSD68_qual2}
\end{figure*}

We additionally visualize the results for CBSD68 images corrupted by AWGN with a noise level $\sigma=25$ and show the PSNR and SSIM in Figure \ref{fig:CBSD68_qual1} and \ref{fig:CBSD68_qual2}. In Figure \ref{fig:CBSD68_qual1}, our method outperforms other methods trained with unpaired dataset by at least +3.44dB and +0.08 in terms of PSNR and SSIM, respectively. LIR and N2V spoil the color and lights, but our method preserves both the color and lights and successfully removes the noise. We also show the challenging example that has repetitive high-frequency patterns hard to distinguish with noise in Figure \ref{fig:CBSD68_qual2}. Our approach removes noise without artifact and also preserves the patterns of the zebra. Although our method is trained under unpaired settings, it shows comparable performance in PSNR and SSIM with the supervised models in Figure \ref{fig:CBSD68_qual2}. Furthermore, compared to methods trained with unpaired dataset, our approach achieves the best performance in both PSNR and SSIM.

\section{Additional Qualitative Results on Real-World Noise}
\label{real}
\subsection{Low-Dose CT}
\begin{figure*}[hp]
    \centering
    \subfigure[LDCT]{\includegraphics[width=.24\textwidth]{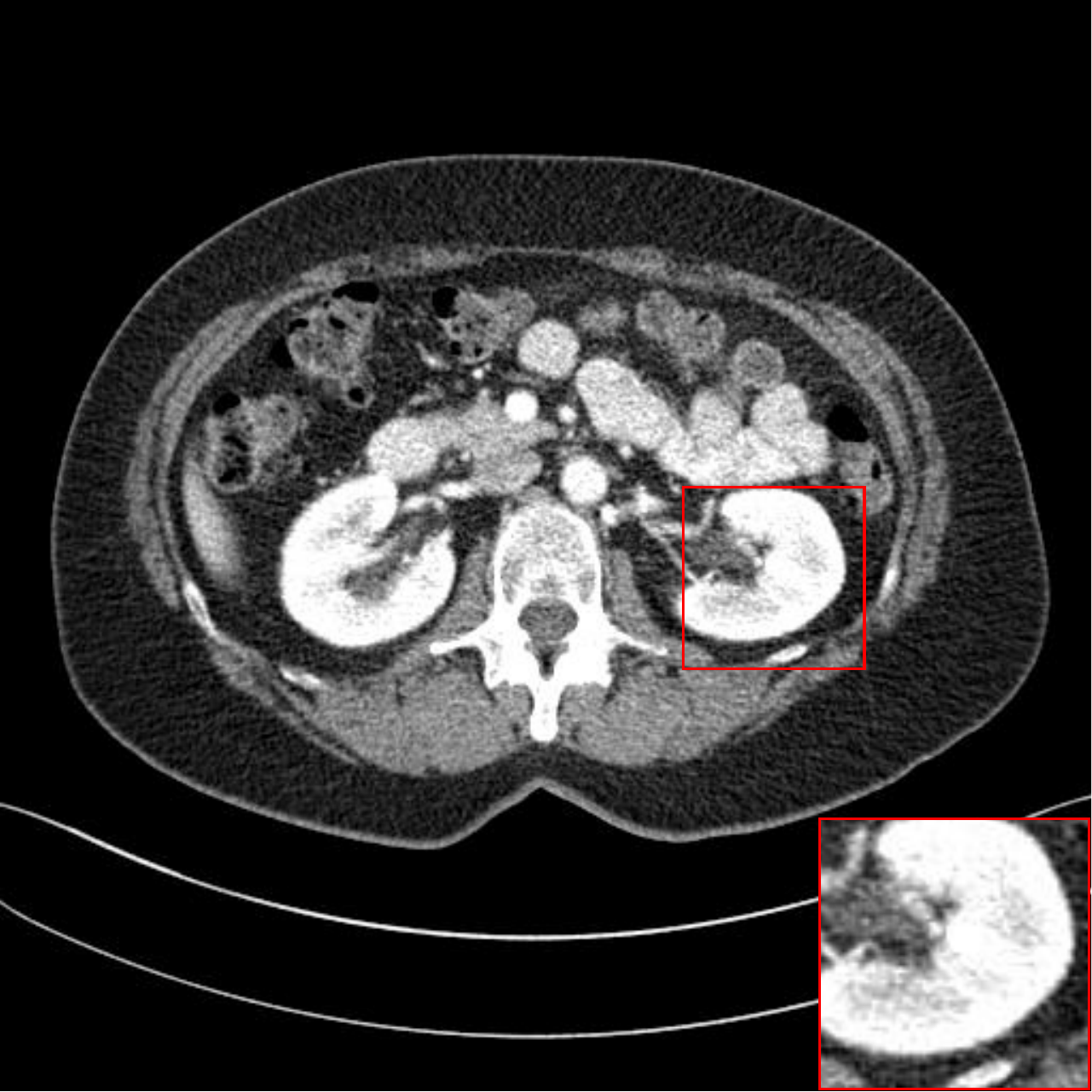}}
    \subfigure[BM3D]{\includegraphics[width=.24\textwidth]{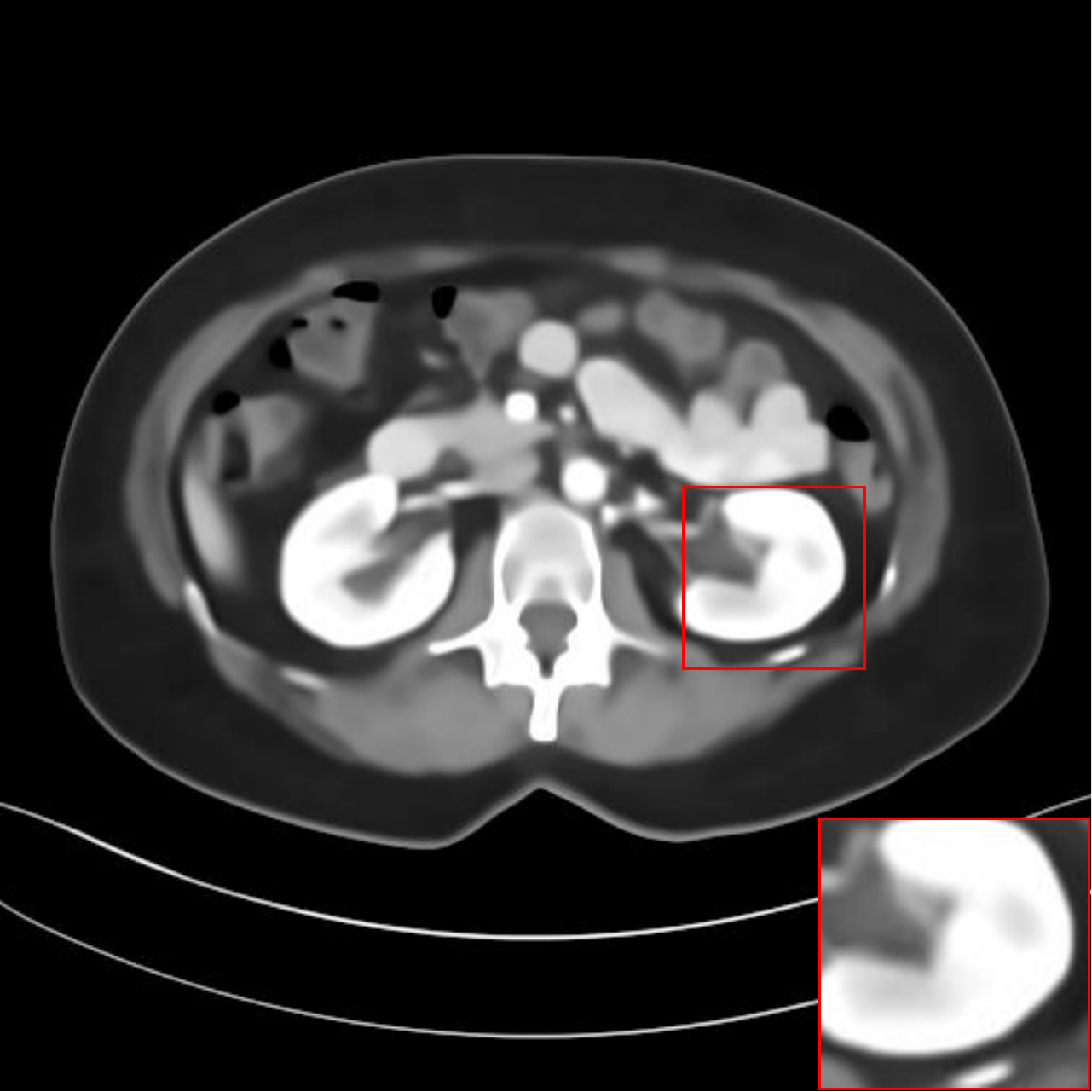}}
    \subfigure[RED-CNN]{\includegraphics[width=.24\textwidth]{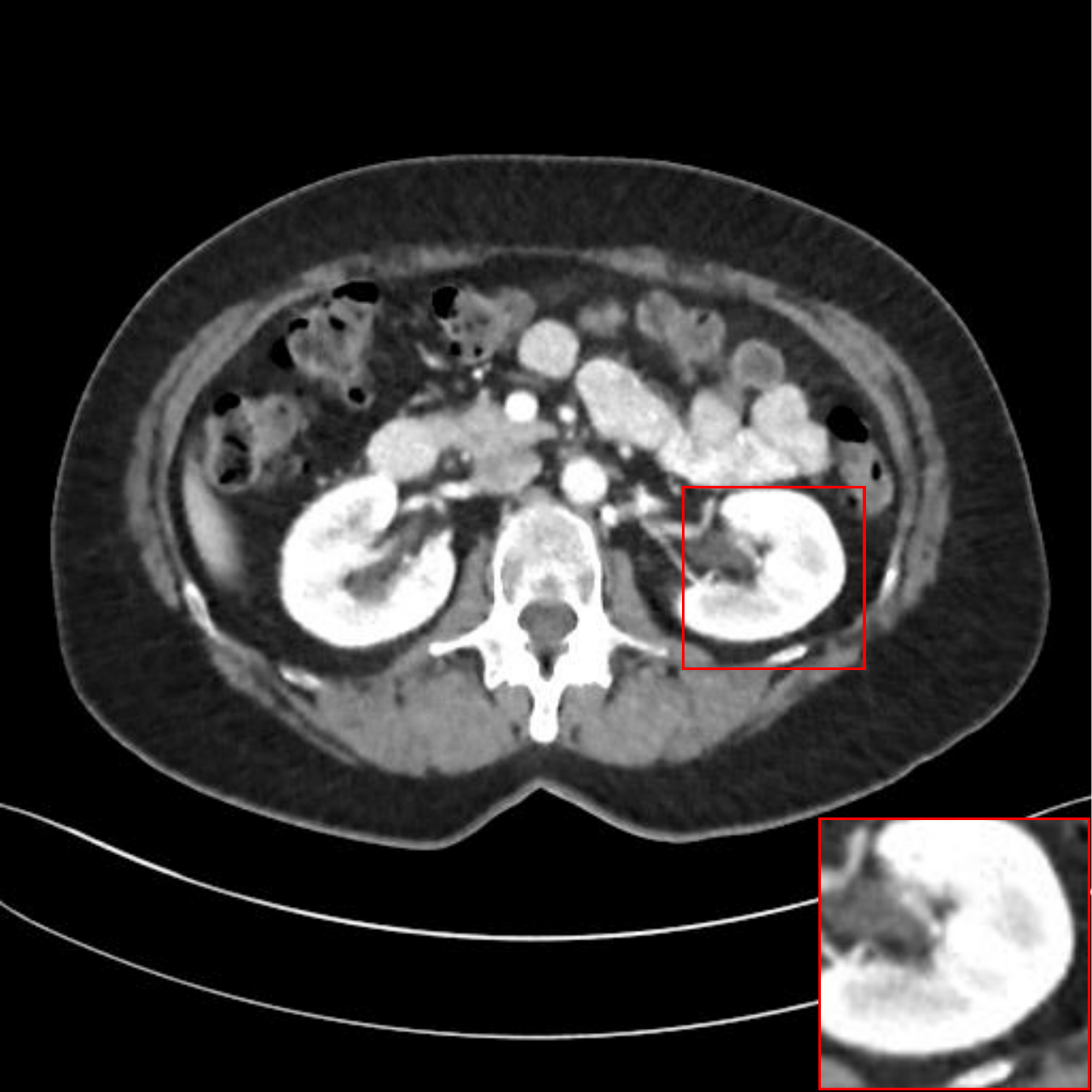}}
    \subfigure[DIP]{\includegraphics[width=.24\textwidth]{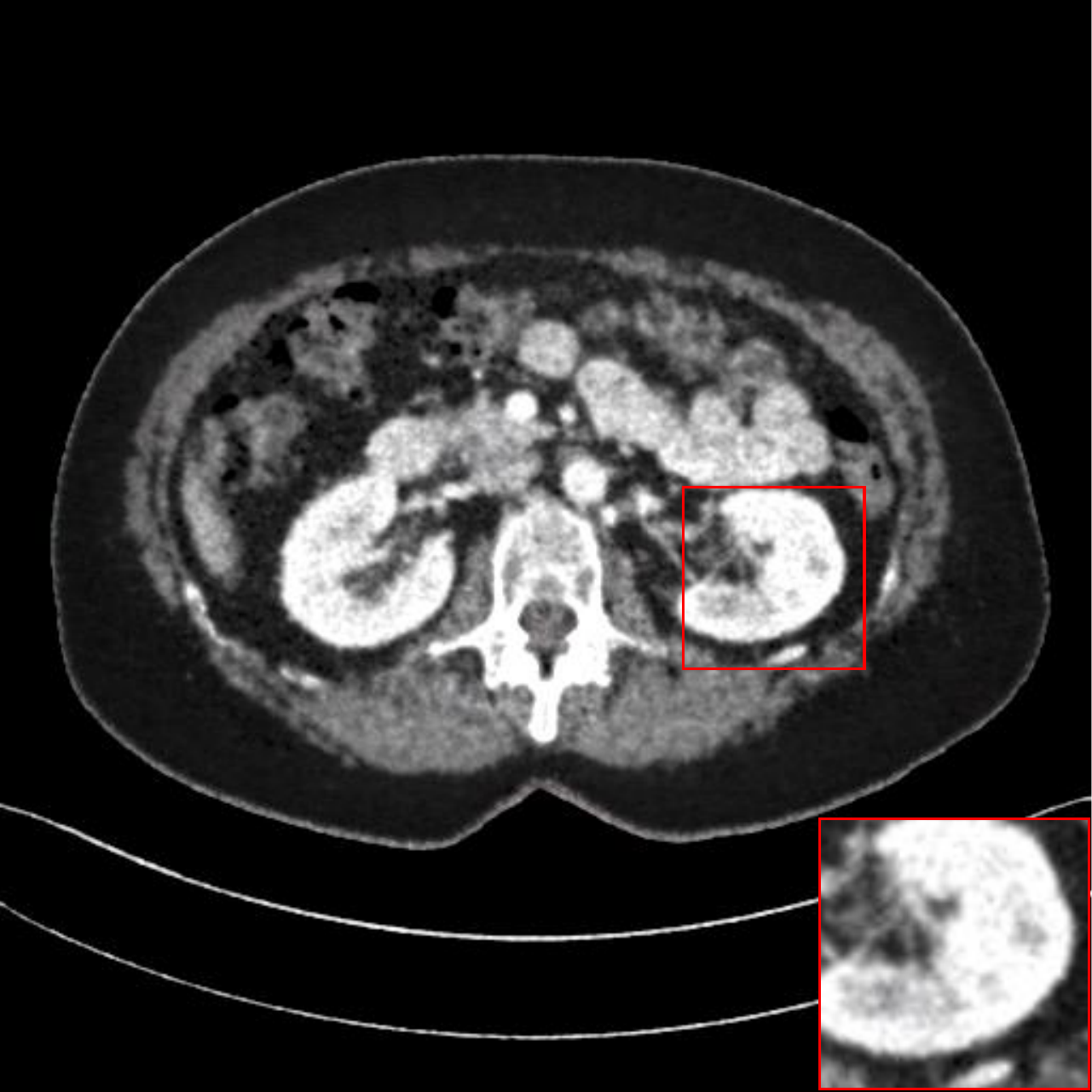}}
    \subfigure[LIR]{\includegraphics[width=.24\textwidth]{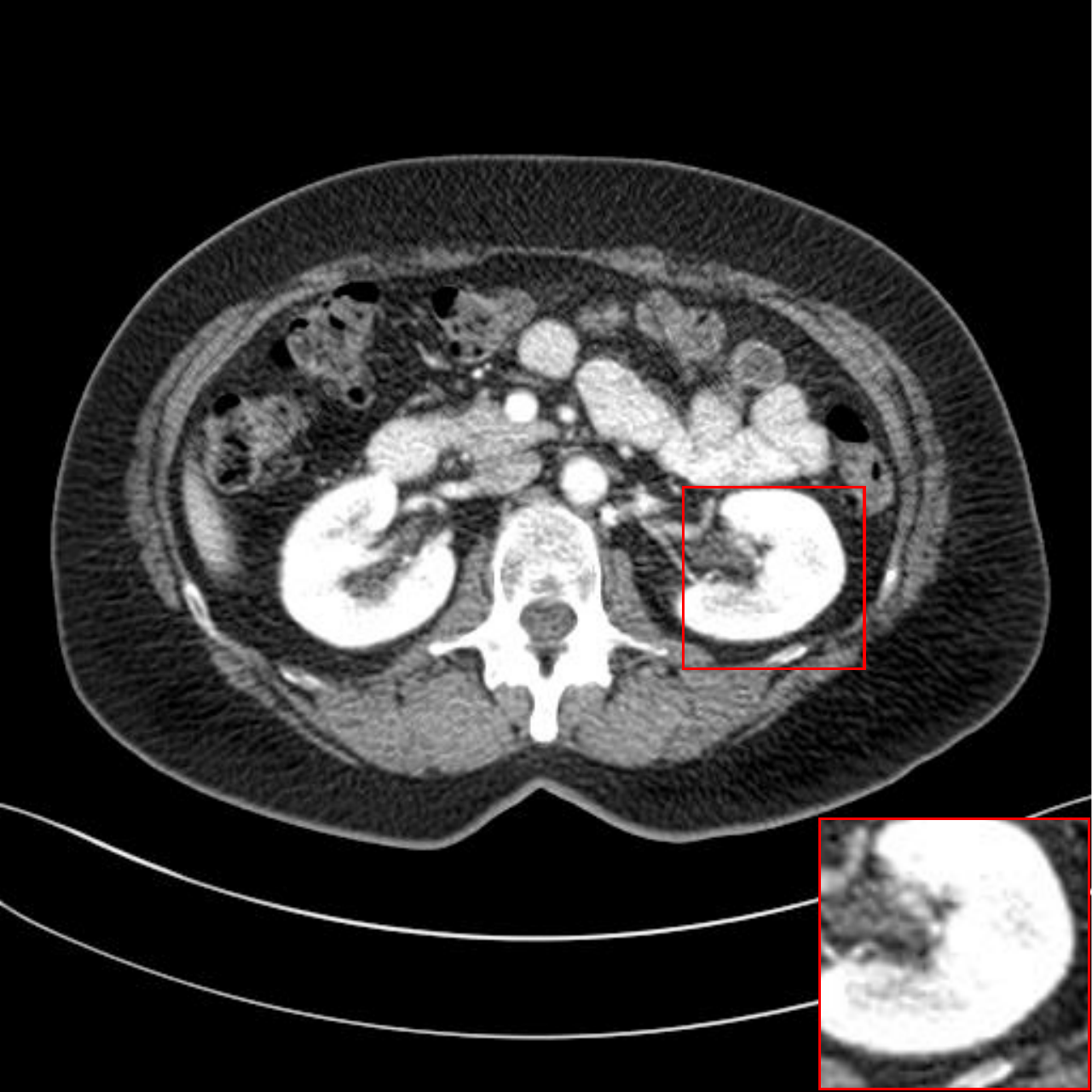}}
    \subfigure[Ours]{\includegraphics[width=.24\textwidth]{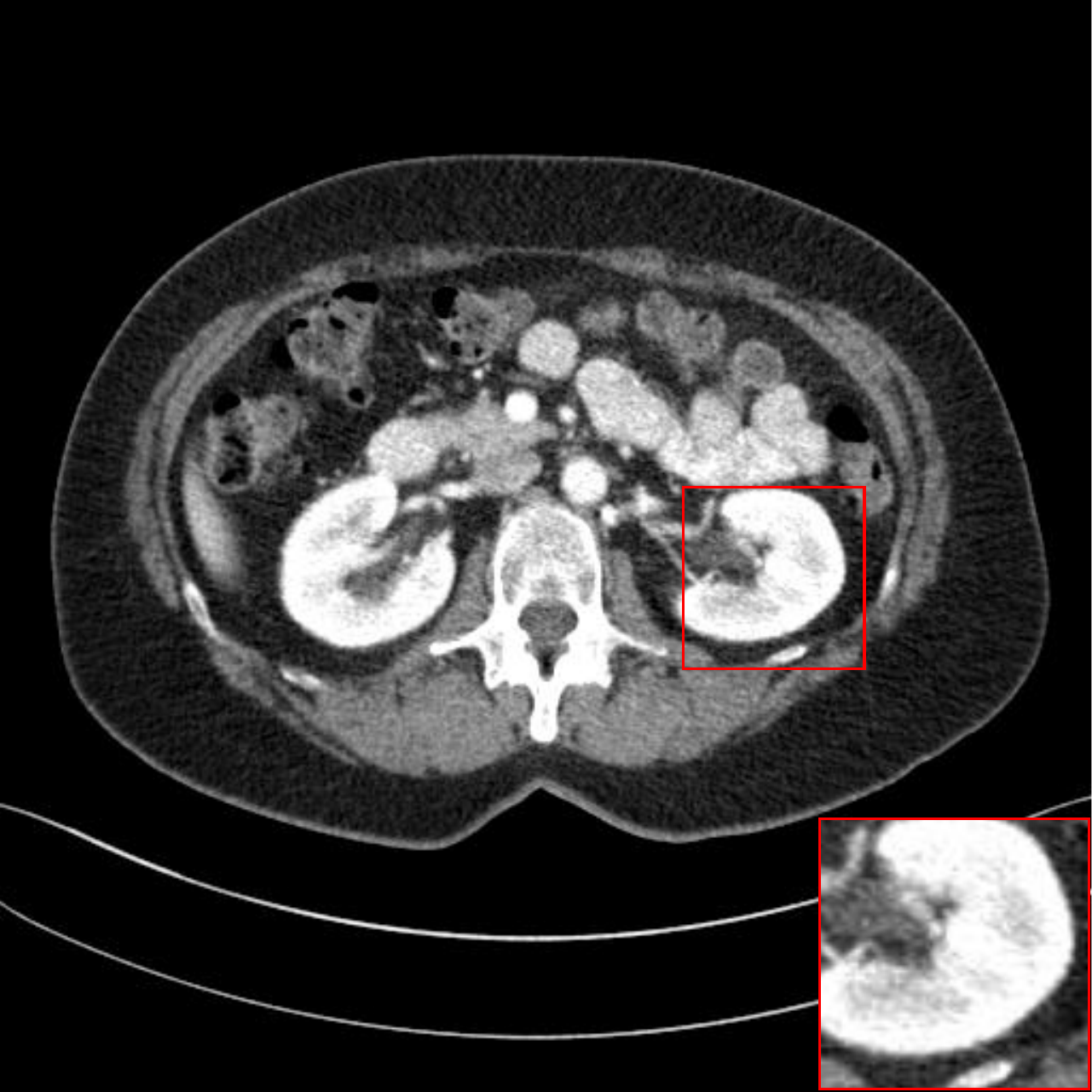}}
    \subfigure[NDCT]{\includegraphics[width=.24\textwidth]{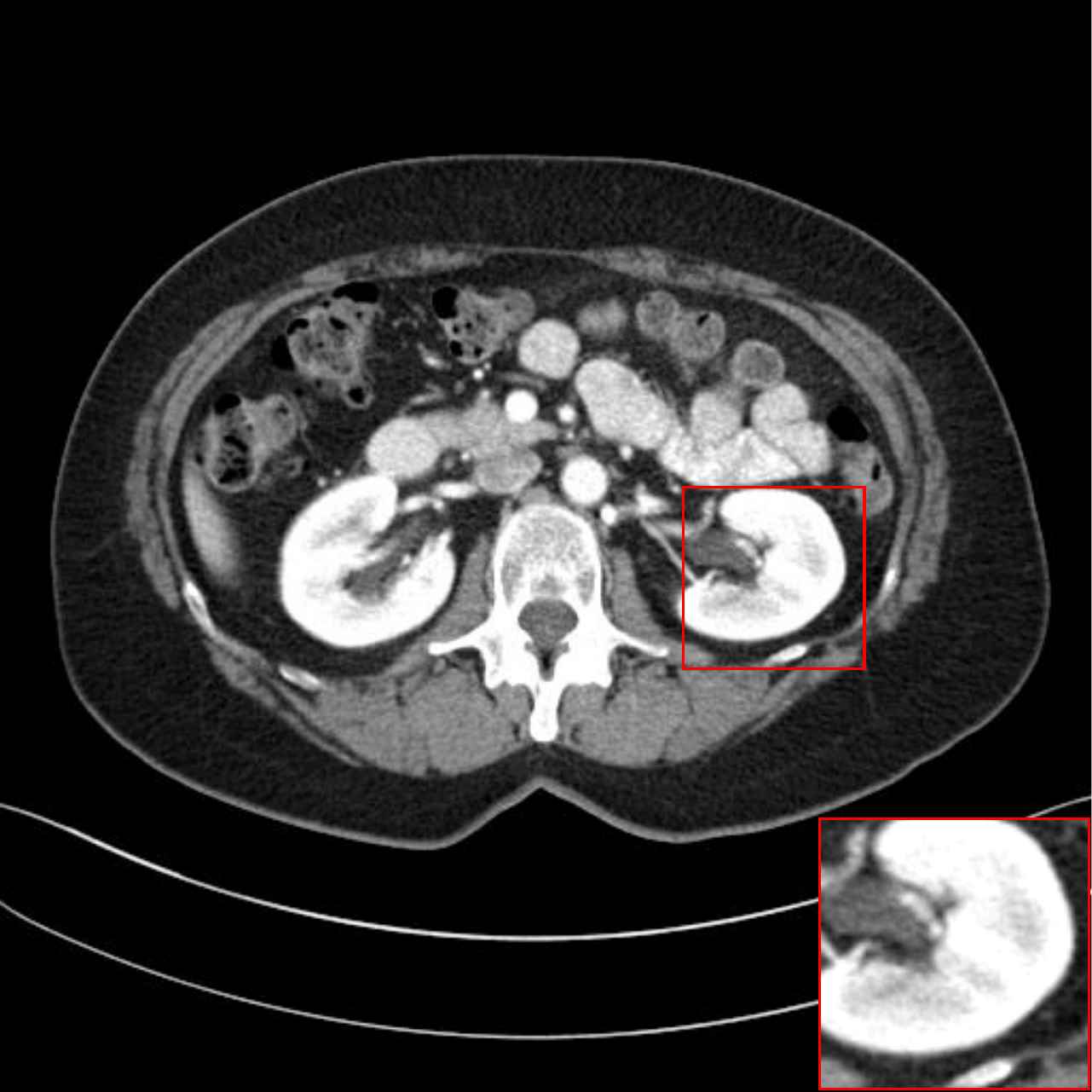}}
    
    \subfigure[LDCT]{\includegraphics[width=.24\textwidth]{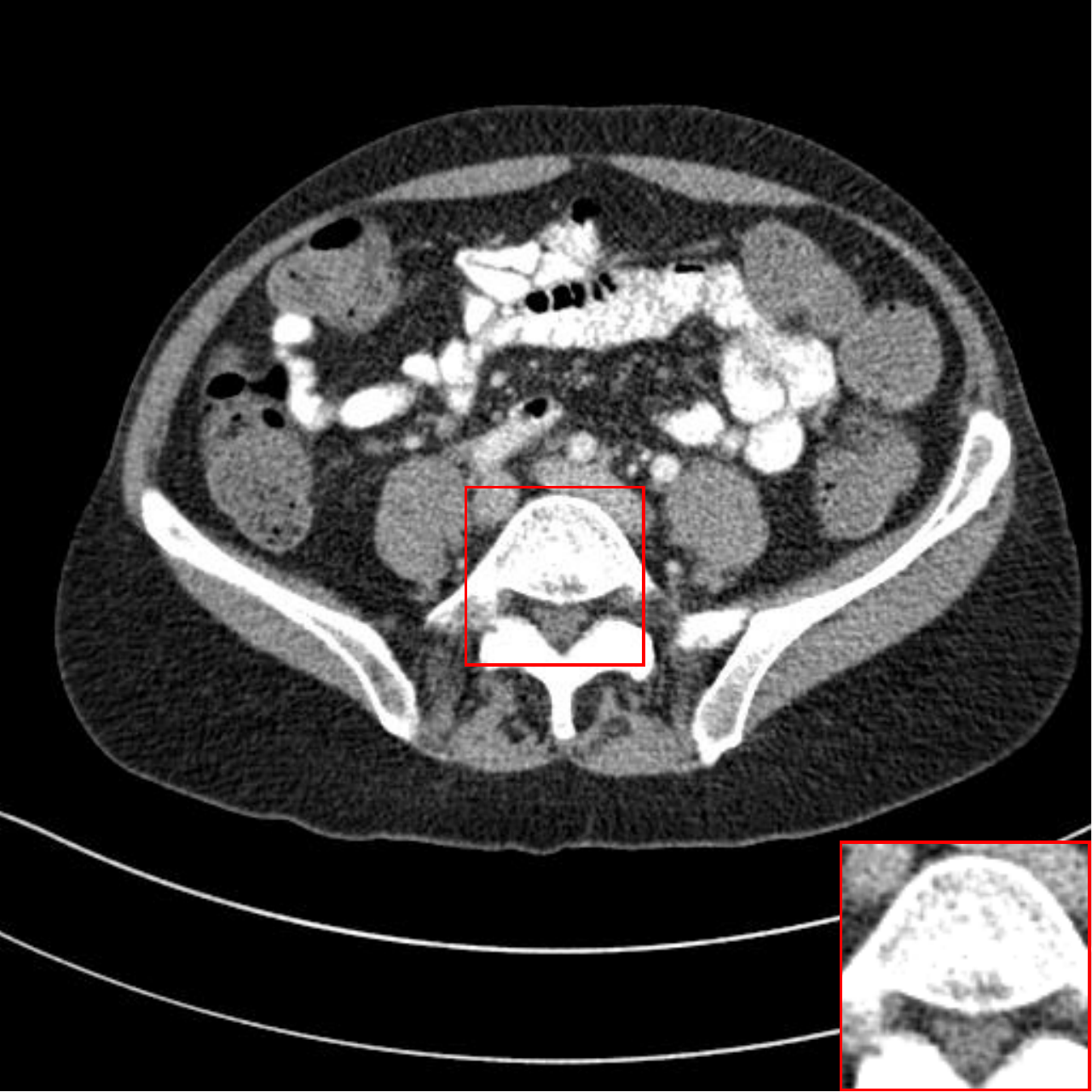}}
    \subfigure[BM3D]{\includegraphics[width=.24\textwidth]{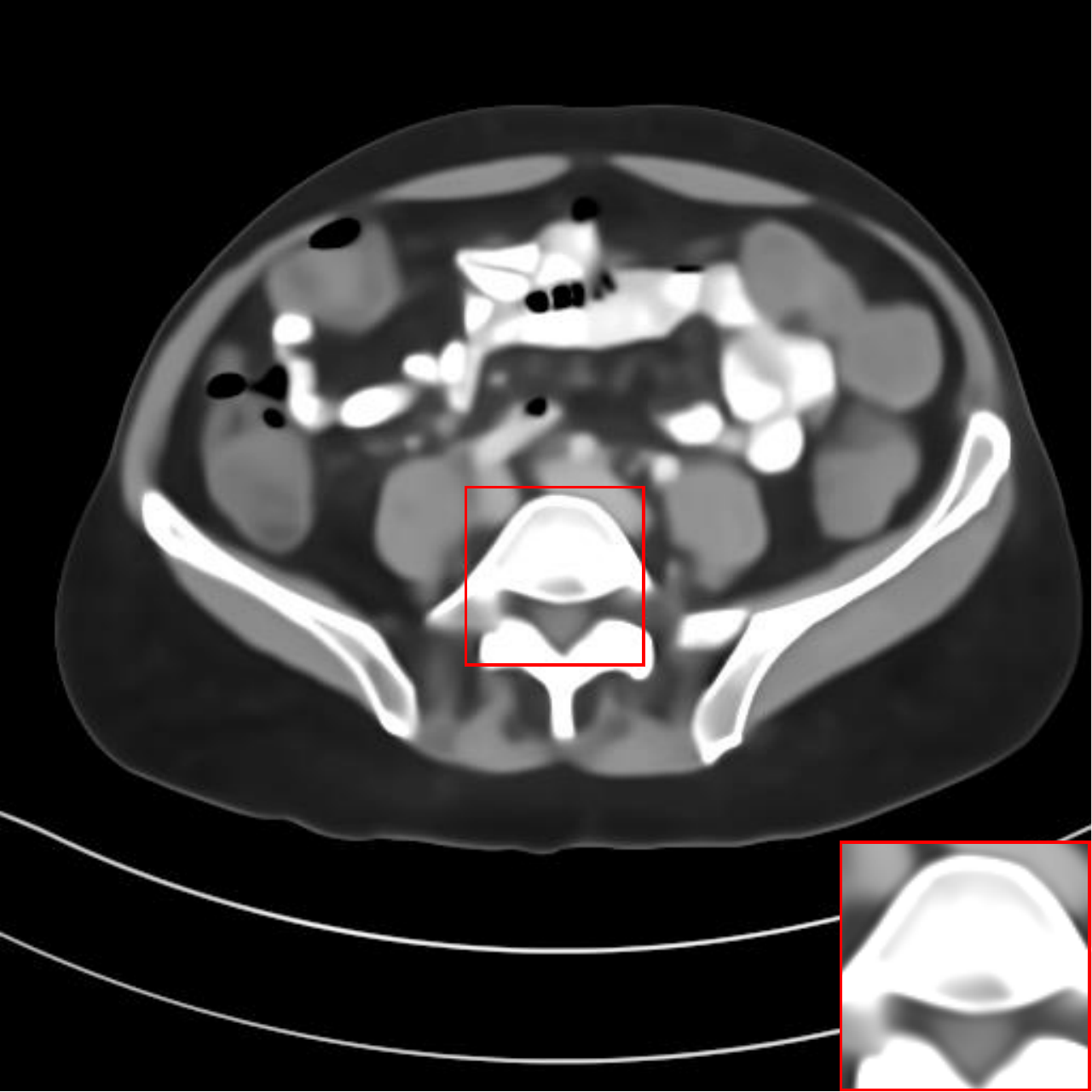}}
    \subfigure[RED-CNN]{\includegraphics[width=.24\textwidth]{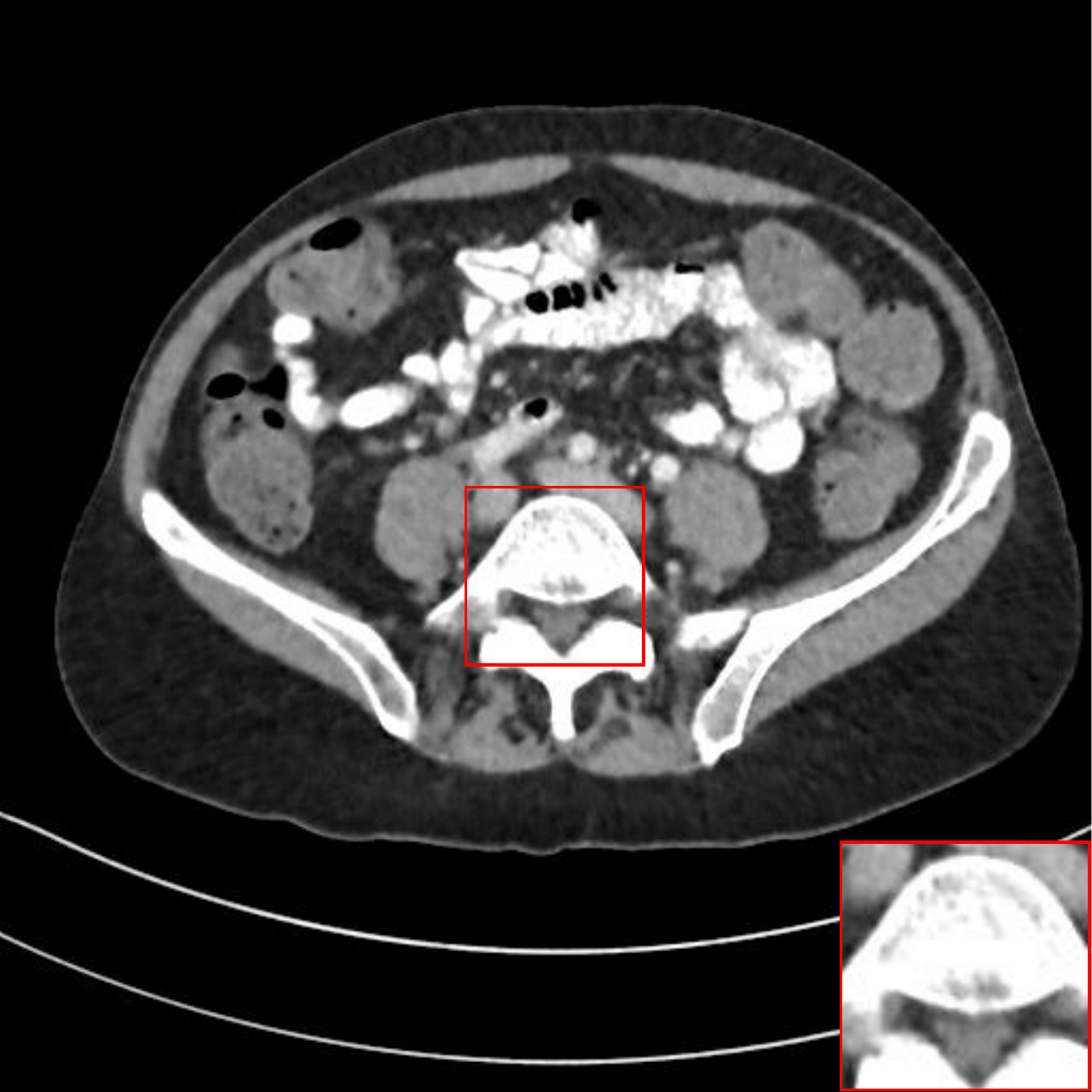}}
    \subfigure[DIP]{\includegraphics[width=.24\textwidth]{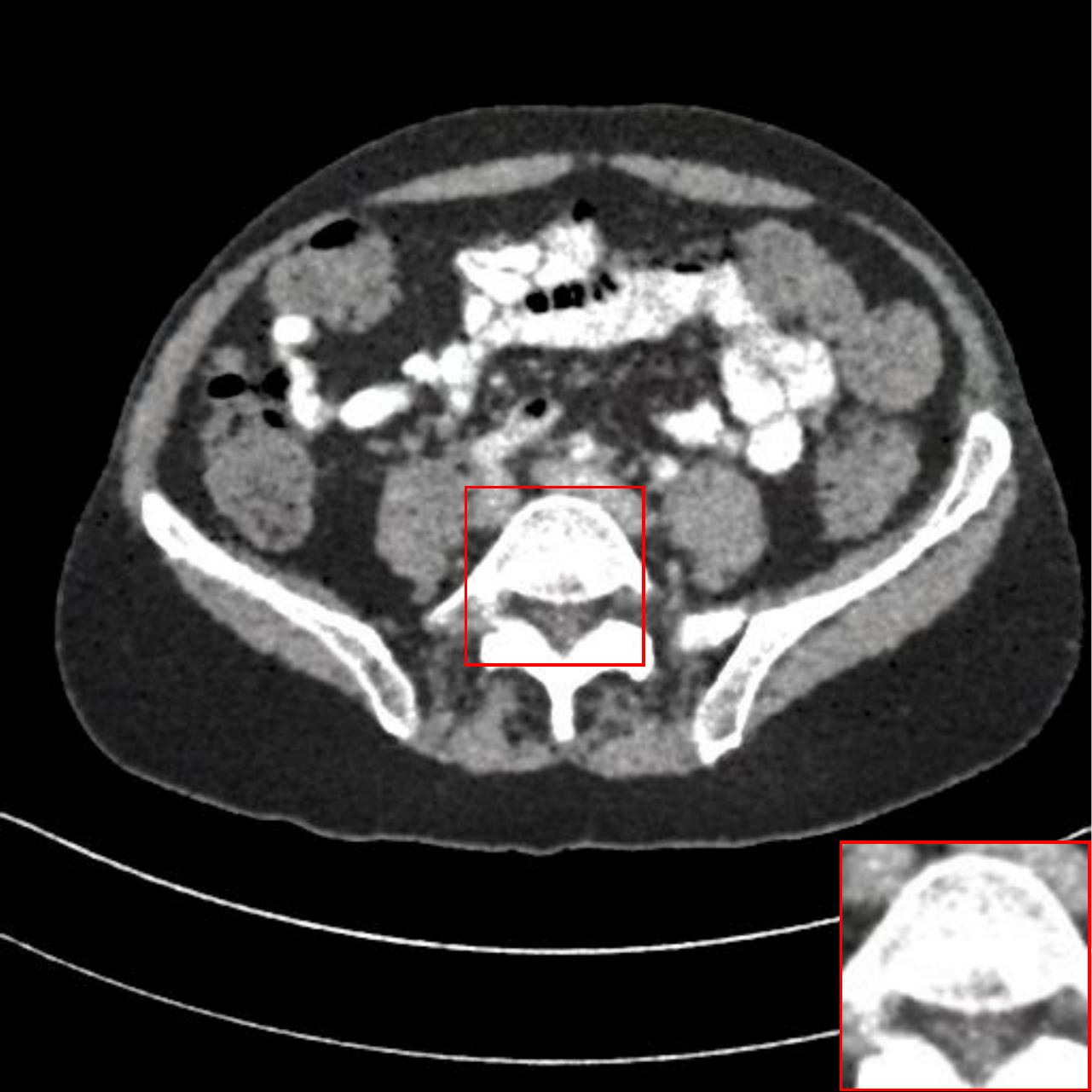}}
    \subfigure[LIR]{\includegraphics[width=.24\textwidth]{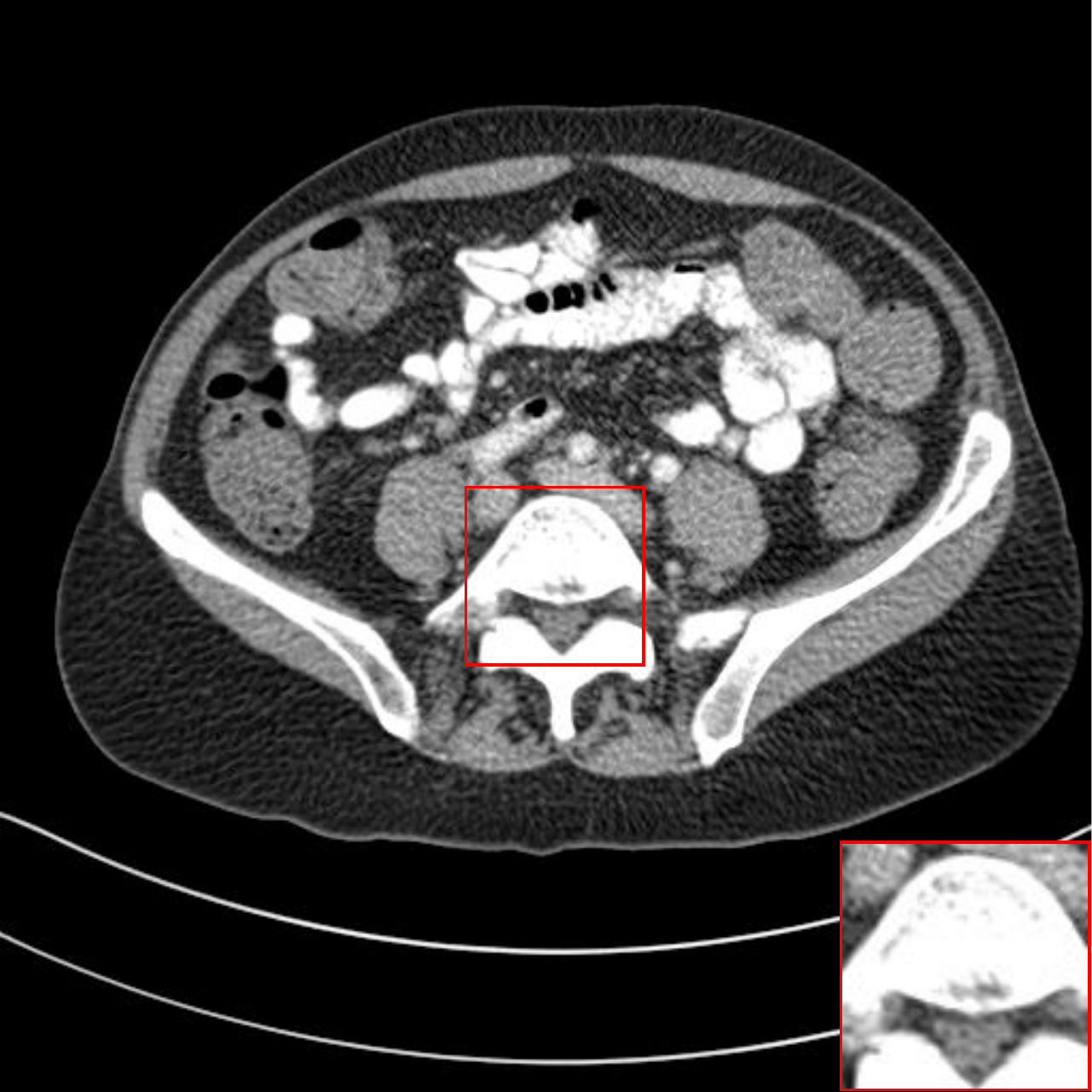}}
    \subfigure[Ours]{\includegraphics[width=.24\textwidth]{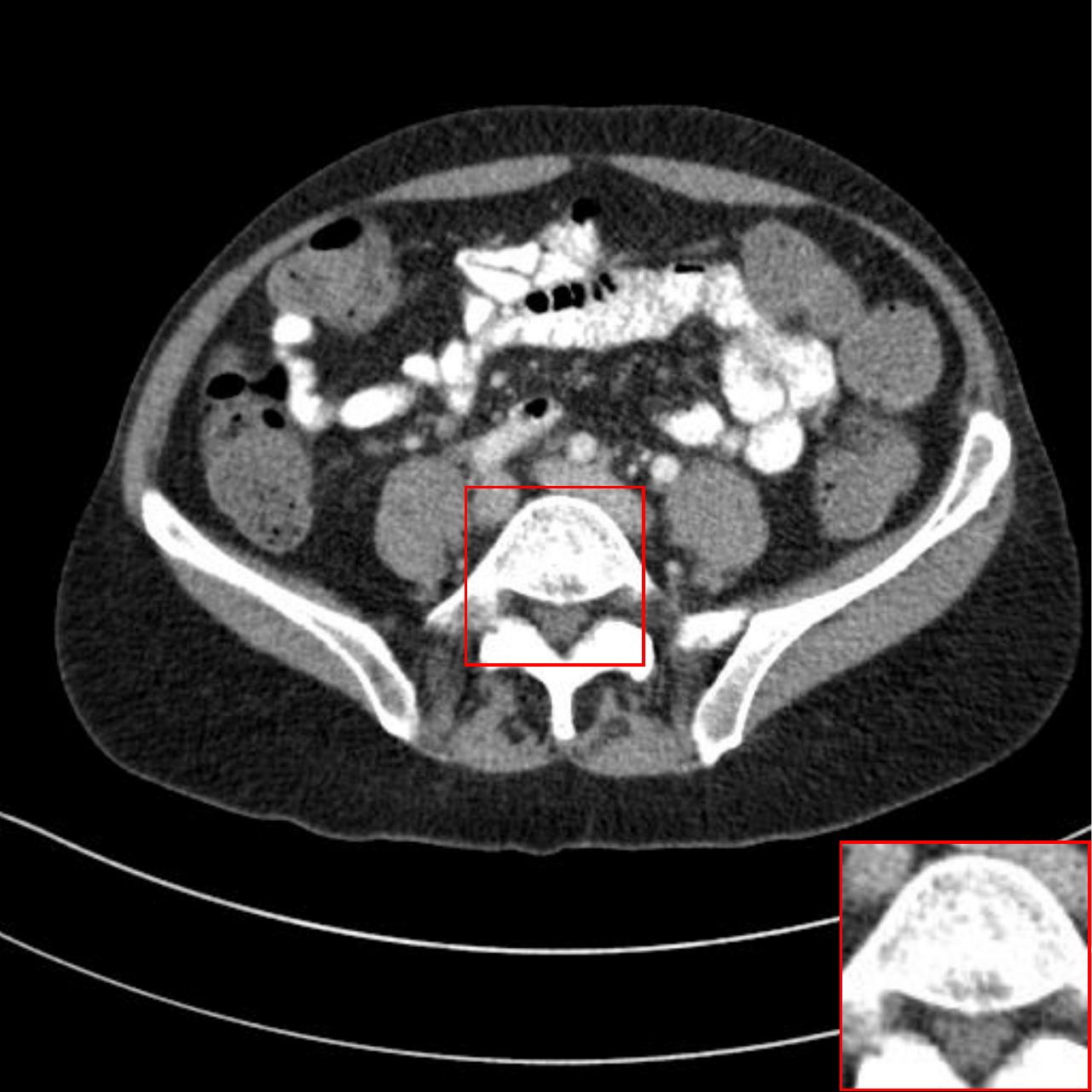}}
    \subfigure[NDCT]{\includegraphics[width=.24\textwidth]{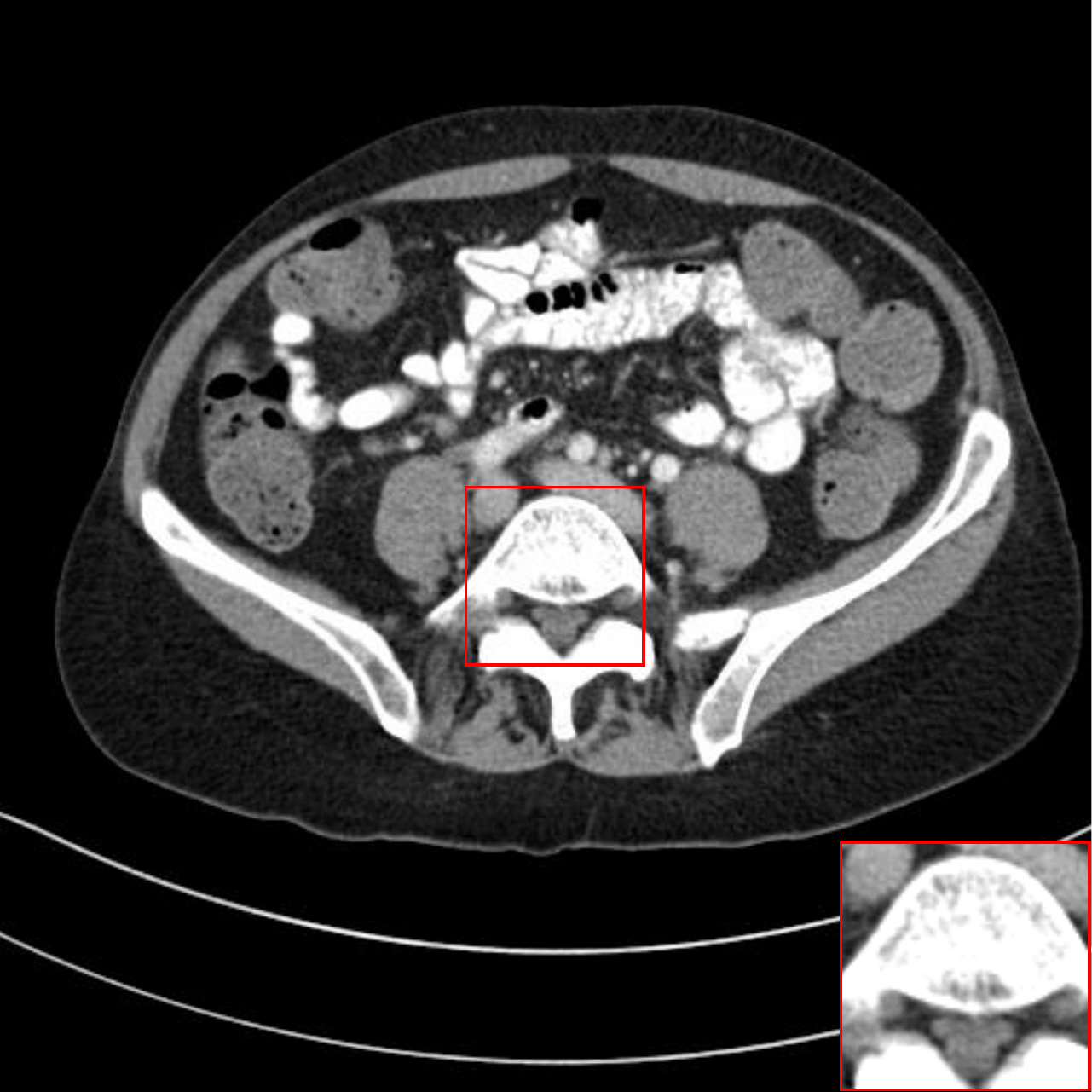}}
    \caption{Qualitative results of our method and other baselines on \textit{Mayo Clinic Low Dose CT dataset}. As shown in the highlighted red box, the reconstructed images by our method have few noise and preserve the details of organs. The display window is $[160, 240]$ HU.}
    \label{fig:CT_supp}
\end{figure*}

In this subsection, we show the additional qualitative results on Low-Dose CT dataset authorized by Mayo Clinic \cite{moen2021low} in Figure \ref{fig:CT_supp}. As shown in Figure \ref{fig:CT_supp}, previous methods tend to lose details and generate blurred results. However, our method removes the noise, while preserving the details of organs. It shows that our method is also practical for medical image denoising.

\subsection{Real Photographs}
\begin{figure*}[hp]
    \centering
    \subfigure[Input]{\includegraphics[width=.24\textwidth]{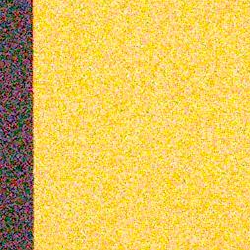}}
    \subfigure[CBM3D]{\includegraphics[width=.24\textwidth]{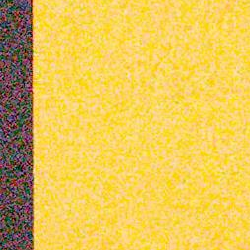}}
    \subfigure[RedNet-30]{\includegraphics[width=.24\textwidth]{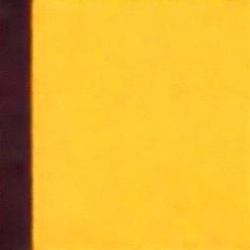}}
    \subfigure[DIP]{\includegraphics[width=.24\textwidth]{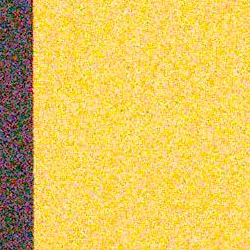}}
    \subfigure[LIR]{\includegraphics[width=.24\textwidth]{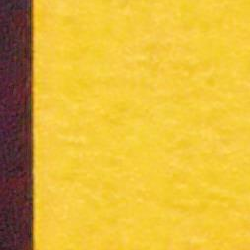}}
    \subfigure[Ours]{\includegraphics[width=.24\textwidth]{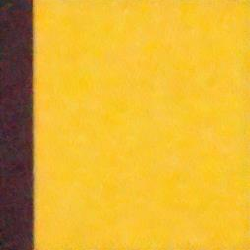}}
    \subfigure[GT]{\includegraphics[width=.24\textwidth]{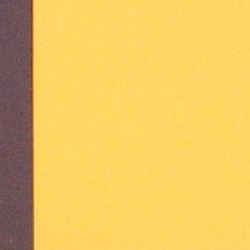}}
    
    \subfigure[Input]{\includegraphics[width=.24\textwidth]{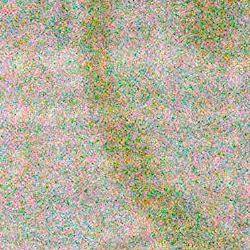}}
    \subfigure[CBM3D]{\includegraphics[width=.24\textwidth]{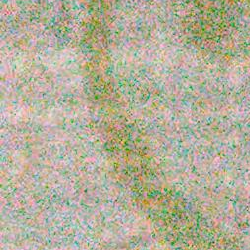}}
    \subfigure[RedNet-30]{\includegraphics[width=.24\textwidth]{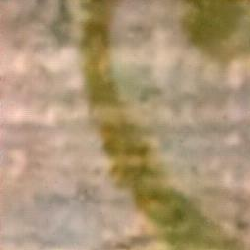}}
    \subfigure[DIP]{\includegraphics[width=.24\textwidth]{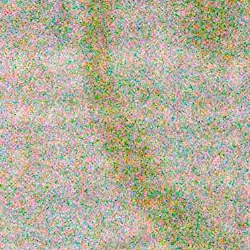}}
    \subfigure[LIR]{\includegraphics[width=.24\textwidth]{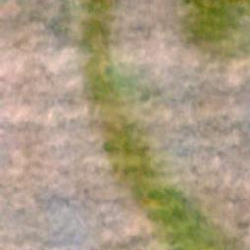}}
    \subfigure[Ours]{\includegraphics[width=.24\textwidth]{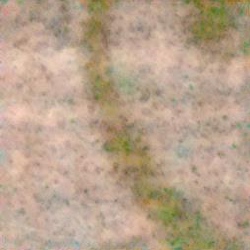}}
    \subfigure[GT]{\includegraphics[width=.24\textwidth]{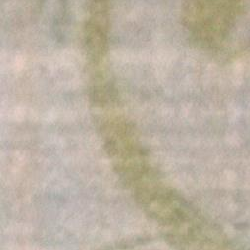}}

    \caption{Qualitative results of our method and other baselines on real noisy data, \textit{SIDD}.}
    \label{fig:SIDD_qual1}
\end{figure*}

\begin{figure*}[hp]
    \centering
    \subfigure[Input]{\includegraphics[width=.24\textwidth]{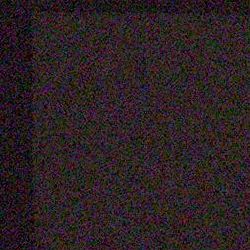}}
    \subfigure[CBM3D]{\includegraphics[width=.24\textwidth]{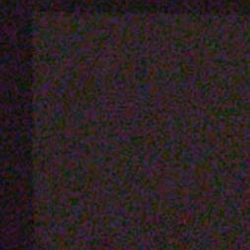}}
    \subfigure[RedNet-30]{\includegraphics[width=.24\textwidth]{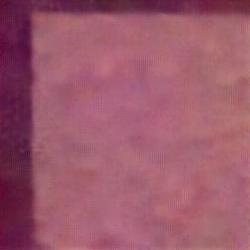}}
    \subfigure[DIP]{\includegraphics[width=.24\textwidth]{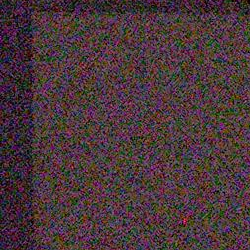}}
    \subfigure[LIR]{\includegraphics[width=.24\textwidth]{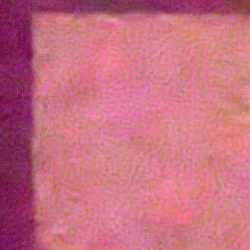}}
    \subfigure[Ours]{\includegraphics[width=.24\textwidth]{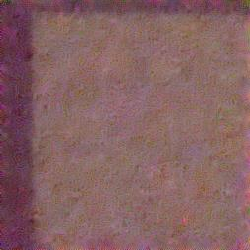}}
    \subfigure[GT]{\includegraphics[width=.24\textwidth]{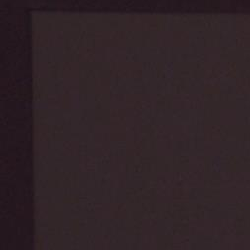}}
    
    \subfigure[Input]{\includegraphics[width=.24\textwidth]{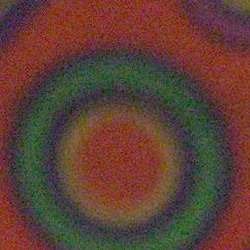}}
    \subfigure[CBM3D]{\includegraphics[width=.24\textwidth]{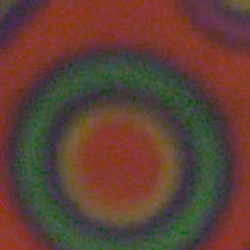}}
    \subfigure[RedNet-30]{\includegraphics[width=.24\textwidth]{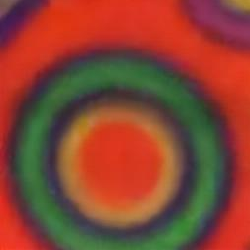}}
    \subfigure[DIP]{\includegraphics[width=.24\textwidth]{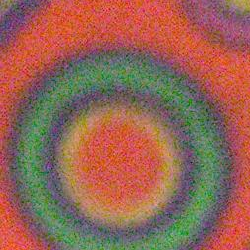}}
    \subfigure[LIR]{\includegraphics[width=.24\textwidth]{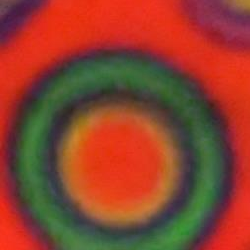}}
    \subfigure[Ours]{\includegraphics[width=.24\textwidth]{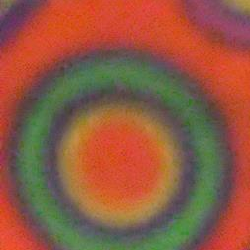}}
    \subfigure[GT]{\includegraphics[width=.24\textwidth]{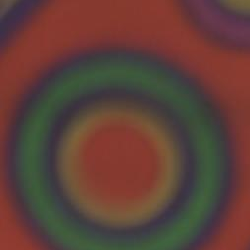}}
    
    \caption{Qualitative results of our method and other baselines on real noisy data, \textit{SIDD}.}
    \label{fig:SIDD_qual2}
\end{figure*}

In this subsection, we visualize the additional qualitative results on SIDD \cite{abdelhamed2018sidd} in Figure \ref{fig:SIDD_qual1} and \ref{fig:SIDD_qual2}. As shown in Figure \ref{fig:SIDD_qual1}, previous methods tend to lose the texture and leave the noise. In contrast, our method removes the noise while preserving the texture compared to other baselines. In Figure \ref{fig:SIDD_qual2}, we observe that our method removes the intense noise while preserving the color of images compared to other baselines.

\section{Additional Ablation Study}
\label{ablation}
We conduct an additional ablation study to demonstrate the validity of the perceptual loss $L_{VGG}$, the cycle consistency loss $L_{CC}$, and the reconstruction loss $L_{Recon}$. First, to verify the effectiveness of the $L_{VGG}$, we only add the $L_{VGG}$. As shown in Table \ref{loss_ablation}, when the $L_{VGG}$ is used, both PSNR and SSIM increase by 0.07dB and 0.0068. It demonstrates that the perceptual loss $L_{VGG}$ helps to improve the performance, preserving the semantics even after the noise has been removed. Next, to verify the contribution of $L_{CC}$, we integrate it with the $L_{VGG}$. We observe that the $L_{CC}$ which enables the one-to-one mapping between noisy and denoised images improves the PSNR and SSIM by 0.08dB and 0.004. Finally, when we integrate the $L_{Recon}$ with the $L_{VGG}$ and the $L_{CC}$, both PSNR and SSIM increase by 0.15dB and 0.0063, thus showing the best results in PSNR and SSIM. Through this experiment, we validate that each of the losses contributes to the performance improvement.

\begin{table}[htbp]
\begin{center}
\begin{tabular}{|ccc|cc|}
\hline
$L_{VGG}$ & $L_{CC}$ & $L_{Recon}$ & PSNR (dB) & SSIM  \\ \hline \hline
\xmark &\xmark &\xmark &25.67 &0.8204 \\ 
\cmark &\xmark &\xmark &25.74 &0.8272 \\
\cmark &\cmark &\xmark &25.88 &0.8312  \\
\cmark &\cmark &\cmark &\textbf{26.03} &\textbf{0.8375} \\\hline
\end{tabular}
\end{center}
\caption{Ablation study. Quantitative results of our method with and without the perceptual loss $L_{VGG}$, the cycle consistency loss $L_{CC}$, and the reconstruction loss $L_{Recon}$ on CBSD68 corrupted by AWGN with a noise level $\sigma=50$. We report the PSNR and SSIM (higher is better). The best results are marked in \textbf{bold}.}
\label{loss_ablation}
\end{table}

\section{Evaluation on Several Noise Types}
\label{several}
\subsection{Structured Noise}
\begin{figure*}[htbp]
    \centering
    \subfigure[Input]{\includegraphics[width=.24\textwidth]{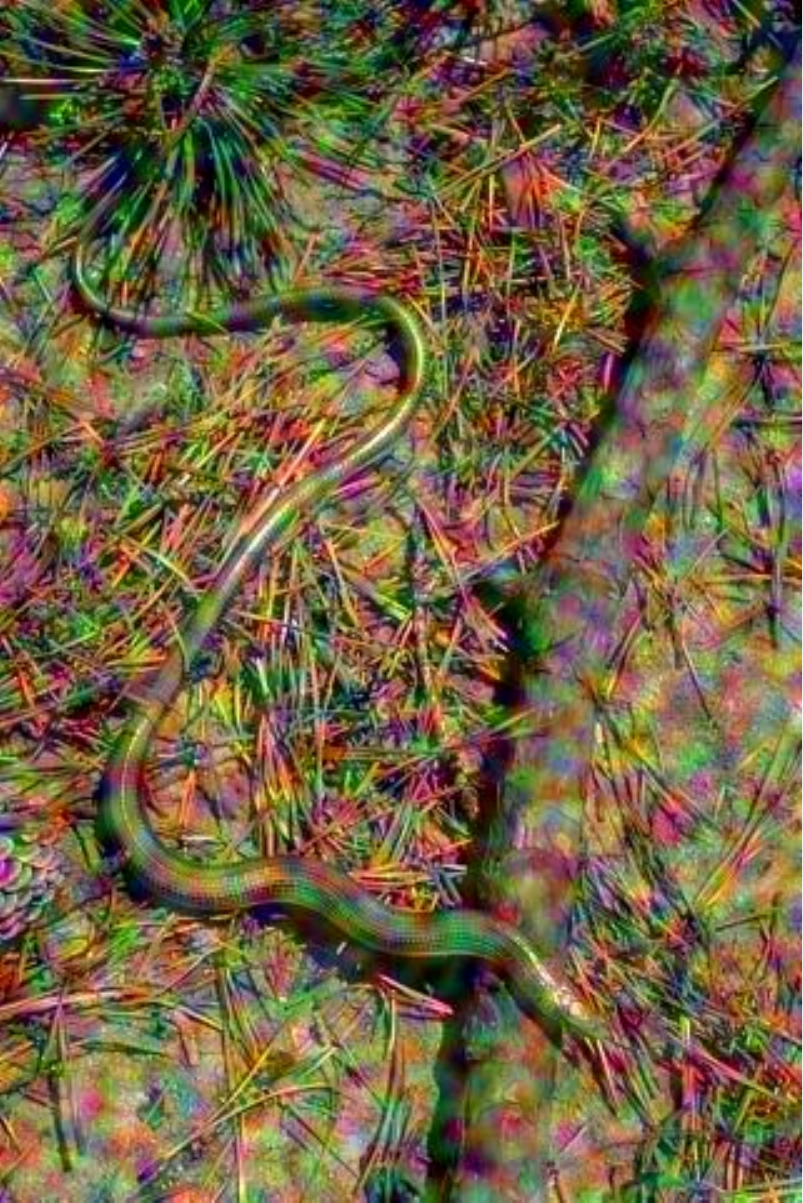}}
    \subfigure[CBM3D]{\includegraphics[width=.24\textwidth]{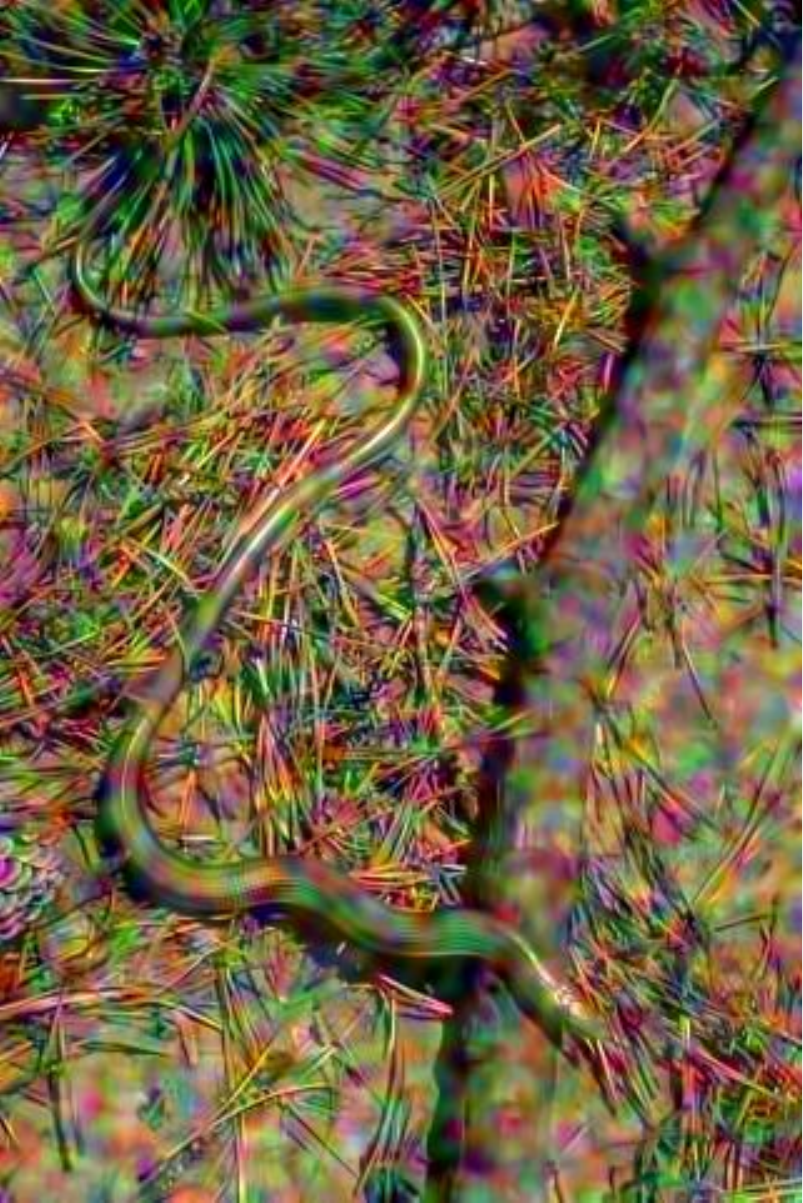}}
    \subfigure[RedNet-30]{\includegraphics[width=.24\textwidth]{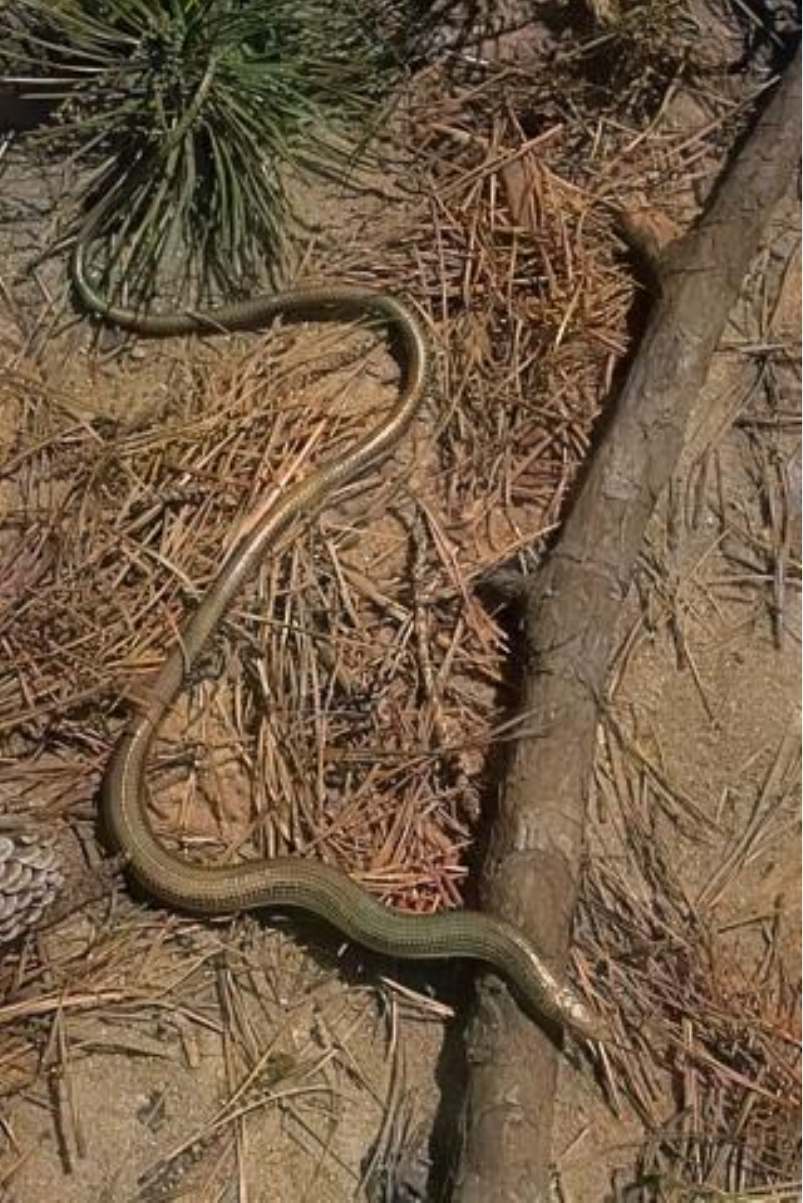}}
    \subfigure[DIP]{\includegraphics[width=.24\textwidth]{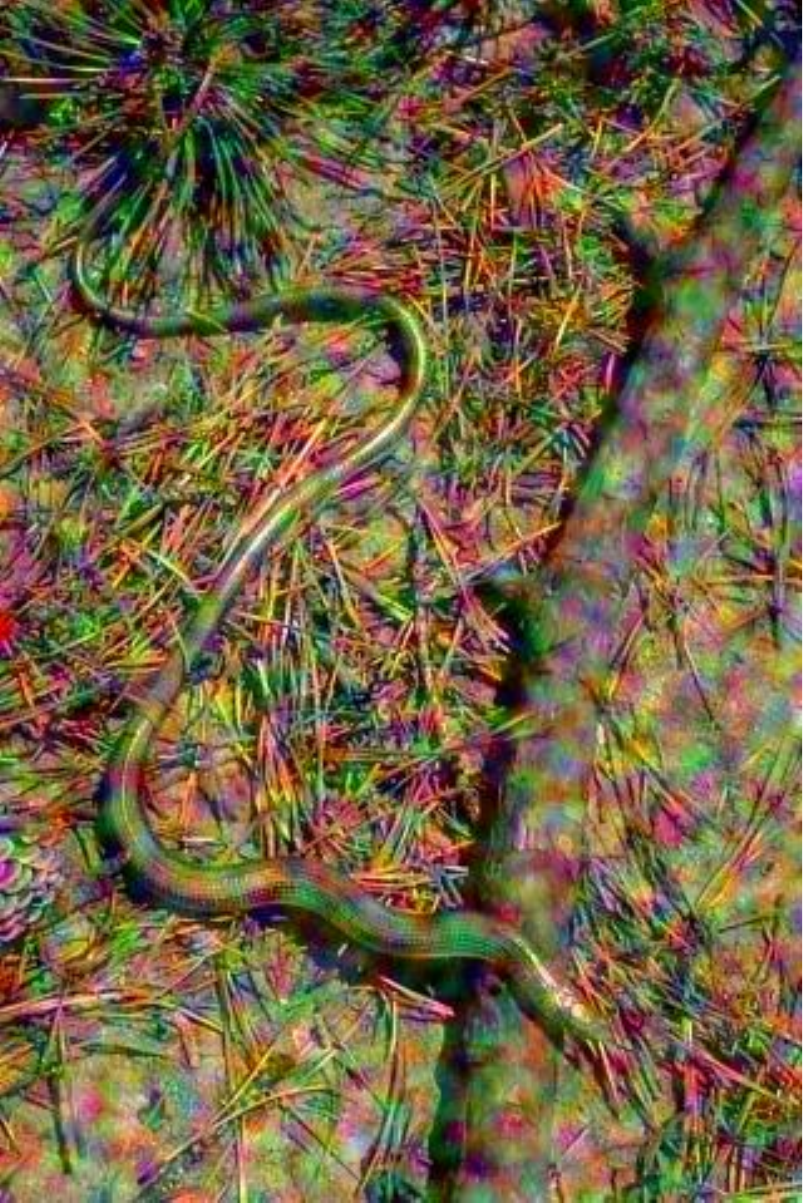}}
    \subfigure[LIR]{\includegraphics[width=.24\textwidth]{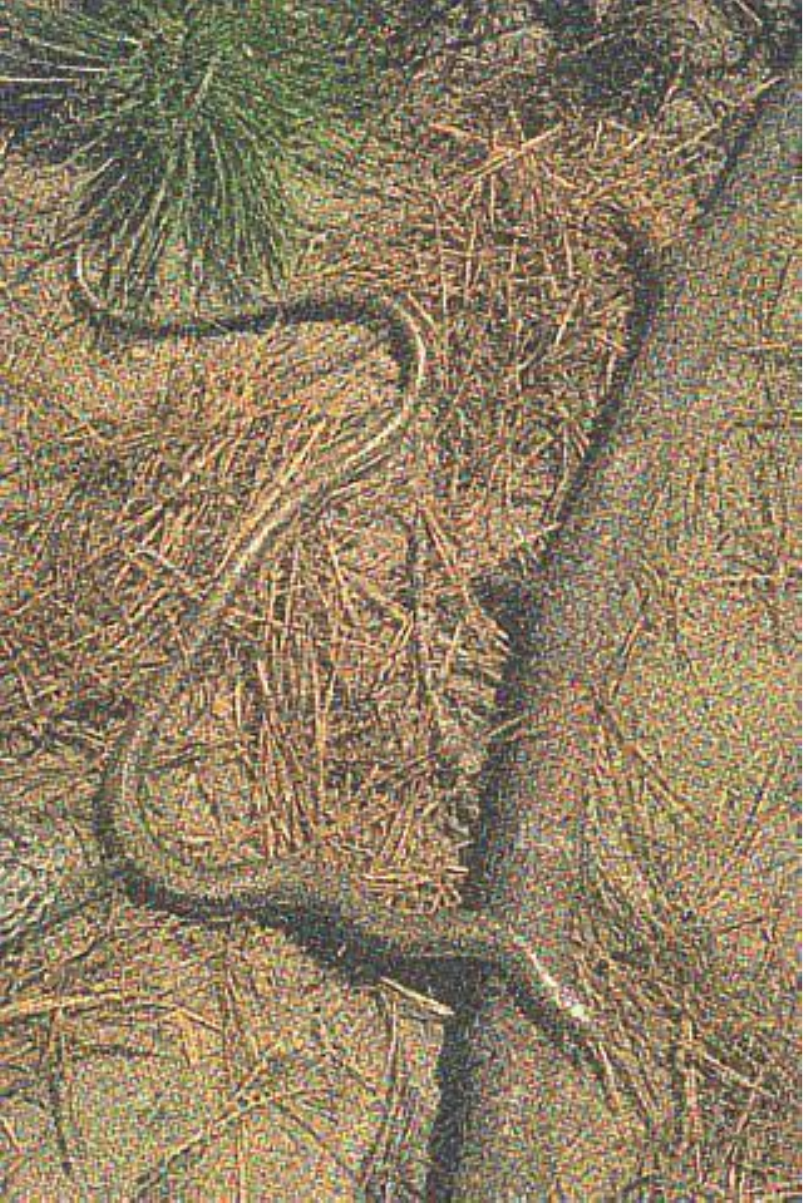}}
    \subfigure[Ours]{\includegraphics[width=.24\textwidth]{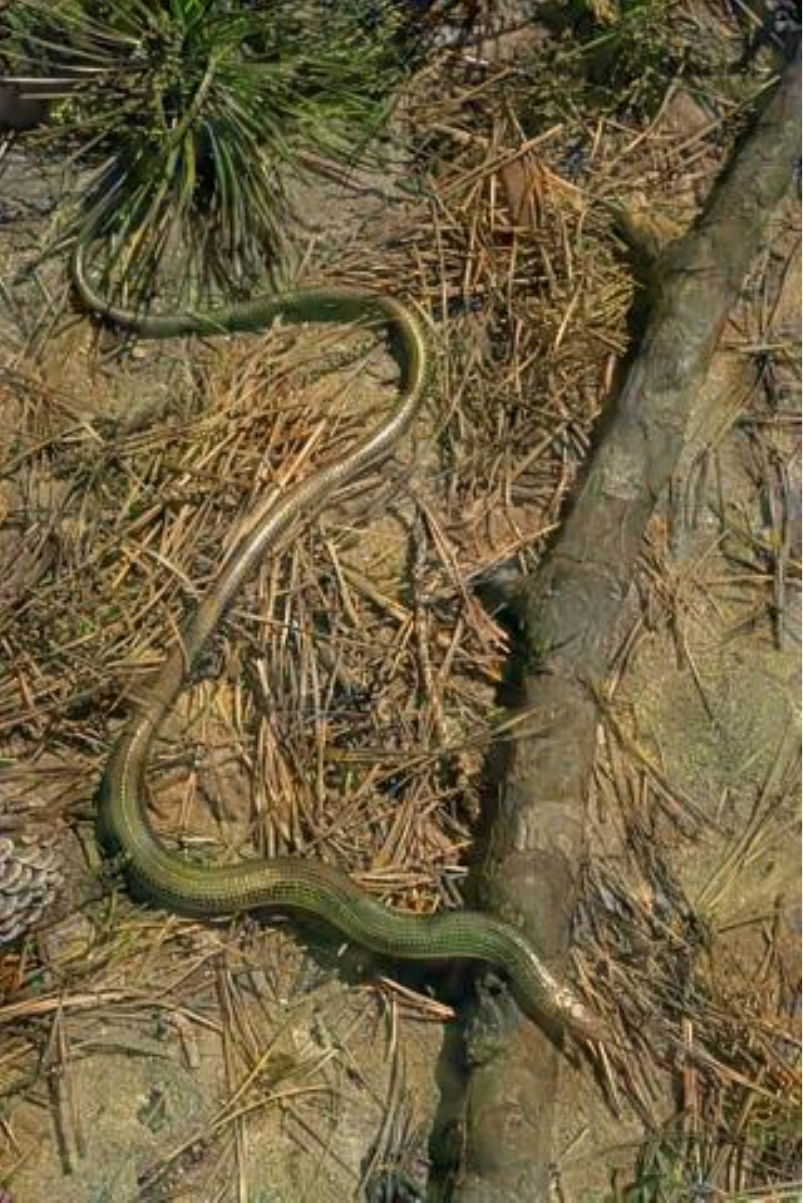}}
    \subfigure[GT]{\includegraphics[width=.24\textwidth]{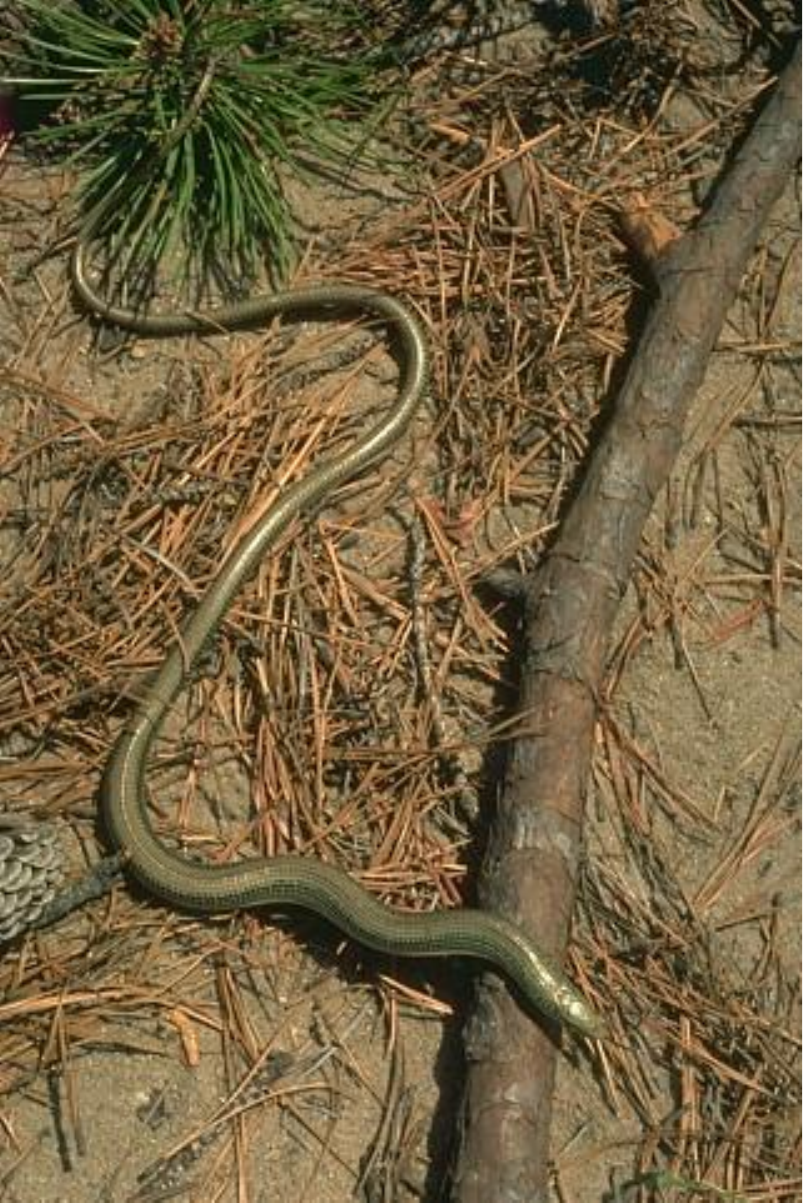}}
    
    \caption{Qualitative results of our method and other baselines on \textit{CBSD68} corrupted by structured noise.}
    \label{fig:structured_qual}
\end{figure*}

In this subsection, we show the results on structured noise. To generate the structured noise, we sample the pixel-wise i.i.d white noise, and convolve it with a 2D Gaussian filter whose a kernel size is $21 \times 21$ and $\sigma$ is $3$ pixel. For the train and evaluation, we follow the same setting as the setting for synthetic noise removal in the main paper. As shown in Figure \ref{fig:structured_qual}, our method is able to remove complex noise compared to BM3D \cite{bm3d} and DIP \cite{ulyanov2018deep}. Furthermore, while LIR \cite{lir} spoil the lights, our method successfully preserves both the color and lights. The quantitative results are summarized in Table \ref{table:struc}. Our method outperforms the traditional and unsupervised methods, achieving the second-best performance in terms of PSNR and SSIM.

\begin{table}[htbp]
\begin{center}
\resizebox{0.65\textwidth}{!}
{%
\begin{tabular}{|c|c|c|ccc|}
\hline
& \multicolumn{1}{c|}{Traditional} & \multicolumn{1}{c|}{Paired setting} &\multicolumn{3}{c|}{Unpaired setting} \\ \hline
\hline
Methods     & CBM3D \cite{bm3d} & RedNet-30 \cite{red30} & DIP \cite{ulyanov2018deep} & LIR \cite{lir} & Ours        \\ \hline
PSNR (dB)   & 20.62 & 28.51          & 20.70     & 16.90       & \textbf{25.18}       \\ \hline
SSIM        & 0.5650 & 0.9588         & 0.7239    & 0.3738      & \textbf{0.9026}      \\ \hline
\end{tabular}}
\end{center}
\caption{The average PSNR and SSIM results of different methods on \textit{CBSD68} corrupted by structured noise. Our results are marked in \textbf{bold}.}
\label{table:struc}
\end{table}

\subsection{Poisson Noise}
\begin{figure*}[htbp]
    \centering
    \subfigure[Input]{\includegraphics[width=.24\textwidth]{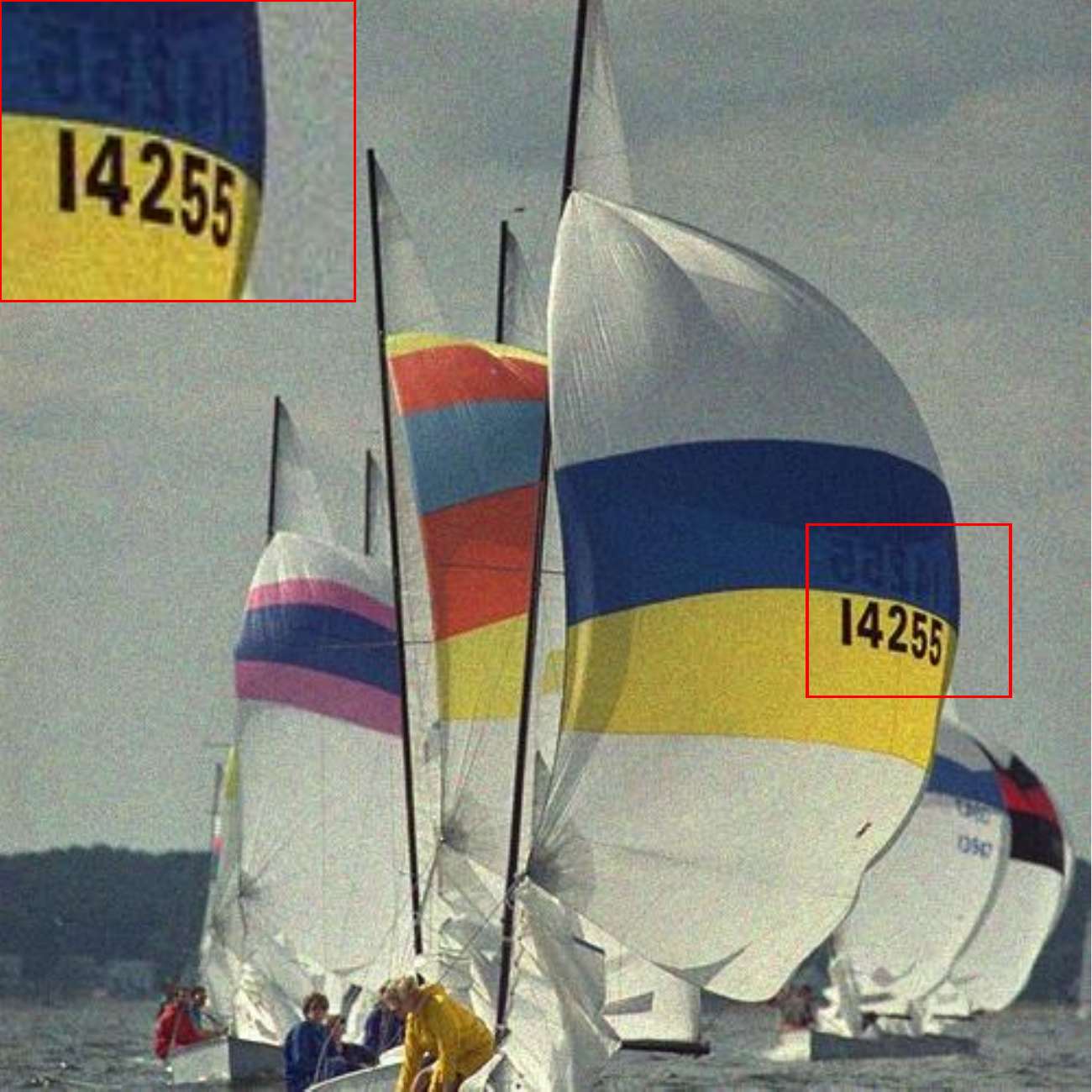}}
    \subfigure[CBM3D]{\includegraphics[width=.24\textwidth]{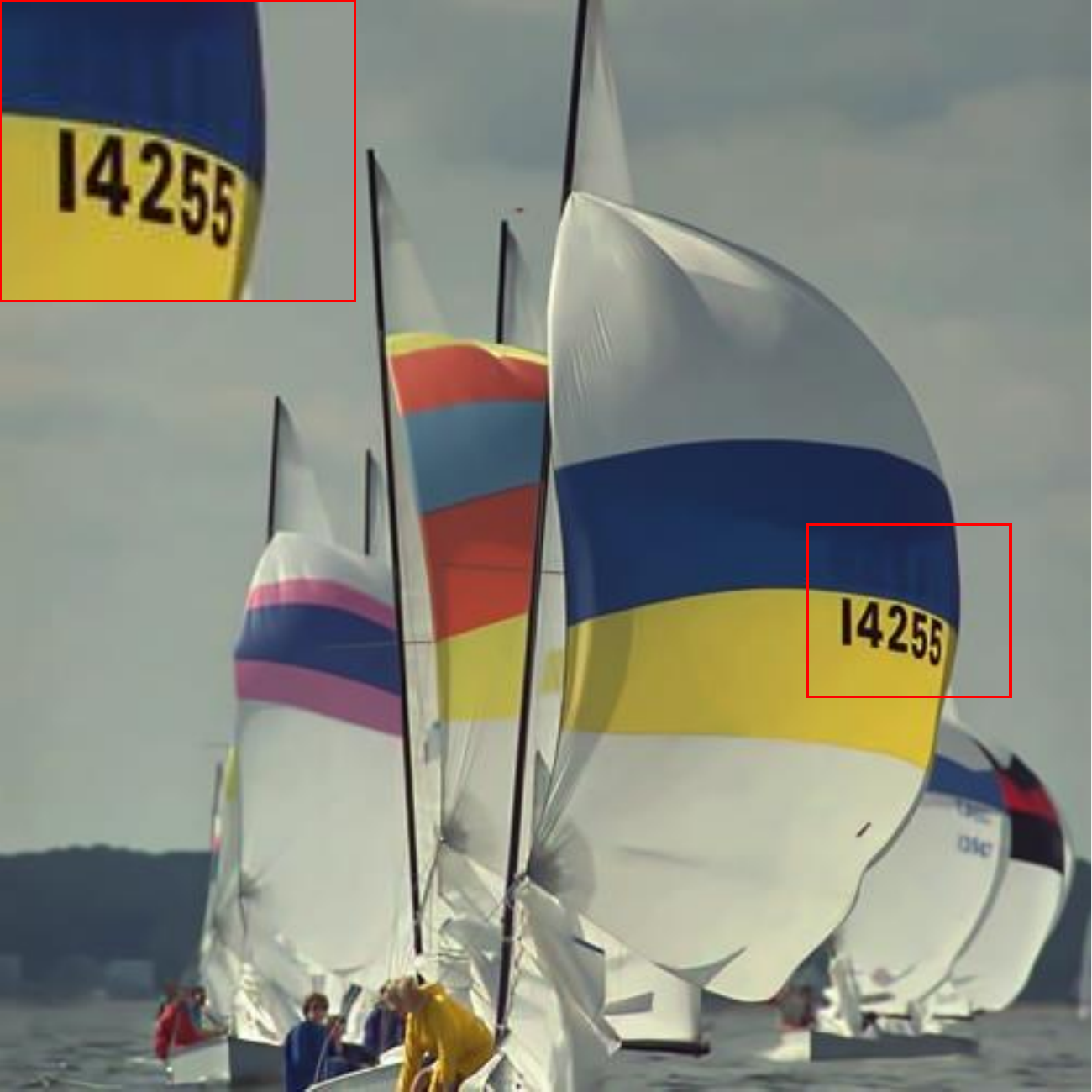}}
    \subfigure[RedNet-30]{\includegraphics[width=.24\textwidth]{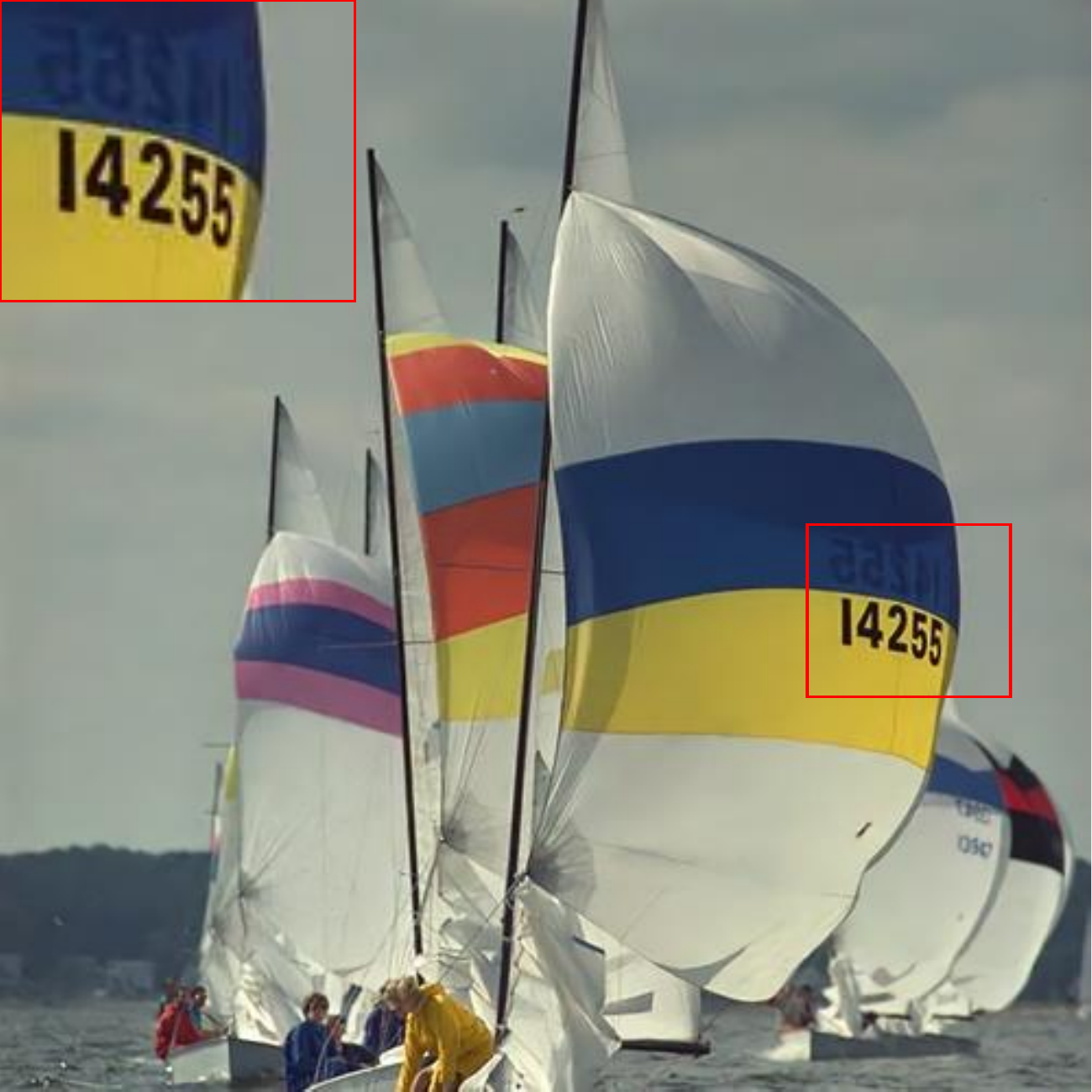}}
    \subfigure[DIP]{\includegraphics[width=.24\textwidth]{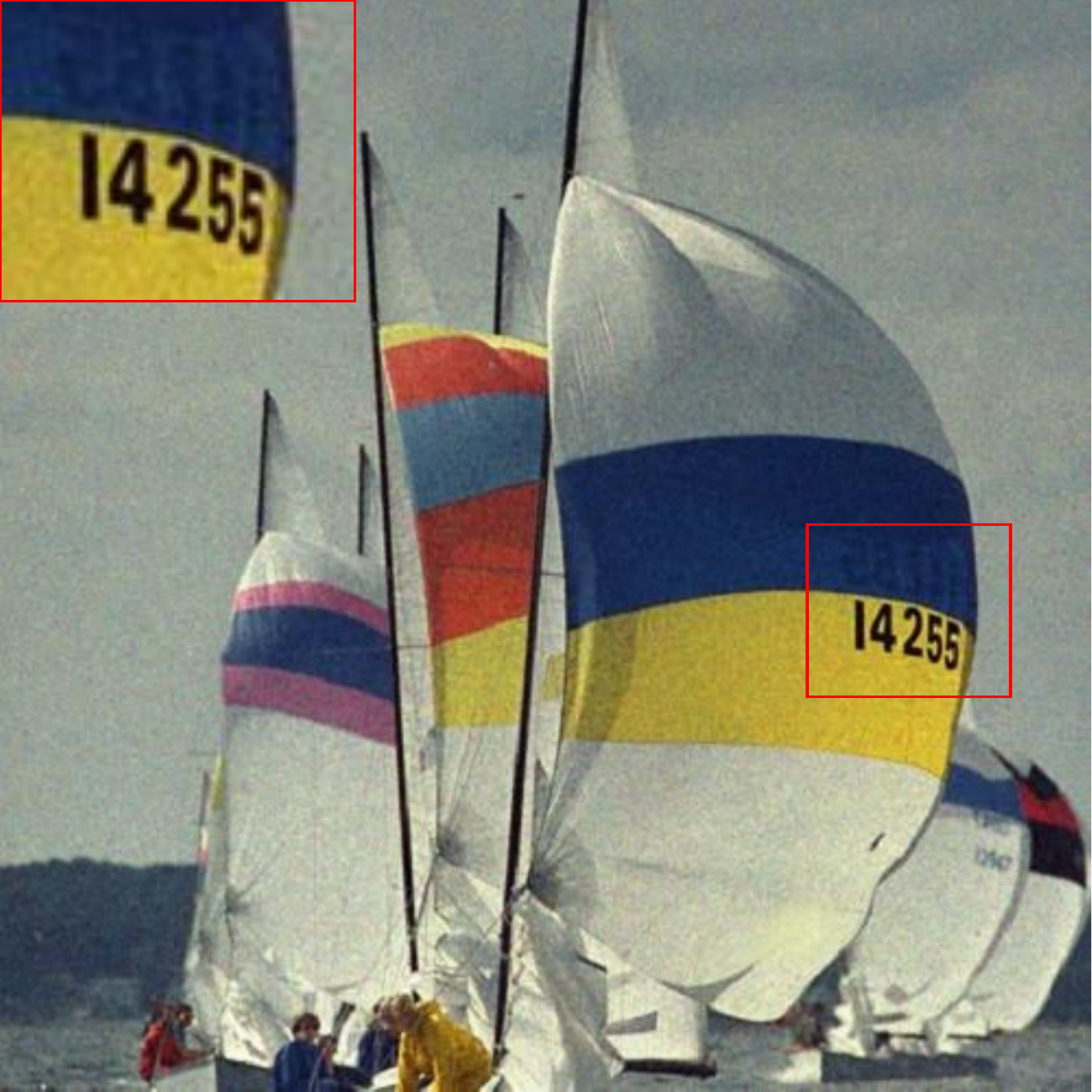}}
    \subfigure[LIR]{\includegraphics[width=.24\textwidth]{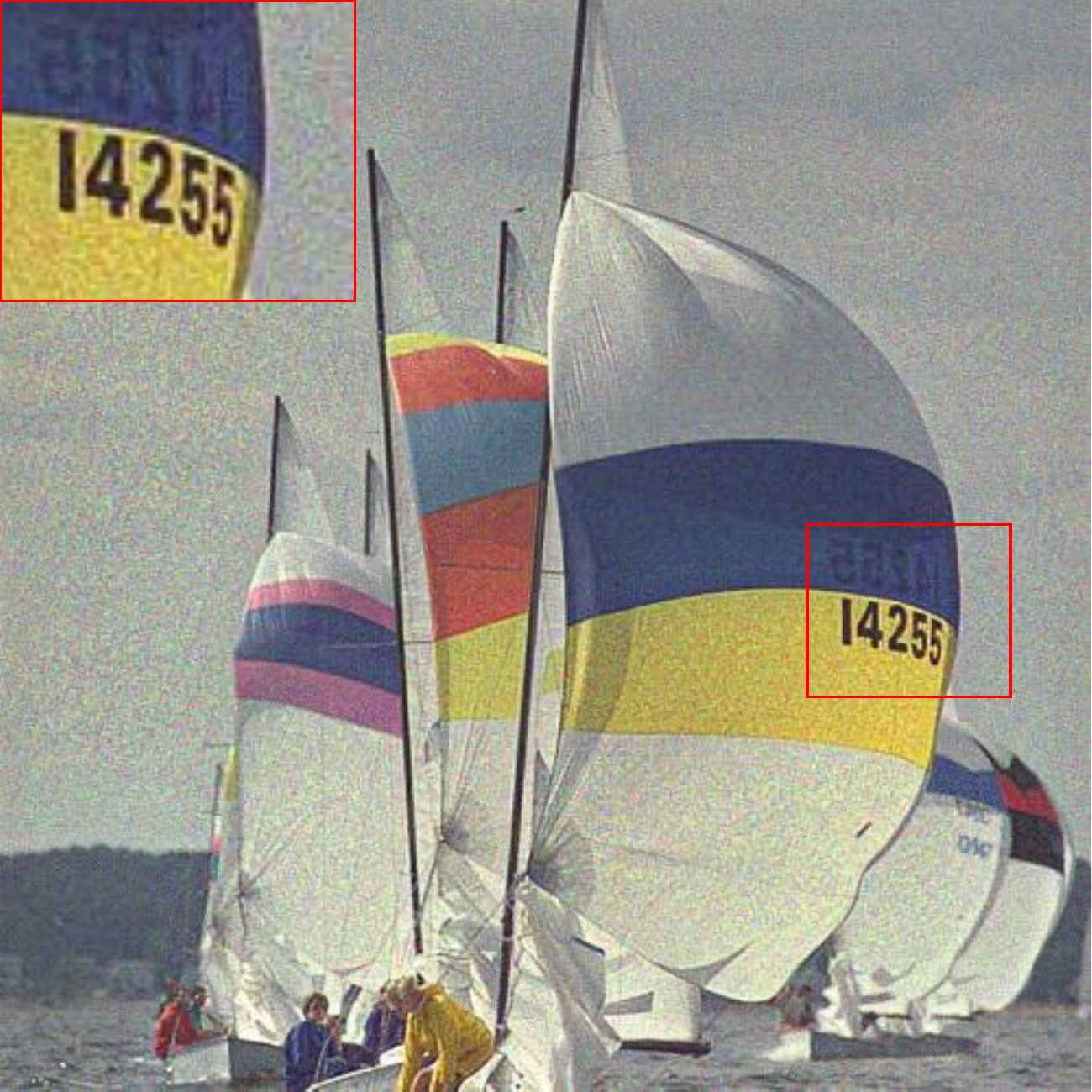}}
    \subfigure[Ours]{\includegraphics[width=.24\textwidth]{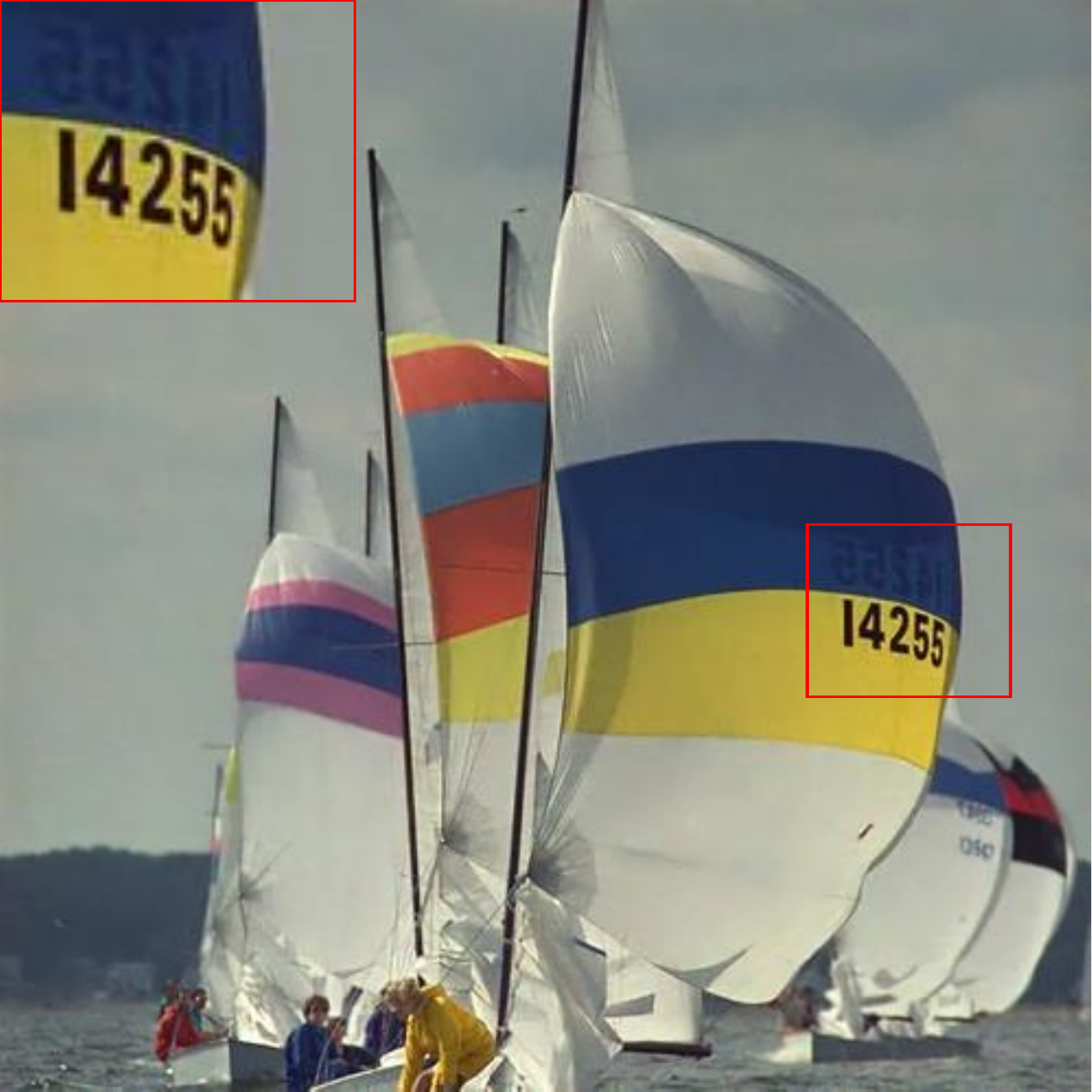}}
    \subfigure[GT]{\includegraphics[width=.24\textwidth]{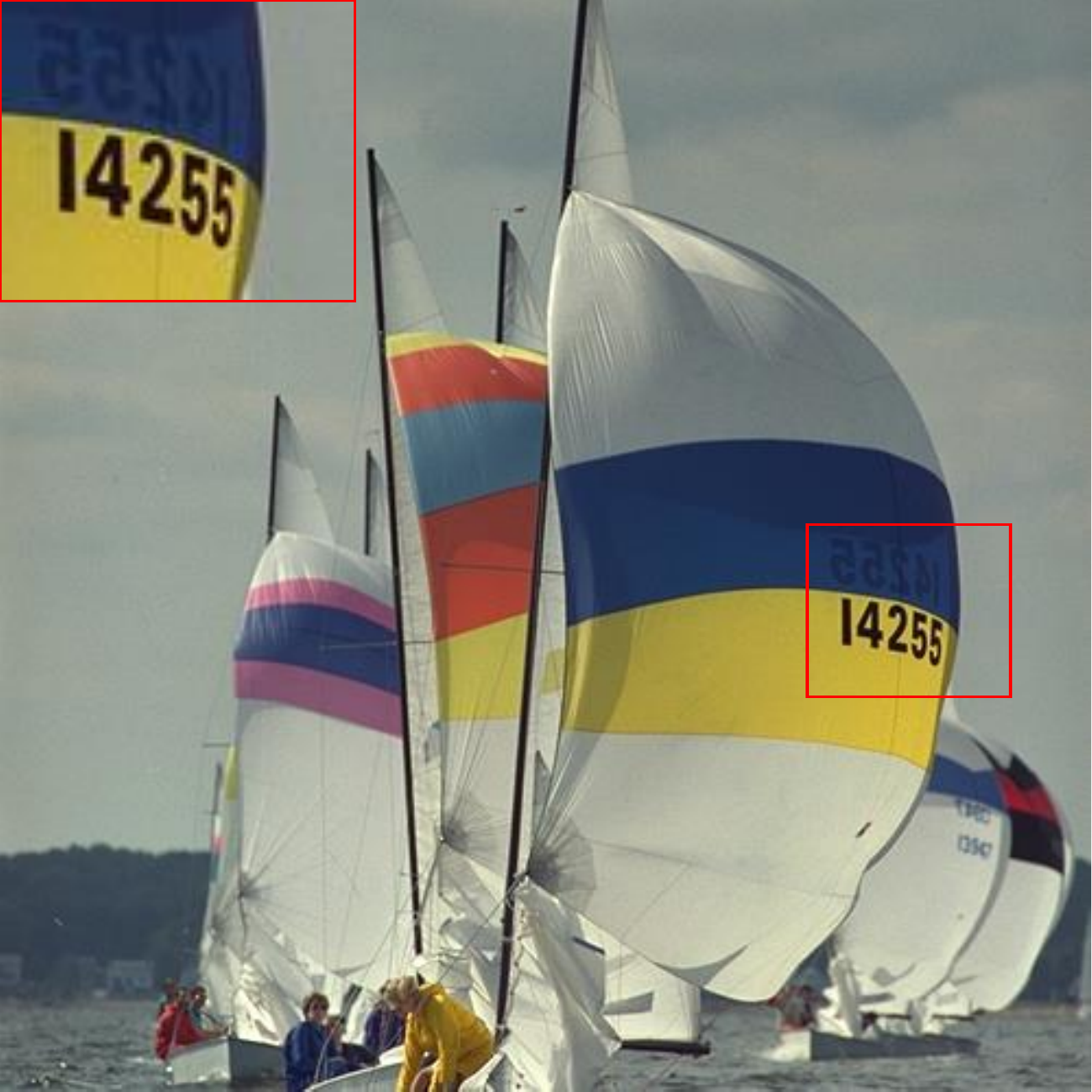}}
    \caption{Qualitative results of our method and other baselines on \textit{Kodak24} corrupted by Poisson noise.}
    \label{fig:poisson_qual1}
\end{figure*}

\begin{figure*}[htbp]
    \centering
    \subfigure[Input]{\includegraphics[width=.24\textwidth]{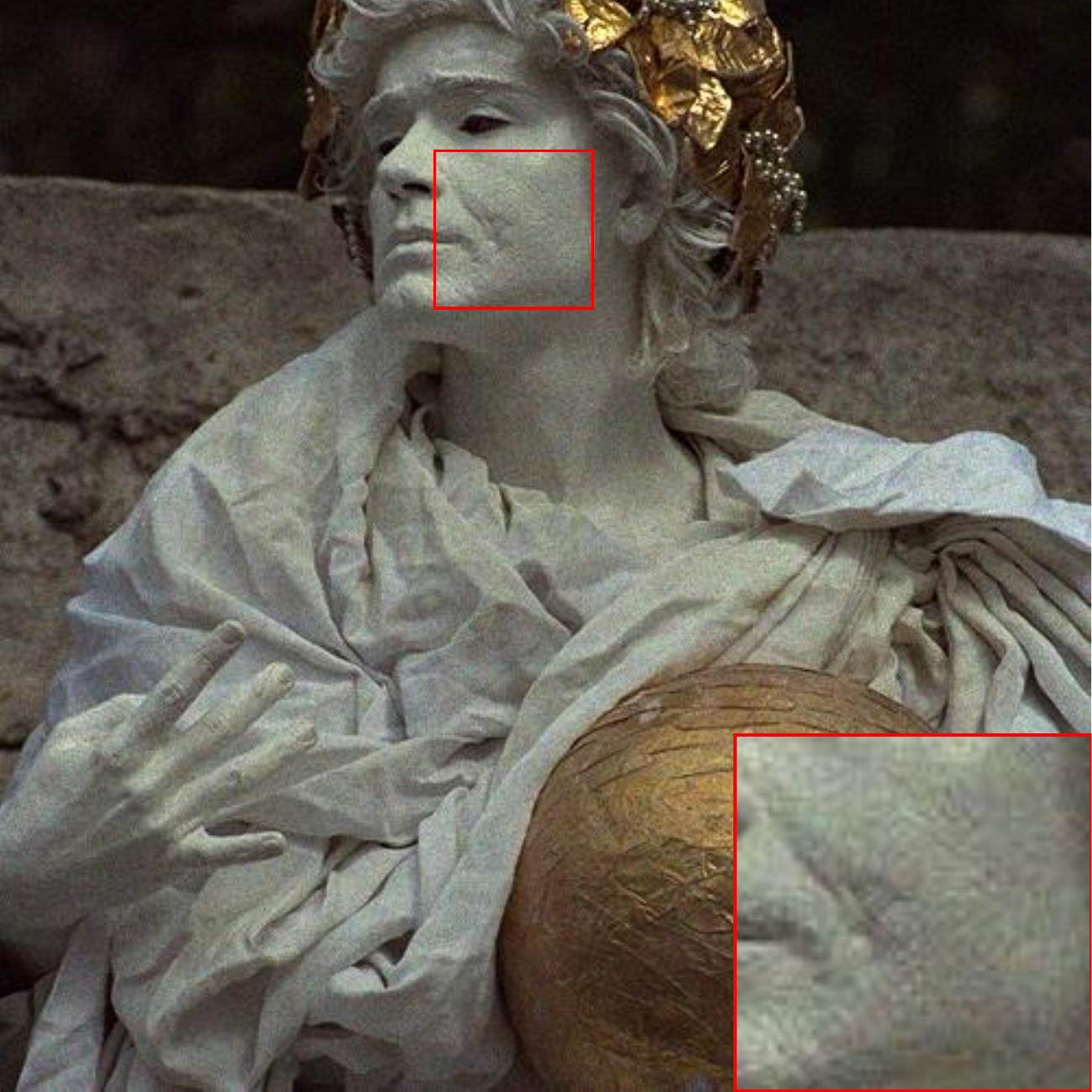}}
    \subfigure[CBM3D]{\includegraphics[width=.24\textwidth]{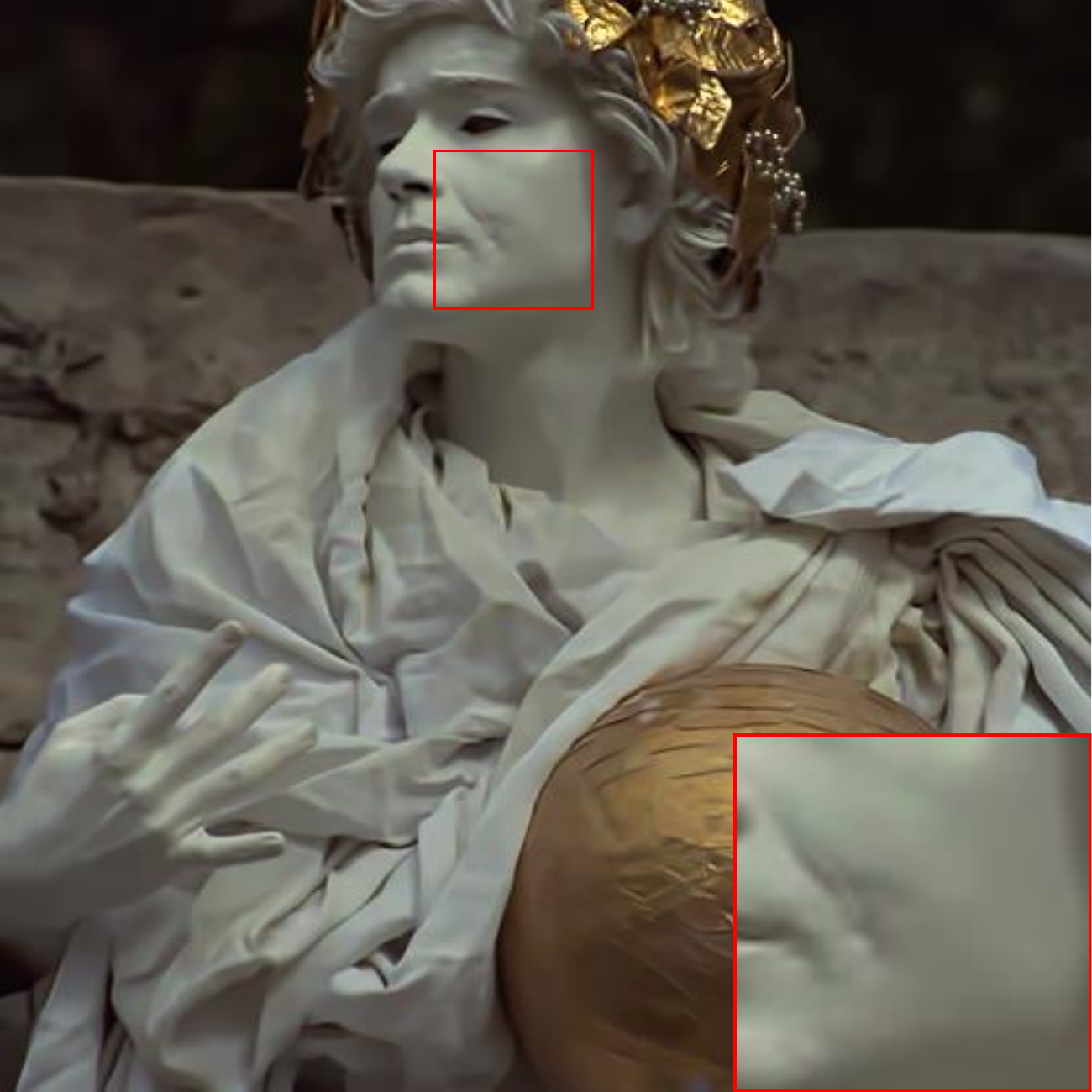}}
    \subfigure[RedNet-30]{\includegraphics[width=.24\textwidth]{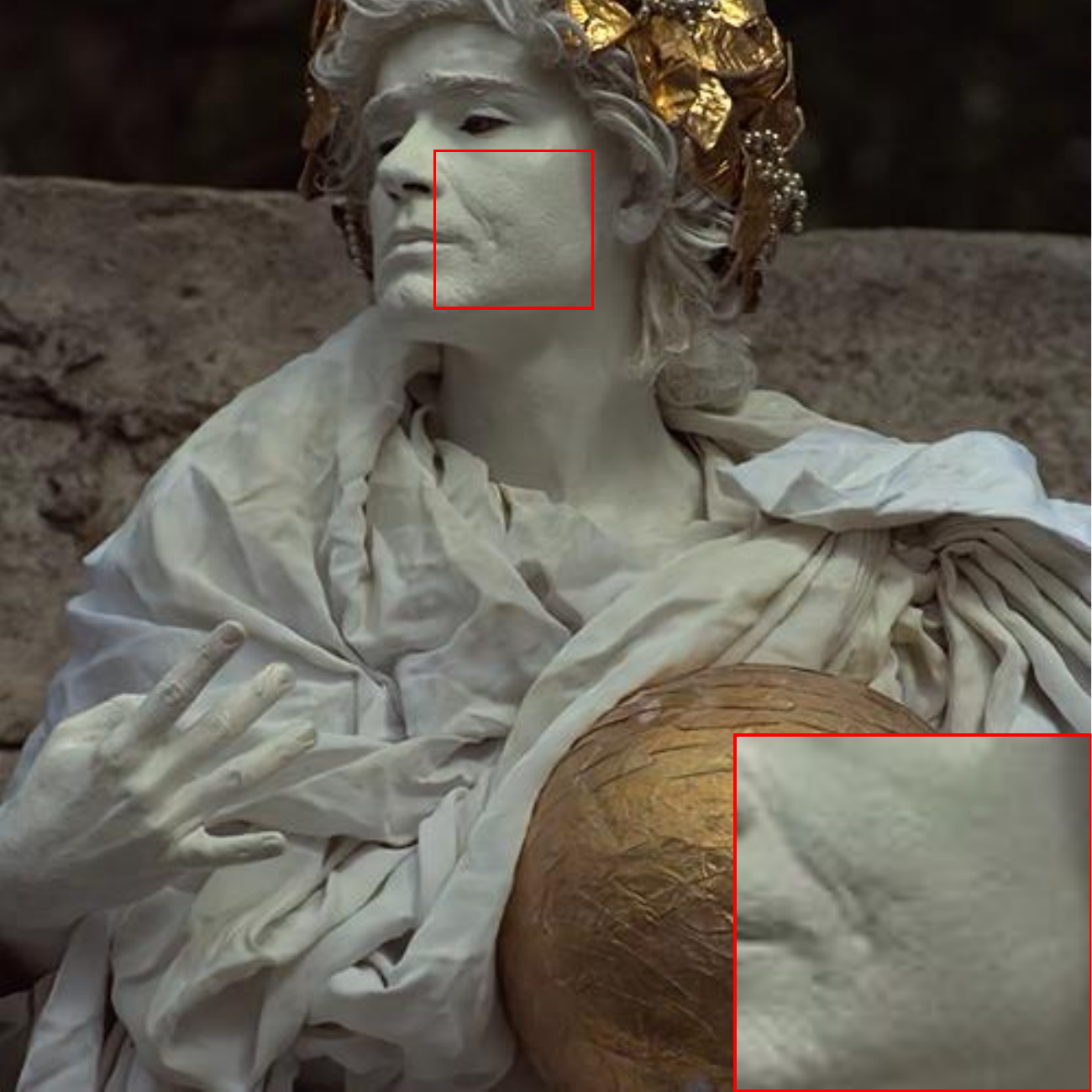}}
    \subfigure[DIP]{\includegraphics[width=.24\textwidth]{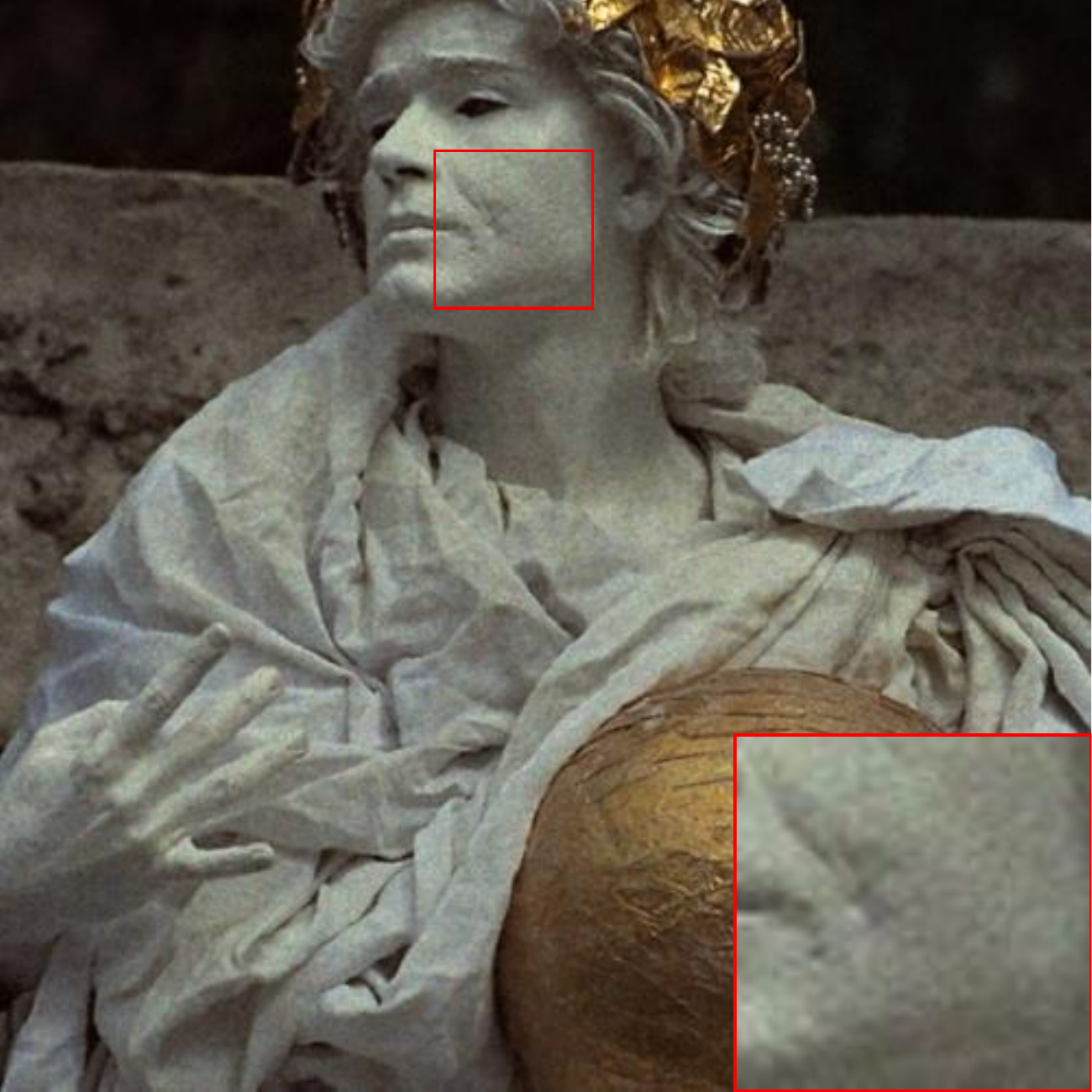}}
    \subfigure[LIR]{\includegraphics[width=.24\textwidth]{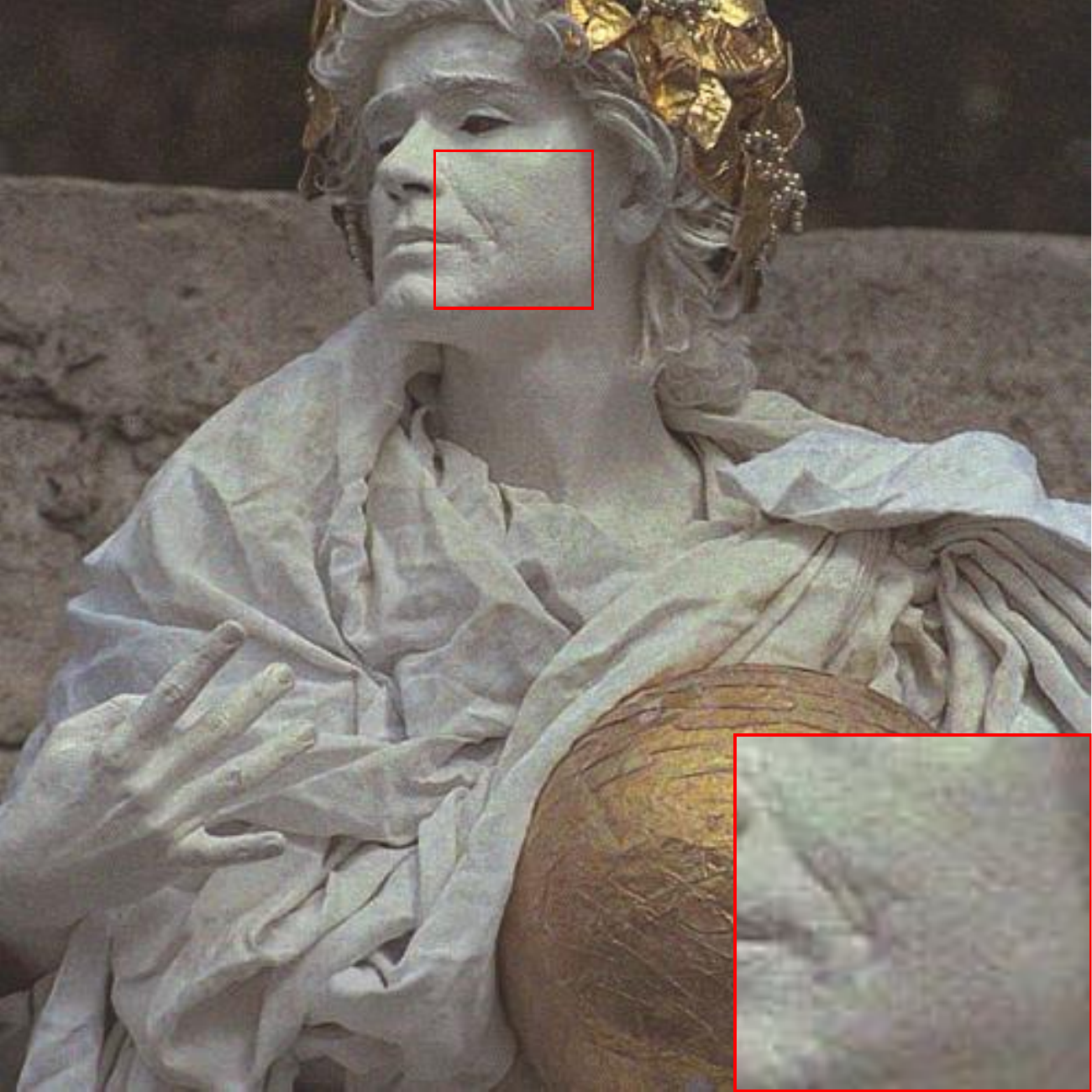}}
    \subfigure[Ours]{\includegraphics[width=.24\textwidth]{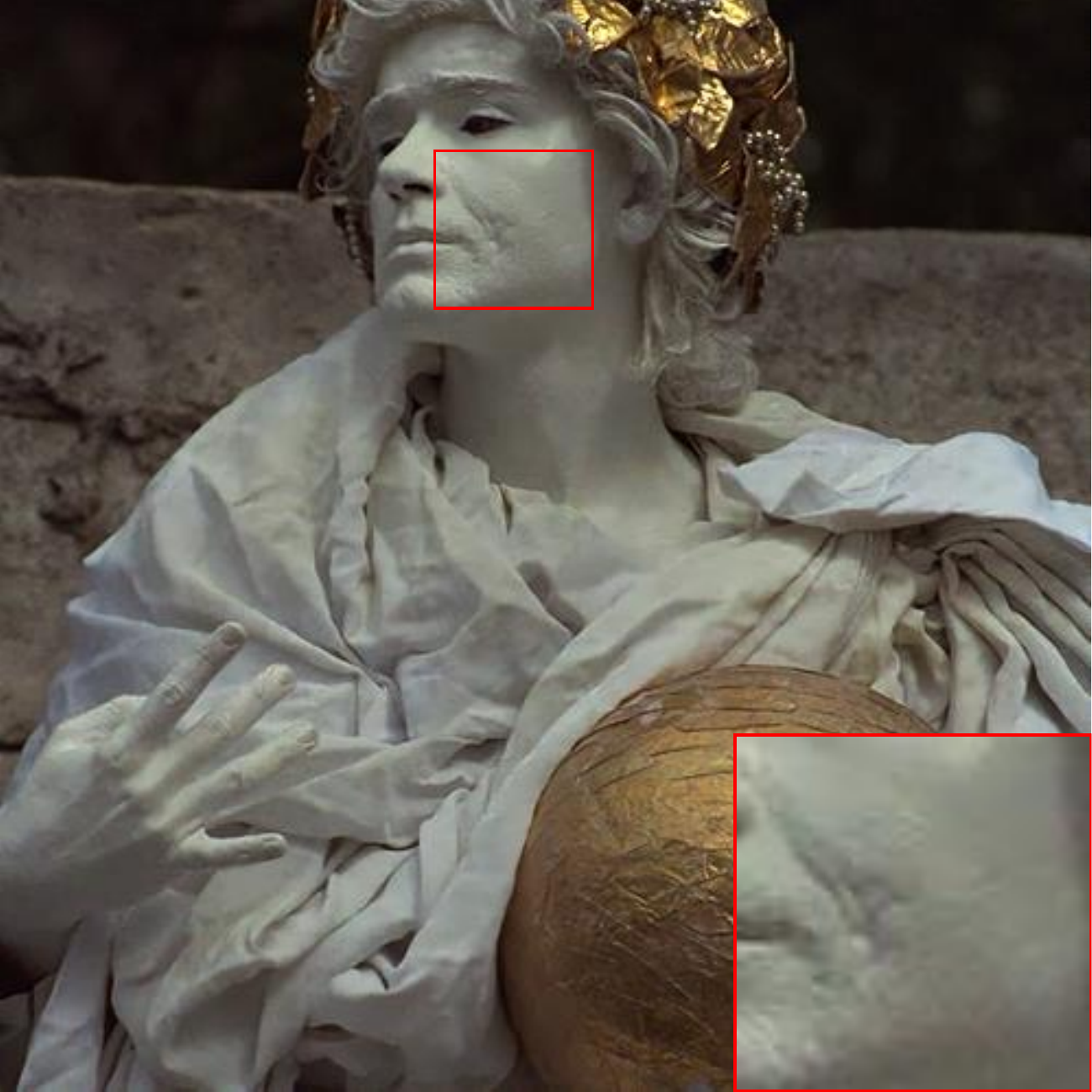}}
    \subfigure[GT]{\includegraphics[width=.24\textwidth]{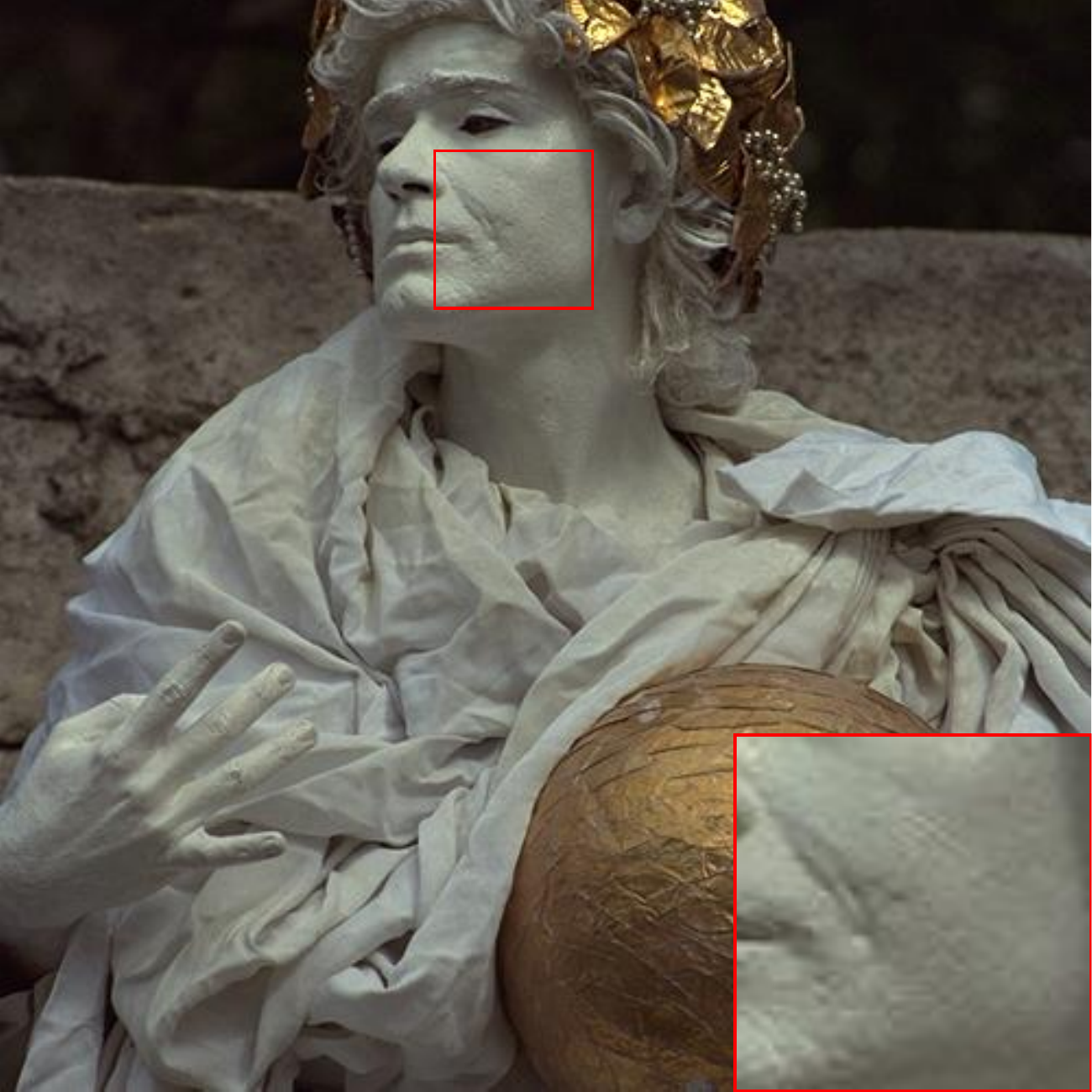}}
    
    \caption{Qualitative results of our method and other baselines on \textit{Kodak24} corrupted by Poisson noise.}
    \label{fig:poisson_qual2}
\end{figure*}

In the comparisons of Poisson noisy images, we use Kodak24 as the test dataset. The images are corrupted by independent Poisson noise from Scikit-image library \cite{van2014scikit}. We train the models following the settings in the main paper. The visualized results of Poisson noise removal are given in Figure \ref{fig:poisson_qual1} and \ref{fig:poisson_qual2}. Our approach shows impressive noise removal results. While LIR and DIP fail to remove the Poisson noise, our method successfully eliminates the noise and preserves the colors. In Table \ref{table:poisson}, our method achieves the best performance in PSNR and the second-best performance in terms of SSIM even when it is trained under the unpaired dataset. It demonstrates that our method has robustness and generalization against various noise types. Note that we do not change any hyper-parameters when trained under several types of noise.

\begin{table}[htbp]
\begin{center}
\resizebox{0.65\textwidth}{!}
{%
\begin{tabular}{|c|c|c|ccc|}
\hline
& \multicolumn{1}{c|}{Traditional} & \multicolumn{1}{c|}{Paired setting} &\multicolumn{3}{c|}{Unpaired setting} \\ \hline
\hline
Methods     & CBM3D \cite{bm3d} & RedNet-30 \cite{red30} & DIP \cite{ulyanov2018deep} & LIR \cite{lir} & Ours        \\ \hline
PSNR (dB)   & 32.36 & 29.59          & 29.59     & 26.20       & \textbf{34.93}       \\ \hline
SSIM        & 0.8694 & 0.9778         & 0.8774   & 0.7741      & \textbf{0.9691}      \\ \hline
\end{tabular}}
\end{center}
\caption{The average PSNR and SSIM results of different methods on \textit{Kodak24 dataset} corrupted by Poisson noise. Our results are marked in \textbf{bold}.}
\label{table:poisson}
\end{table}

\end{document}